\documentclass[twocolumn]{aastex61}

\newcommand\aastex{AAS\TeX}

\shorttitle{\aastex\ Atmosphere models from high-resolution observations}
\shortauthors{Cristaldi et al.}

\begin{document}

\title{1D ATMOSPHERE MODELS FROM INVERSION OF {Fe\,{\footnotesize I}} 630 nm OBSERVATIONS \\
WITH AN APPLICATION TO SOLAR IRRADIANCE STUDIES}

\correspondingauthor{Alice Cristaldi}
\email{alice.cristaldi@oaroma.inaf.it}

\author[0000-0002-9281-046X]{Alice Cristaldi}
\affil{INAF Osservatorio Astronomico di Roma \\
via Frascati 33 \\
Monte Porzio Catone, 00078, Italy}

\author[0000-0003-2596-9523 ]{Ilaria Ermolli}
\affil{INAF Osservatorio Astronomico di Roma \\
via Frascati 33 \\
Monte Porzio Catone, 00078, Italy}
\nocollaboration

\nocollaboration

\begin{abstract}

Present-day semi-empirical models of solar irradiance (SI) variations employ spectra computed on one-dimensional atmosphere models (1D models) representative of various solar surface features to reconstruct SI  changes measured on timescales greater than a day. Various recent studies have, however, pointed out that the spectra synthesized on 1D  models do not reflect the radiative emission of the inhomogenous atmosphere revealed by high-resolution solar  observations. We aimed to derive observational-based atmospheres from such  observations and test their accuracy for SI  estimates. 
We analysed spectro-polarimetric data of the {Fe\,{\footnotesize I}}  630 nm line pair on photospheric regions representative of the granular, quiet Sun pattern (QS)  and of  small- and large-scale magnetic features, both bright and dark with respect to the QS. The data were taken on 2011 August 6,  with the CRISP at the Swedish Solar Telescope, under excellent seeing conditions. We derived atmosphere models of the observed regions from data inversion with the SIR code. We studied the sensitivity of results to spatial resolution and temporal evolution, and discussed the obtained atmospheres  with respect to several 1D models. The atmospheres derived from our study agree well with most of the compared 1D models, both qualitatively and quantitatively (differences are within 10\%), but for pore regions.  Spectral synthesis computations  on the  atmosphere obtained from the QS observations  return SI between 400 nm and 2400 nm that agrees, on average, within  2.2\% with standard reference measurements, and within -0.14\% with the SI computed on the quiet Sun atmosphere employed by the most advanced semi-empirical model of SI variations.

\end{abstract}

\keywords{Sun: atmosphere --- Sun: magnetic fields --- Sun: photosphere --- Sun: faculae, plages --- Sun: sunspots}

\section{Introduction} \label{sec:intro}

The Solar Irradiance (SI) is the fundamental source of energy entering the Earth's system. Accurate knowledge of its variations is thus crucial
to understand the externally driven changes to the system, and, in particular, to 
 the regional and global Earth's 
climate  \citep[see e.g.][]{Haigh_2007,Solanki_etal2013ARAA}.
Regular monitoring of the total SI (TSI\footnote{The spectrally integrated solar radiative flux incident at the top of Earth's atmosphere at the mean distance of one astronomical unit.}) and of the spectral SI (SSI) in  the ultraviolet (UV), carried out since 1978 with satellite measurements, has shown  
that the SI varies 
on timescales from tens of seconds to decades, and on all spectral bands. 
In particular, available measurements show 
TSI variations 
 of $\approx$0.1\%  in phase with the 11-yr solar cycle, and of up 
$\approx$0.3\%  on the timescales of solar rotation. 
 It is worth noting that, on the whole,   30-60\% of the TSI variations over the solar cycle are produced at UV wavelengths  \citep{Lean_etal1997,Krivova_etal2006} that, over the same period, change up to  100\% and even more  \citep[e.g.][and references therein]{Frohlich_2013,Rottman_2006,Kopp_2016}. 
UV SSI variations  can have a significant impact on the Earth's climate system. Indeed, the SSI below 400 nm takes  an active part in governing the chemistry and dynamics of the Earth's upper stratosphere and mesosphere, by affecting production, dissociation, and heating processes of ozone, oxygen and other components; this also implies changes in winds and atmospheric circulation  \citep[e.g.][and references therein]{Solanki_etal2013ARAA}.

In order to accurately estimate effects of SI variations on the Earth's system, climate models  require long and precise series of SI data. Due to short duration and difficult  calibration of the available measurements,   satellite records however still suffer uncertainties, e.g. on  the  TSI trends measured on timescales longer than the 11-yr cycle and on the SSI changes occurring at some spectral bands 
 \citep[][]{Ermolli_etal2013,Solanki_etal2013ARAA}. 
In addition to improve  our understanding of the physical processes responsible for the measured SI changes, 
precise models of SI can also support the analysis of the existing SI records for Earth's climate studies, by allowing to interpret, complement, and extend available data series. 

Models that ascribe variations in SI at timescales greater than a day to solar surface magnetism are particularly successful in reproducing existing SI observations \citep[e.g.][]{Domingo_etal2009}.  
There are two classes of such models, called proxy and semi-empirical \citep[][]{Ermolli_etal2013,Yeo_etal2014,Yeo_etal2014b}. 
The former class of  SI models combine proxies  of solar surface magnetic features using regressions to match observed TSI changes. The proxies  most frequently used are the photometric sunspot index and the 
chromospheric Mg\,{\footnotesize II} index, to describe the sunspot darkening and facular brightening, respectively. 
The semi-empirical SI models   reproduce SI variations by summing up the contributions to SI of the different  features observed on the solar disc in time.  For each time and observed feature, they employ the surface area and position covered by the feature at the given time, and its time-invariant brightness as a function of wavelength and  position on the solar disc. The latter quantity is calculated from the spectral synthesis performed under some assumptions on  semi-empirical, one-dimensional, plane-parallel, static atmosphere models (hereafter 1D models) representative of the observed  feature \citep[see, e.g.][]{Ermolli_etal2013,Yeo_etal2014,Yeo_etal2014b}. 
Examples of the 1D models  employed in SI reconstructions are the ones presented by \citet[][]{Vernazza_etal1981}, \citet[][]{Fontenla_etal1993,Fontenla_etal1999,Fontenla_etal2009,Fontenla_etal2011,Fontenla_etal2015}, \citet[][]{Kurucz_1993,Kurucz_2005}, and \citet[][]{Unruh_etal1999}.

Present-day, most advanced  semi-empirical SI models   \citep[e.g. SATIRE-S,][]{Yeo_etal2014b} replicate more than  95\% of the TSI variability measured over cycle 23 and most of the SSI changes detected on rotational timescales, especially between 400 and 1200 nm. 
Despite the excellent match of modeled  to measured SI, current semi-empirical SI models still need 
improvements to overcome some limitations due to e.g. application of  free parameters and of simplifying assumptions. Besides, from computations of the radiative transfer (RT) in atmospheres resulting from magneto-hydrodynamic  simulations,
it was shown that ``a one-dimensional atmospheric model that reproduces the mean spectrum of an inhomogeneous atmosphere necessarily does not reflect the average physical properties of that atmosphere and is therefore inherently unreliable'' \citep{Uitenbroek_Criscuoli_2011}. This casts doubts on  the accuracy of 1D models employed in SI reconstructions, particularly  to account for  the radiant properties of the  small-scale features observed on the solar disc \citep[][]{Uitenbroek_Criscuoli_2011,Criscuoli_2013,Yeo_etal2014,Yeo_etal2014b}.

In this paper, we 
derive atmosphere models of various solar photospheric features from inversion of spectro-polarimetric observations, and discuss the results obtained  with respect to the 1D models most widely employed in SI reconstructions, and other 1D models  derived from spectro-polarimetric data. 
 In the following sections we describe the observations and data analysed in our study, and the methods applied  (Sect. 2). Then we present the results derived from the data inversion (Sect. 3) and discuss them with respect to 1D models in the literature (Sect. 4). Finally, we investigate the accuracy of using the obtained models in SI reconstructions (Sect. 5), discuss the results obtained from our study and draw our conclusions (Sect. 6).

\section{Data and Methods}
\subsection{Observations}

  \begin{figure*}
\centering
{
\includegraphics[scale=.7]{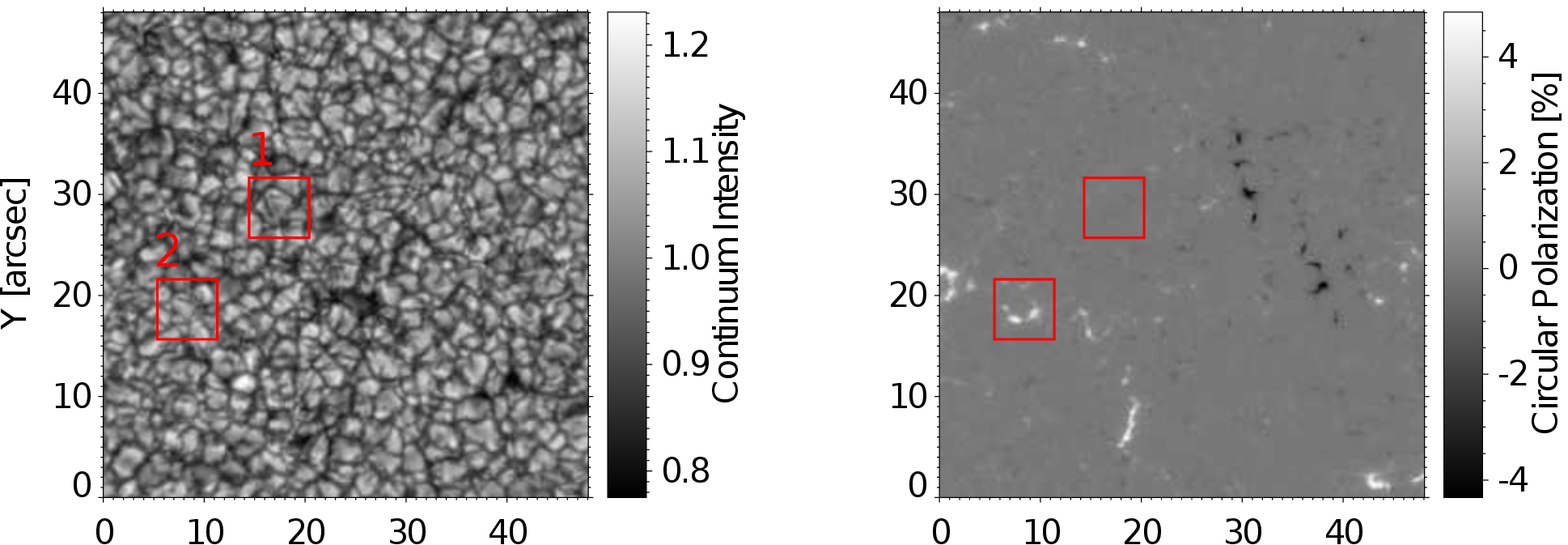}\\
\includegraphics[scale=.7]{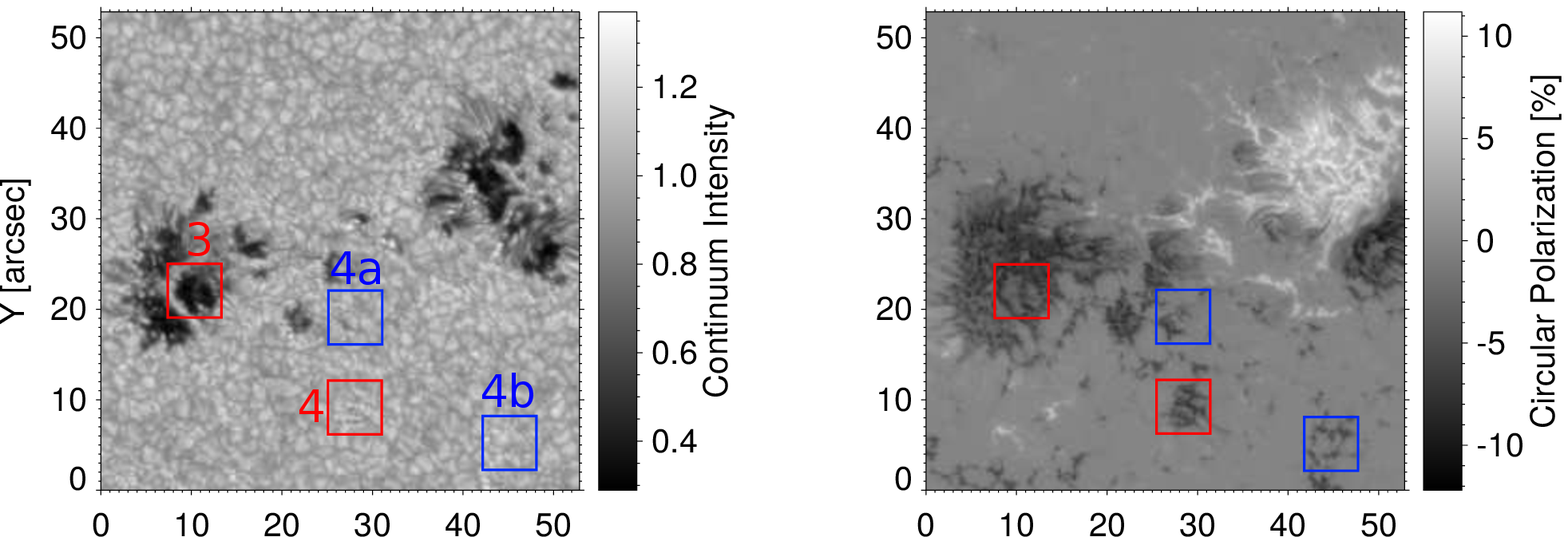}\\
\includegraphics[scale=.7]{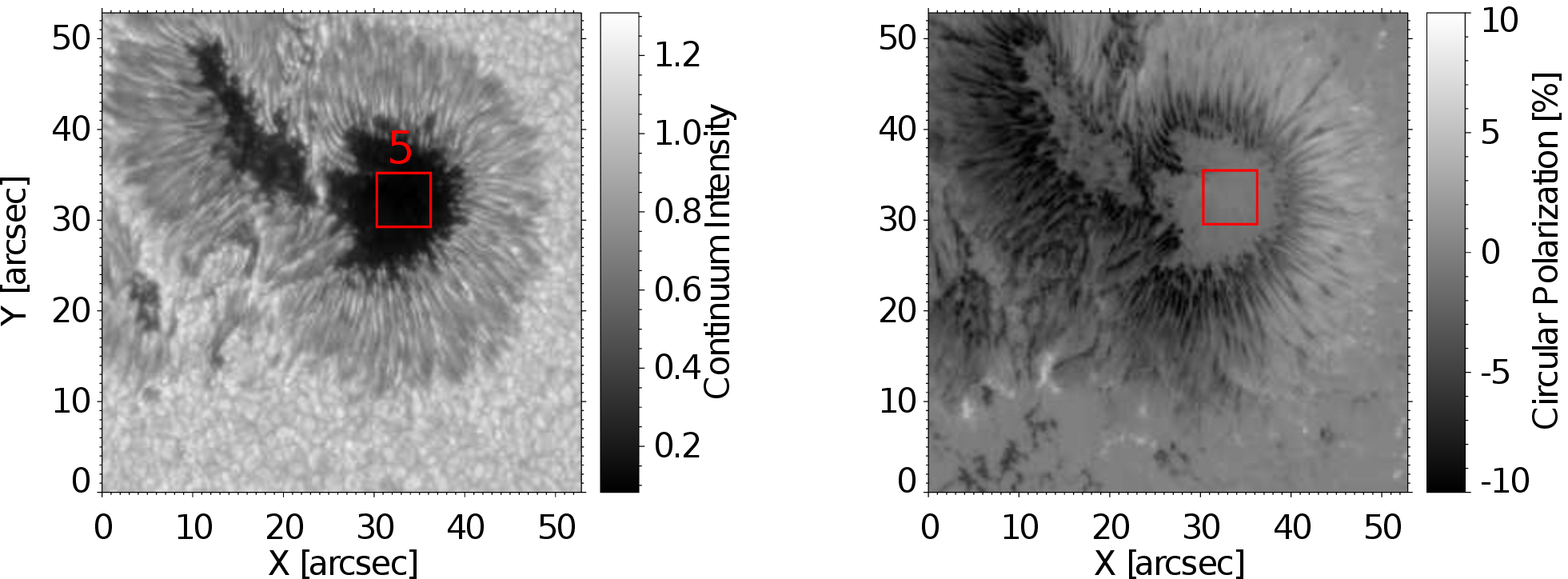}
}
\caption{Example of the observations and subFOVs analysed in our study. Continuum image (left) and circular polarization map (right) of the studied QS (top), active region AR1 (middle) and mature-spot AR2 (bottom). The red boxes in each panel show the inverted subFOVs, representative of unmagnetized (quiet, QS), bright points (BPs), plage (PL), pore (PO), and umbral (UM) regions, labelled  1, 2, 3, 4, and 5, respectively. Blue boxes in AR1 mark  two more plage regions (labeled 4a and 4b) also analysed in our study and discussed in Sect. 5.}
\label{fig1_fov}
 \end{figure*}

The data analysed in our study were acquired on 2011 August 6, from 07:57 UT to 10:48 UT, with the CRisp Imaging Spectropolarimeter \citep[CRISP,][]{Scharmer2008} at the  Swedish 1-m Solar Telescope \citep[SST,][]{Scharmer_etal2003b}. They consist of  full-Stokes spectro-polarimetric measurements derived from a  30 wavelengths scan of the  photospheric {Fe\,{\footnotesize I}} doublet, from 630.12 nm to 630.28 nm, over a  field-of-view (FOV) of $\approx$57$\times$57 arcsec$^2$, at three disc positions. The 30 wavelengths  spectro-polarimetric scans  were taken with a cadence of 28 s and a spectral sampling of $\approx$0.0044 nm.  
The above data are complemented with simultaneous and co-spatial chromospheric broadband images taken at the core of the {Ca\,{\footnotesize II}} H line at 396.9 nm; in this study, these data were employed to check our identification of the bright magnetic regions described in the following. The observations were assisted by the adaptive optics system of the SST \citep[][]{Scharmer_etal2003b}, under excellent seeing conditions. 

The pixel scale  of the analysed observations is $\approx$0.059 arcsec/pixel. 
 The polarimetric sensitivity of the analysed data, which was estimated as the standard deviation of the Stokes Q, U, and V profiles in the continuum, is $< $3.3$\times$10$^{-3}$ of the continuum intensity for all the Stokes parameters.   

The observations targeted  a quiet Sun (QS) region at disc center, the  active region (AR) NOAA 11267 (AR1) consisting of two sunspots of opposite polarity at disc position [S17, E24, cosine of the heliocentric angle $\mu$=0.84], and a mature spot in AR NOAA 11263 (AR2) at disc position [N16, W43, $\mu$=0.76].
The data of the three above regions were also analysed by \citet[][]{Stangalini_etal2015}, \citet[][]{Cristaldi_etal2014}, and \citet[][]{Falco_etal2016}, respectively. More details about the analysed observations can be found  in the above papers. 

The  observations were processed with the standard reduction pipeline  \citep[CRISPRED,][]{Delacruzrodriguez_etal2015}, to compensate data for the dark and flat-field response of the CCD devices, and for instrument- and telescope-induced polarisations. They  were also restored for seeing-induced  degradations, by using the 
 Multi-Object Multi-Frame Blind Deconvolution technique \citep[MOMFBD,][and references therein]{vanNoort_etal2005}.  
 
We analysed all the data available for the three observed regions, that means series of  79, 101, and 117  sequences of measurements taken over 47, 37, and 56 minutes for the QS, AR1, and AR2 regions, respectively. 
We extracted sub-arrays (hereafter referred to as subFOV) of 100$\times$100 pixels  representative  of quiet Sun regions (QS), small-scale bright magnetic regions such as bright points and network (BPs),  large-scale, bright regions with strong magnetic field as plages (PL), small-scale and large-scale dark magnetic regions as pores (PO) and umbrae (UM), respectively.  Each analysed subFOV represents a $\approx$6$\times$6 arcsec$^2$ region on the solar disc. This  region is of the same order as the elementary area considered when identifying bright and dark solar features in full-disc observations employed in  semi-empirical SI models. Indeed, the spatial resolution of analysed data ranges from $\simeq$ 8 to 1 arcsec, for earlier ground-based and more recent space-borne observations.


Figure  \ref{fig1_fov}  shows examples of the QS, AR1, and AR2 observations analysed in our study. For each region, we show the measured continuum intensity and   signed circular polarization (CP)  maps. The latter quantity has been computed, following \citet[][]{Requerey_etal2014}, as:
 $$CP = \frac{1}{10\langle{I_c}\rangle} \sum_{i=1}^{10} \epsilon_i V_i$$ $$\epsilon = [+1,+1,+1,+1,+1, -1, -1, -1, -1, -1]$$ 
where $\langle{I_c}\rangle$ is the continuum intensity averaged over the subFOV, V is the Stokes-V profile and \textit{i} runs over the 10 spectral points closer to the core of the {Fe\,{\footnotesize I} line at 630.25 nm. 
In the weak field regime, the CP can be considered as a proxy for the longitudinal component of the magnetic field \citep{Landi_Landolfi_2004}.

 The red  boxes in the various panels of Fig. \ref{fig1_fov} show the subFOVs considered in the following to represent  the physical properties of QS, BPs, PL, PO, and UM regions. 
  The blue boxes in the middle panels of  Fig. \ref{fig1_fov}  show two more PL  regions also analysed in our study and discussed in Sect. 5.  

\subsection{Semi-empirical 1D atmosphere models}
\label{1D}

\begin{deluxetable*}{llll}
\tablecaption{1D atmosphere models considered for comparison.\label{tab:table1}}
\tablecolumns{4}
\tablenum{1}
\tablewidth{0pt}
\tablehead{
\colhead{Model label} & \colhead{Reference} & \colhead{Atmosphere model} & \colhead{Observed region} \\}
\startdata
  HSRA & Gongerich et al. (1971) & average quiet Sun & QS\\
  VAL-C      & Vernazza et al. (1981) & average quiet Sun & QS\\
 Maltby & Maltby et al. (1986) & quiet Sun & QS\\
 FAL-(A, C)-93 & Fontenla et al. (1993) &faint   cell interior, average cell interior & QS\\
 FAL-(A, C)-99 &  Fontenla et al. (1999) & faint   cell interior, average cell interior & QS \\
 FAL-(C)-06 &   Fontenla et al. (2006) & quiet Sun  cell interior & QS \\
 FAL-(A, B)-11 &    Fontenla et al. (2011) &dark QS internetwork, QS internetwork & QS \\
  FAL-(A, B)-15 & Fontenla et al. (2015) & network, enhanced network, plage, bright plage  & BPs, PL\\
 FAL-(F, P)-93 & Fontenla et al. (1993) & network, enhanced network, plage, bright plage  & BPs, PL\\
 FAL-(E, F, H, P)-99 & Fontenla et al. (1999) & network, bright network, plage, bright plage  & BPs, PL\\
 FAL-(E, F, H, P)-06 & Fontenla et al. (2006) & network, active network, plage, bright plage  & BPs, PL\\
 FAL-(D, F, H, P)-11 & Fontenla et al. (2011) & network, enhanced network, plage, bright plage  & BPs, PL\\
  FAL-(D, F, H, P)-15 & Fontenla et al. (2015) & QS network lane, enhanced network, plage, very bright plage  & BPs, PL\\
 SOLANNT & Solanki (1986)  & network & PL \\
  SOLANPL & Solanki et al. (1992) & plage & PL \\
COOL  & Collados et al. (1984) & cool (large) spot & PO, UM\\
HOT  & Collados et al. (1984) & hot (small) spot & PO, UM\\
Maltby-M  & Maltby et al. (1986) & umbral core &  PO, UM\\
FAL-S-99 & Fontenla et al. (1999) &  umbra &  PO, UM\\
FAL-(S, R)-06 & Fontenla et al. (2006) &  umbra, penumbra &  PO, UM\\
FAL-(S, R)-11 & Fontenla et al. (2011) &  umbra, penumbra &  PO, UM\\
 FAL-(S, R)-15 & Fontenla et al. (2015) &  umbra, penumbra &  PO, UM\\
\enddata
\end{deluxetable*}

To the purpose of discussing the results derived from the above observations, 
we analyzed several sets of 1D models 
presented in the literature.
In particular, we considered the atmosphere model presented by \citet[][]{Vernazza_etal1981} to represent QS regions (VAL-C),
and the sets of models by \citet[][]{Fontenla_etal1993,Fontenla_etal1999,Fontenla_etal2006,Fontenla_etal2011,Fontenla_etal2015} to describe various solar features, from the faint granular cell interior (FAL-A) and average cell (FAL-C) in QS, to network (FAL-E), enhanced network (FAL-F), plage (FAL-H), bright plage (FAL-P), penumbral (FAL-R), and umbral (FAL-S) regions. These latter models are employed in e.g.  the SRPM semi-empirical SI reconstructions \citep[][ and references therein]{Fontenla_etal2015}.  Notice that the  \citet{Fontenla_etal2015} models are neither discussed nor displayed in the following, since  their difference with respect to  previous models by \citet{Fontenla_etal2011} is not appreciable at the scale of the  plots and at the range of atmospheric heights considered in our study. We also analysed other available 1D models obtained  from inversion of spectro-polarimetric observations. In particular, we considered the SOLANNT and SOLANPL flux-tube models by \citet[][]{Solanki_1986} and \citet[][]{Solanki_etal1992} for network and plage regions, respectively, and the COOL and HOT models by \citet[][]{Collados_etal1994} for large and small spots, in the order given;  all these models are available in the SIR code described below.  Finally, we tested the results derived from our study also with respect to the Harvard-Smithsonian Reference Atmosphere \citep[HSRA,][]{Gingerich_etal1971} and the model by \citet[][]{Maltby_etal1986} for average QS  regions, and the M-model by \citet[][]{Maltby_etal1986}  for spots.
Table 1 
summarises all the 1D models analysed  in our study.

\newpage
\subsection{Stokes inversions}
\label{inv4}
We performed full-Stokes spectro-polarimetric local thermodynamic equilibrium (LTE) inversions of the available data for the selected subFOVs with the SIR code \citep[Stokes Inversion based on Response functions,][]{Ruizcobo_deltoro1992,Bellotrubio_2003}. 
We applied the code simultaneously to measurements of  the {Fe\,{\footnotesize I} lines at 630.15 nm and 630.25 nm, by excluding from  the calculation the Stokes-I measurements in the red wing of the {Fe\,{\footnotesize I} line at 630.25 nm affected by telluric blends. The SIR code utilizes  the atomic parameters taken from the VAL-D database \citep{Piskunov_1995}. 
For each analysed subFOV, we first normalized the measurements  to the 
average continuum intensity measured on a nearby QS region, defined as the region with CP signal lower than 3 times the standard deviation of the entire CP map.
 
 \begin{figure}
\centering
{
\includegraphics[scale=.44, trim=.7cm 8.cm 0cm 7cm,clip=true]{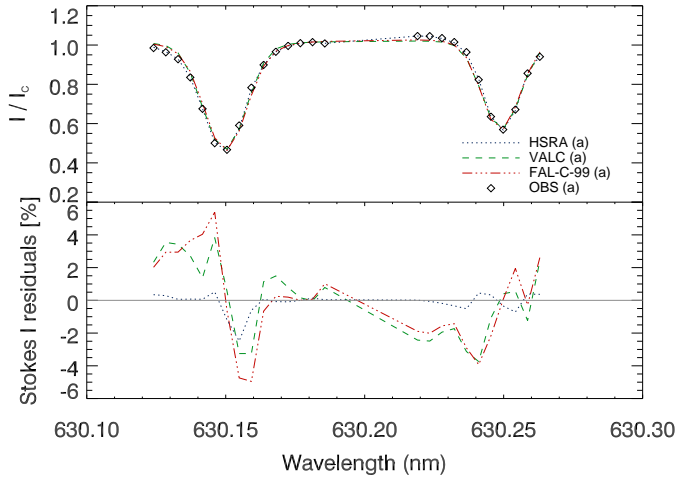}
}
\caption{Top panel: Stokes-I spectra from measurements (OBS) of QS regions and those from SA data inversion using different starting guess models, specifically the HSRA, VAL-C, and FAL-C-99 models. Bottom panel: relative difference between synthetic and observed spectra. }
\label{fig11_c4}
 \end{figure}

 \begin{figure}[h!]
\centering
{
\includegraphics[scale=.44, trim=.7cm 8.cm 0cm 7cm,clip=true]{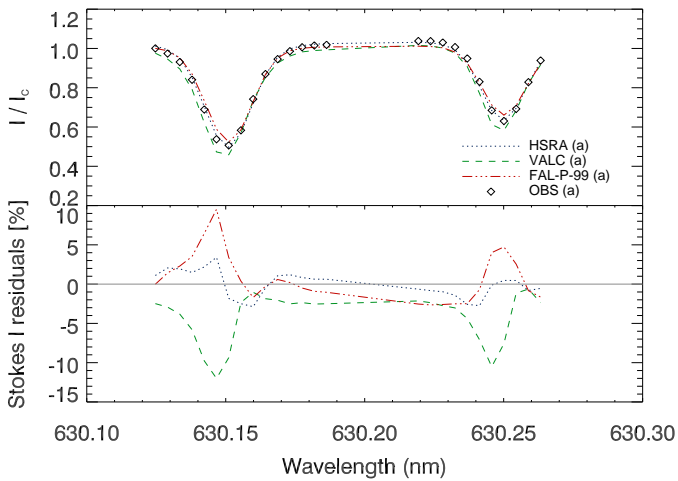}
}
\caption{As in Fig. \ref{fig11_c4} but for measurements (OBS) and SA data inversion results concerning the PL region. The HSRA, VAL-C, and FAL-P-99 models have been tested here as guess models.}
\label{fig14_c4}
\end{figure}

We performed the data inversion  by considering:   a) the mean spectra obtained from the spatial-average of the Stokes measurements taken over each analysed subFOV, and b) the individual Stokes measurements in each pixel of the analysed subFOV. 
In the latter case, we then spatially-averaged the results from the data inversion of the subFoV. These two methods are hereafter referred to as SA and FR, respectively; in the figures, results from SA and FR are labelled  (a) and (b), respectively. 
When applying SA, the spatial information in the analysed data is lost to  the advantage of an increased signal-to-noise ratio of the  Stokes data  to be inverted. When applying FR, the analysis takes advantage of the full spatial resolution of the analysed observations. The SA and FR computations were applied to investigate the effects  due to analysis method on the obtained results, and to spatial inhomogeneities due to smaller-scale features in the observed atmosphere.  SA and FR computations were applied to all analysed subFOVs. Besides, for the  PO data, the SA and FR computations were also applied by considering only the pixels belonging to the dark region in the subFoV. In particular, we a\-na\-ly\-sed the pixels characterized by I$_c<$0.4,  where I$_c$ is the normalized continuum intensity. In the figures, results from these latter  calculations are labeled (a)-dark and (b)-dark, respectively. 

We inverted the data  by assuming that the modeled atmosphere consists of one component with physical quantities that do not vary with atmospheric height, but temperature. 
This  assumption is justified by the  lack of asymmetries  in the analysed line profiles, which manifest  the presence of more than one atmosphere in the analysed resolution element or gradients in some physical parameters. Moreover, our assumption is also based on our aim of comparing the obtained results with 1D  models mostly constructed from spatially unresolved observations. 

We performed the data inversion by applying  two computational cycles.
 In the first  cycle, the temperature was allowed  to vary within 2 nodes, while the other quantities, specifically the  line-of sight (LOS) velocity, the magnetic field strength, the field inclination and  azimuth,  and the microturbulent velocity  were assumed to be constant with height. In the second cycle, we slightly increased the degrees of freedom, by allowing the temperature to vary within 3 nodes. According to \citet{Ruizcobo_deltoro1992} and \citet{Socasnavarro_2011}, the slight increase of nodes in the second cycle helps the code to improve convergence of calculation and to obtain a more stable solution. Since we performed one-component inversions, the magnetic filling factor is unity. We set the height-independent macroturbulent velocity to 2 km/s. 
Besides, we 
modeled the stray-light contamination on the data by averaging Stokes-I computed on subFOV regions with low polarization degree, which was defined as: 
$$ \Pi \equiv \frac{I_{pol}}{I} = \frac{\sqrt{Q^2 + U^2 + V^2}}{I} $$

We performed the data inversion by using various initial guess models. In particular, we considered the HSRA and models by \citet[][]{Vernazza_etal1981}, \citet[][]{Maltby_etal1986}, \citet[][]{Fontenla_etal1993,Fontenla_etal1999}, \citet{Solanki_1986}, and  \citet{Collados_etal1994}, as well as some their modified versions. Based on the best fitting and minimal residual between observed and inverted profiles,  we assumed the following starting guess models: 
for QS data, we adopted the HSRA; for BPs and PL data, we employed the same model but modified with a constant magnetic field strength value of 200 G and 800 G, respectively; 
for 
 PO and UM data, we assumed the HOT and COOL  models proposed by \citet{Collados_etal1994}. We modified these latter models by keeping the magnetic field strength constant with height and assigning 2000 G and 2500 G to the HOT and COOL  model, respectively. 
 
It is worth noting  that the SIR code performs  the data inversion under LTE assumption. Although almost all the {Fe\,{\footnotesize I}} lines show deviation from LTE conditions, it was shown 
\citep{Shchukina_2001} 
that lines synthesized under LTE conditions do not sensitively differ from the ones obtained under NLTE, especially if iron abundance is lower than $7.50 \pm   0.10$ dex, as it was in our calculations  (7.46 dex).
 
Figures  \ref{fig11_c4} and \ref{fig14_c4} show examples of results obtained by using the different starting guess models when inverting QS and PL data. In particular, the top panel of each figure shows the observed Stokes-I spectra and the synthetic ones derived from SA inversion with the various tested models. The bottom panel of each figure shows the relative difference between synthetic and observed profiles, expressed in percentage values.  With the guess models employed in our study (with the worse guess models tested in our study), for QS and BPs regions these residuals are within $\pm$1\% (6\%); for PO and UM regions they are within $\pm$4\% (6\%), and for the PL regions within $\pm$3\% (10\%).

\section{Results} 
 \label{res}
 
 \subsection{Atmospheric models}

\begin{figure*}
\centering
{
\includegraphics[scale=.3, trim=1.8cm 2.2cm  1.cm  .2cm,clip=true ]{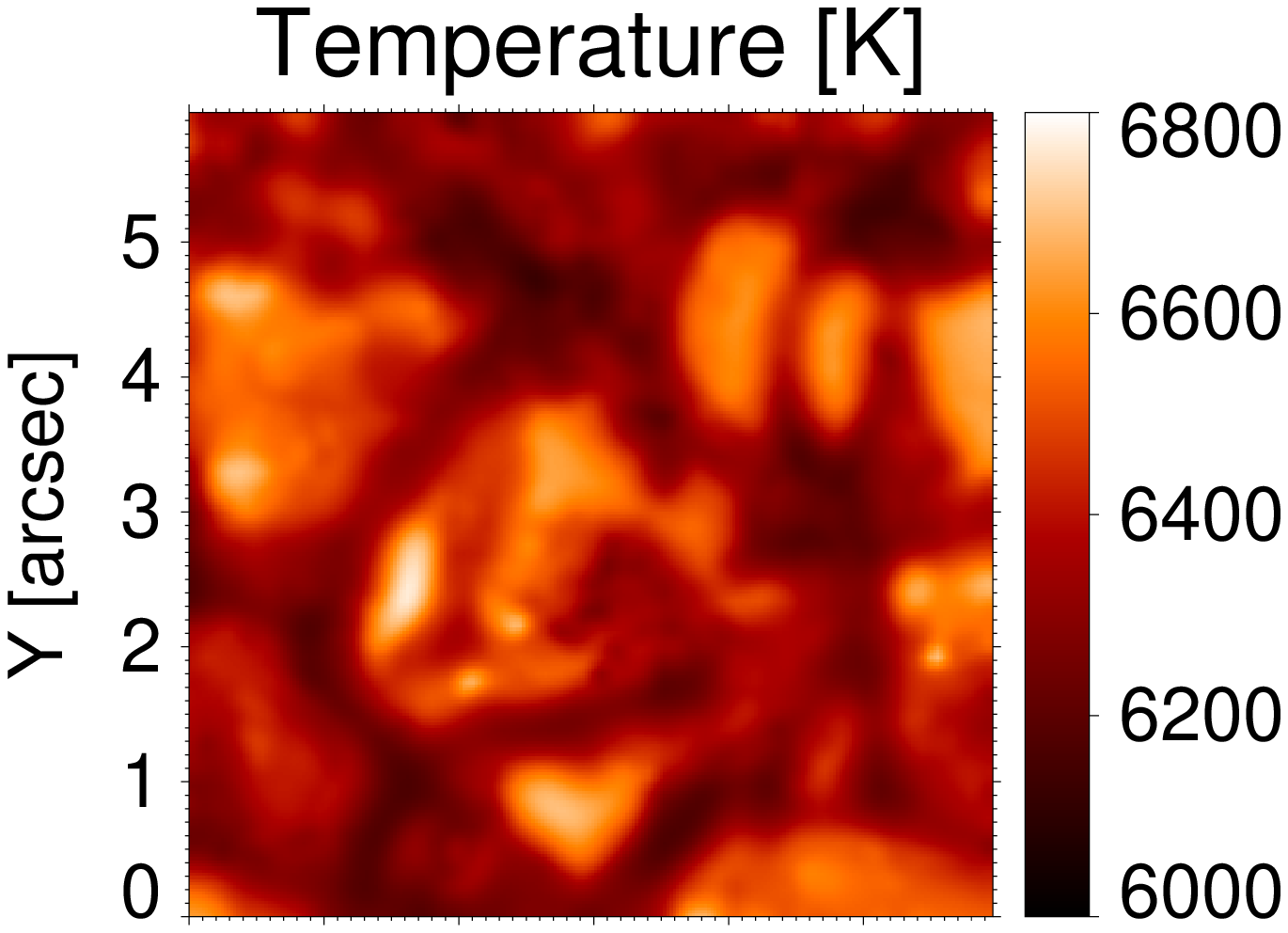}
\includegraphics[scale=.3, trim=3.6cm 2.2cm  1.6cm  .2cm,clip=true ]{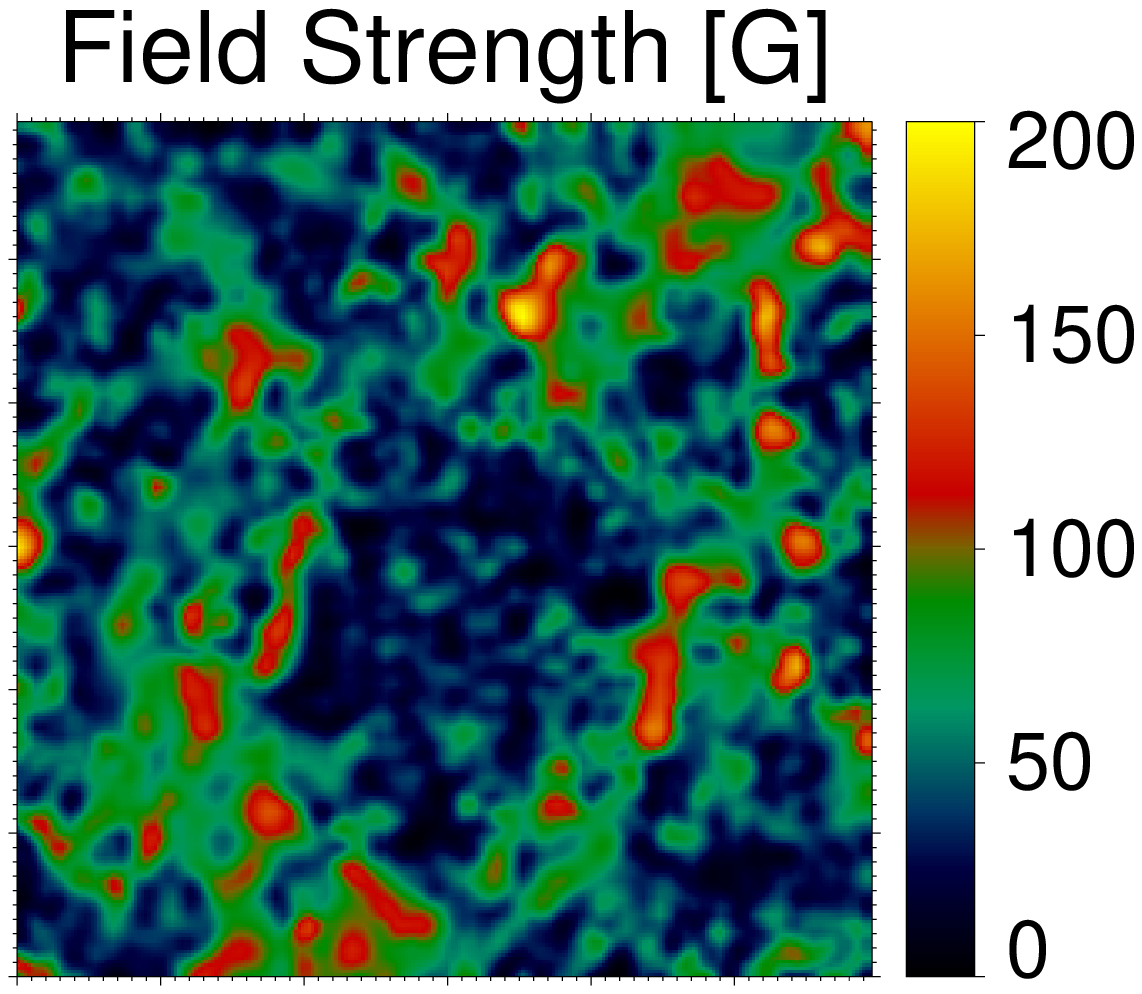} 
\includegraphics[scale=.3, trim=3.6cm 2.2cm  1.6cm  .2cm,clip=true]{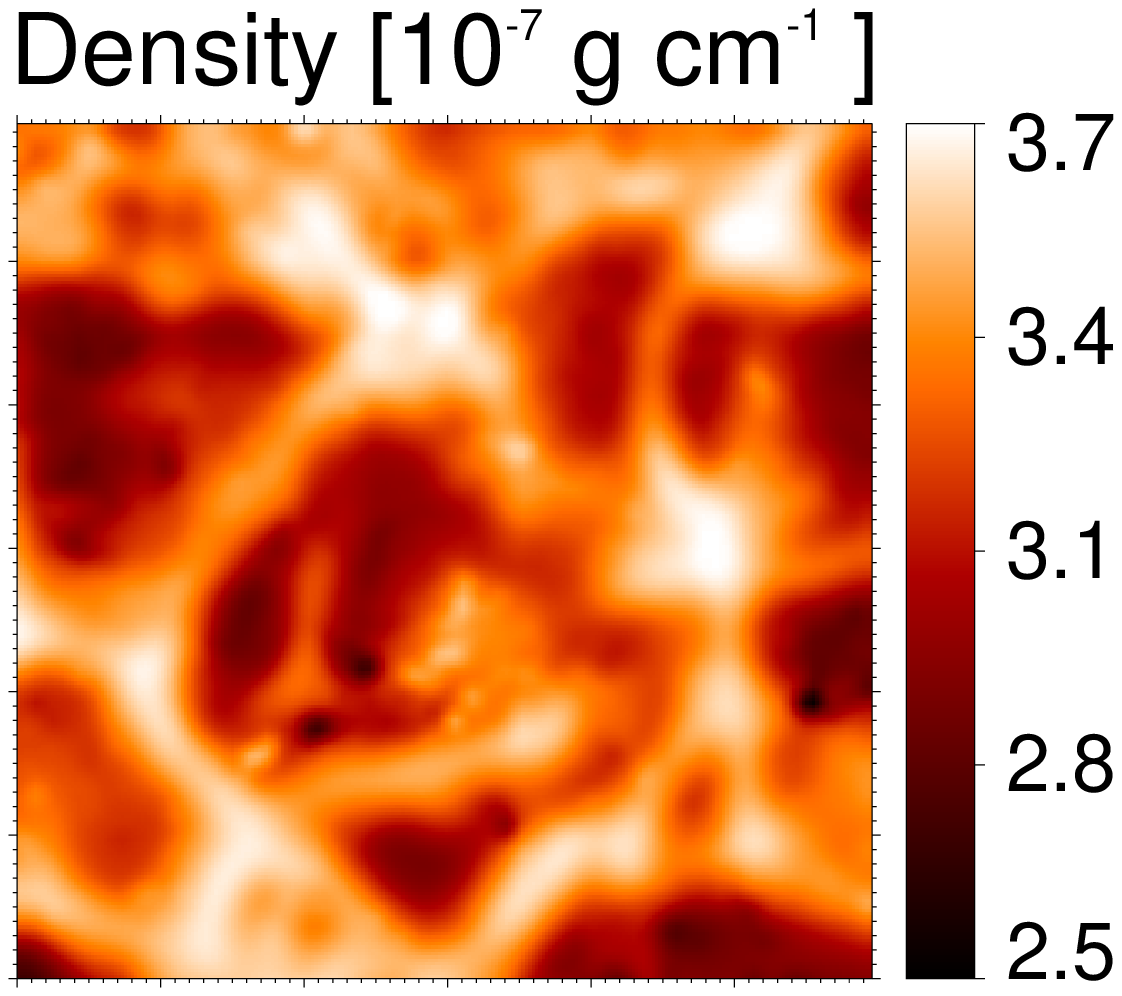}
\includegraphics[scale=.3, trim=3.6cm 2.2cm  1.6cm  .2cm,clip=true ]{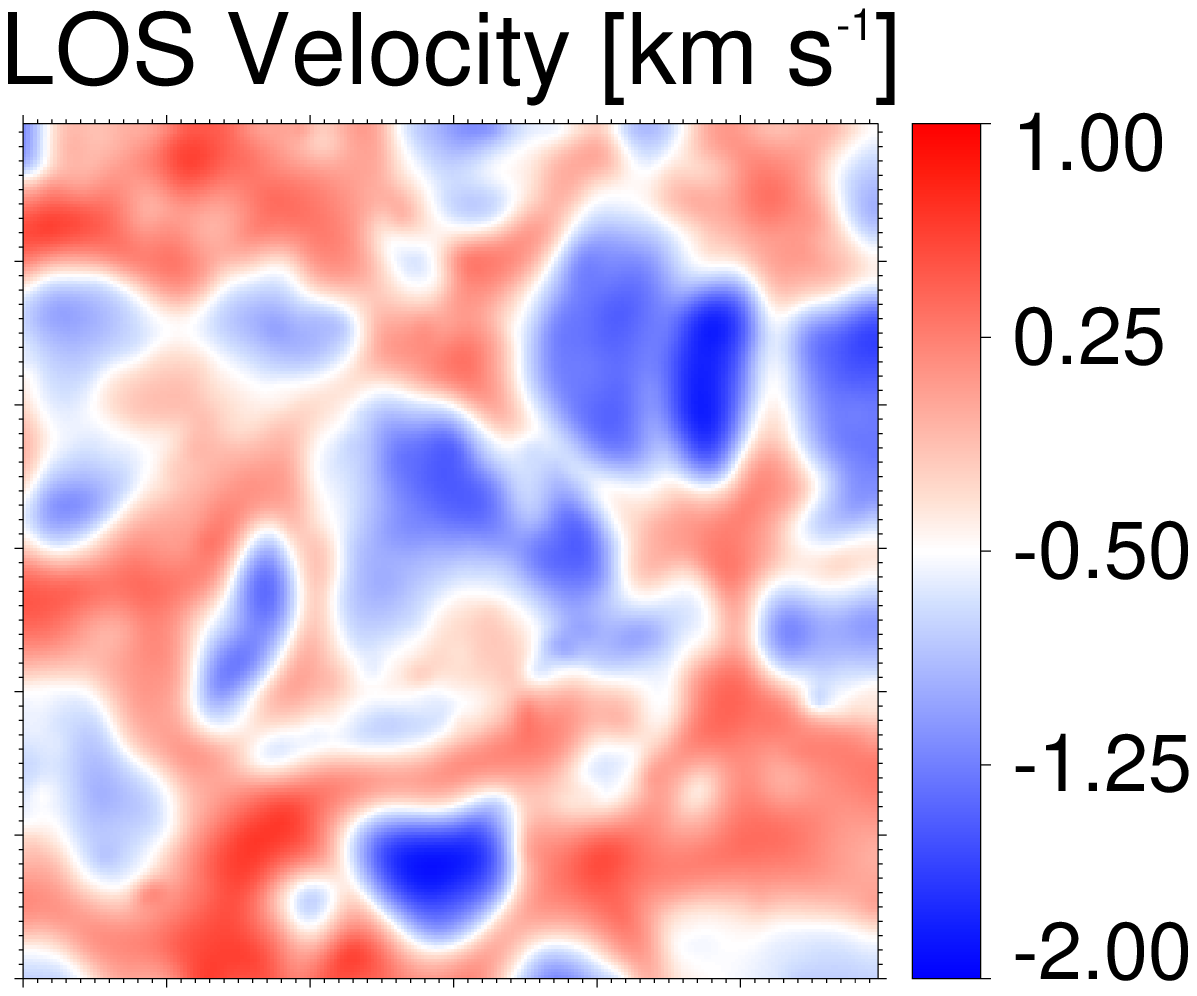} \\
\includegraphics[scale=.3, trim=1.8cm 2.2cm  1.cm  1.2cm,clip=true ]{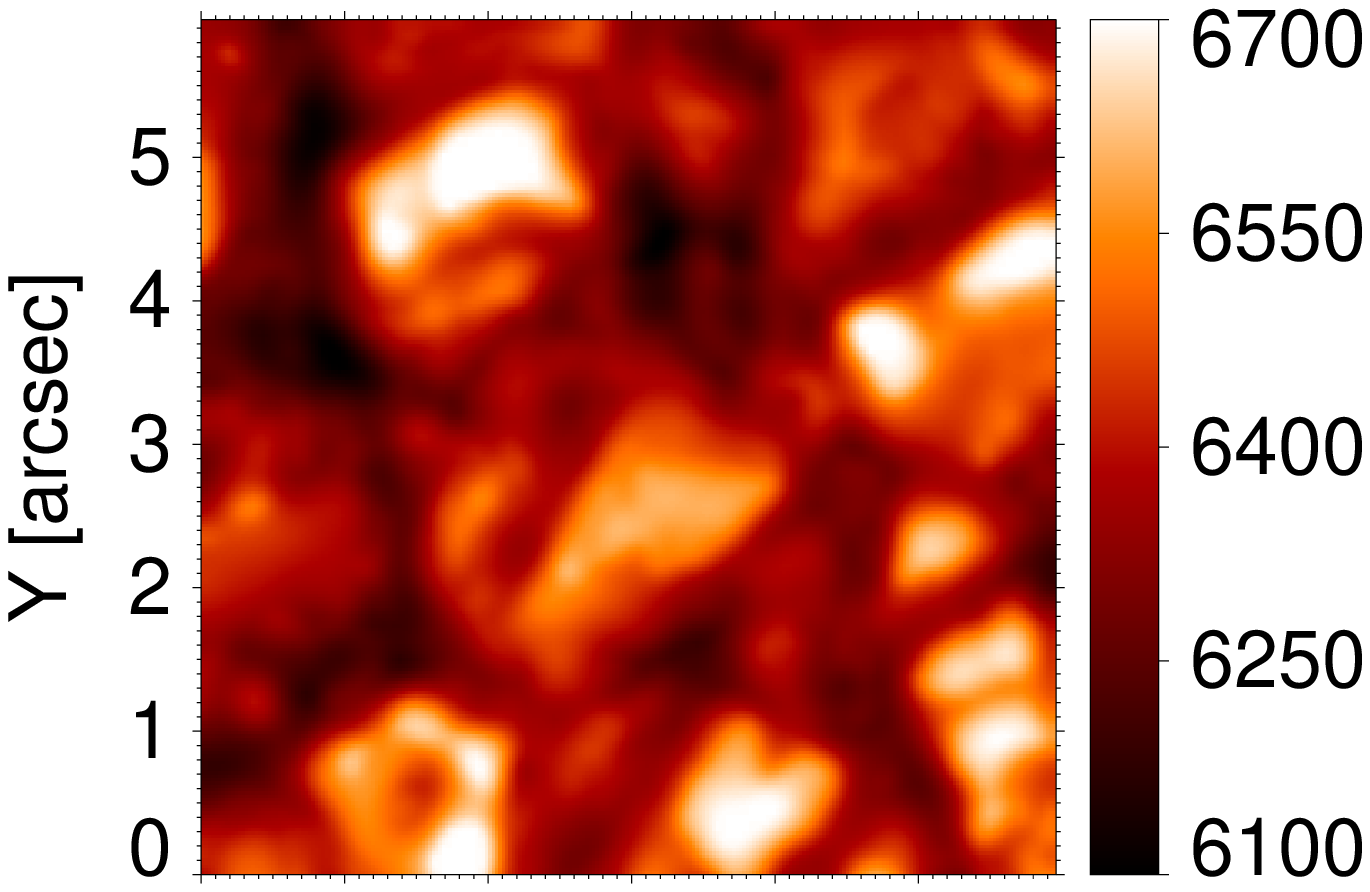} 
\includegraphics[scale=.3, trim=3.6cm 2.2cm  1.6cm   1.2cm,clip=true ]{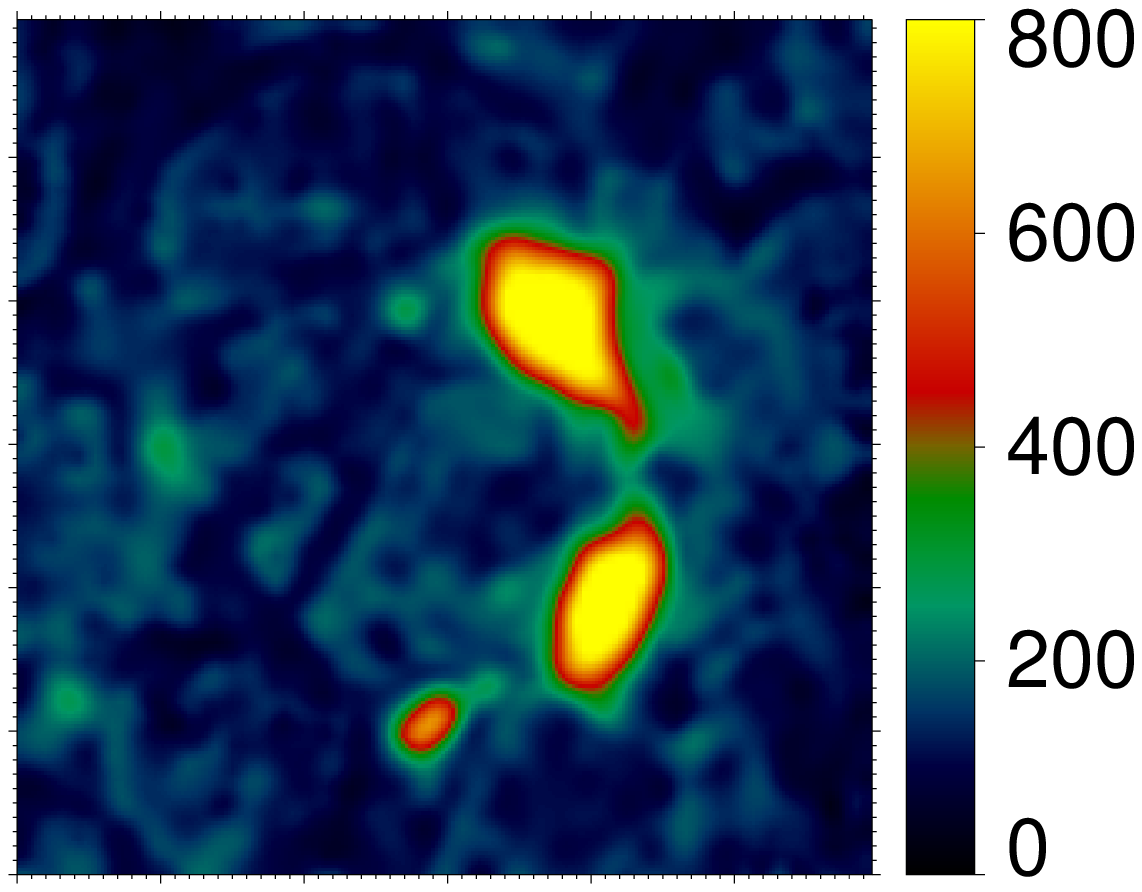} 
\includegraphics[scale=.3, trim=3.6cm 2.2cm  1.6cm  1.2cm,clip=true ]{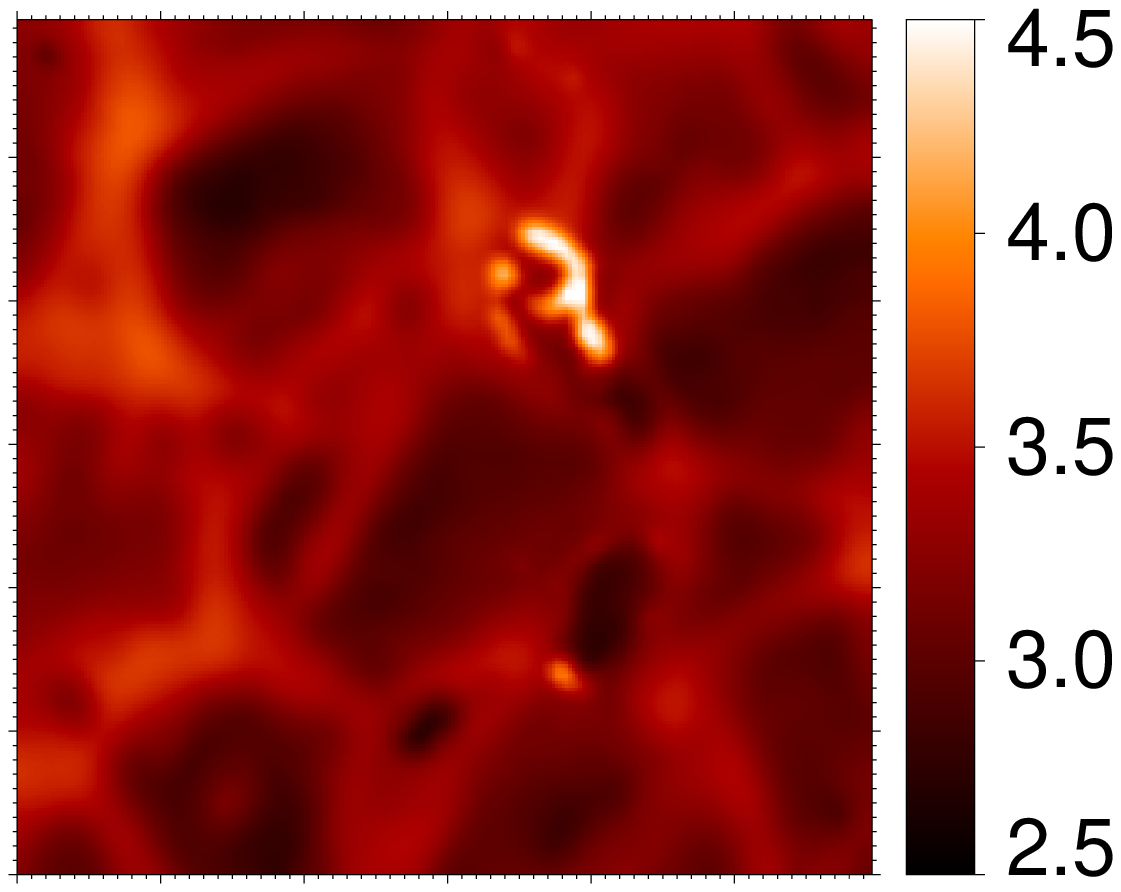} 
\includegraphics[scale=.3, trim=3.6cm 2.2cm  1.6cm   1.2cm,clip=true ]{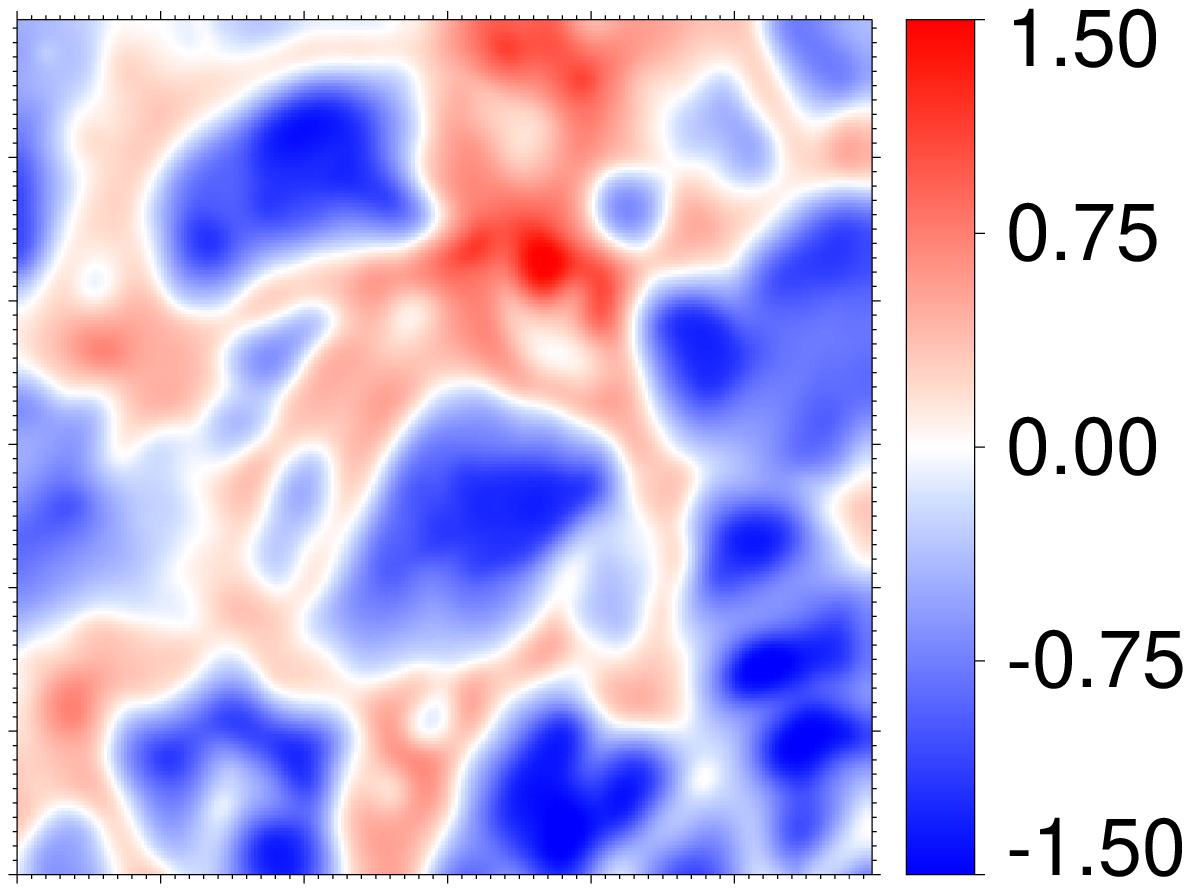}\\
\includegraphics[scale=.3, trim=1.8cm 2.2cm  1.cm  1.2cm,clip=true  ]{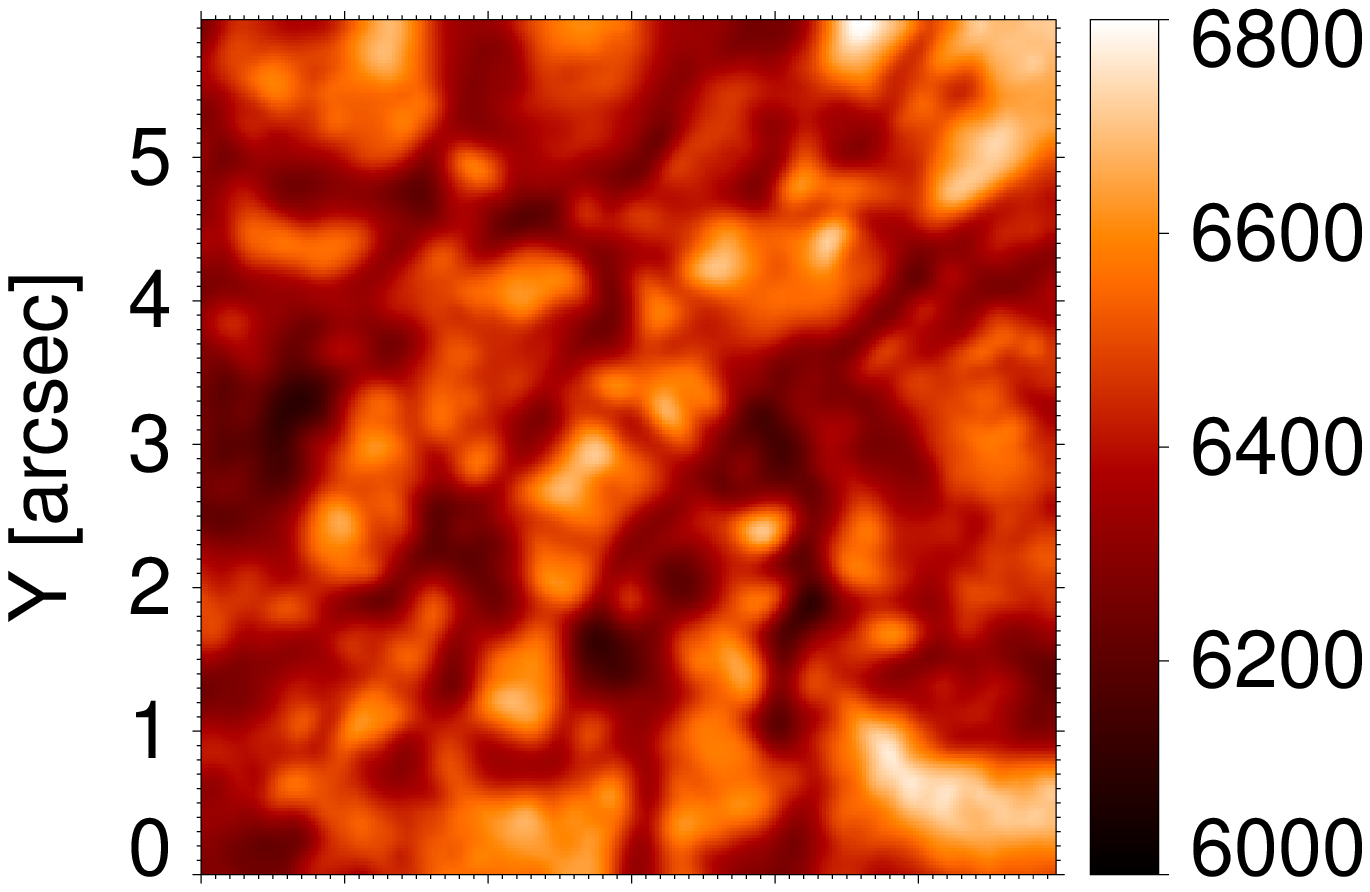} 
\includegraphics[scale=.3, trim=3.6cm 2.2cm  1.6cm  1.2cm,clip=true ]{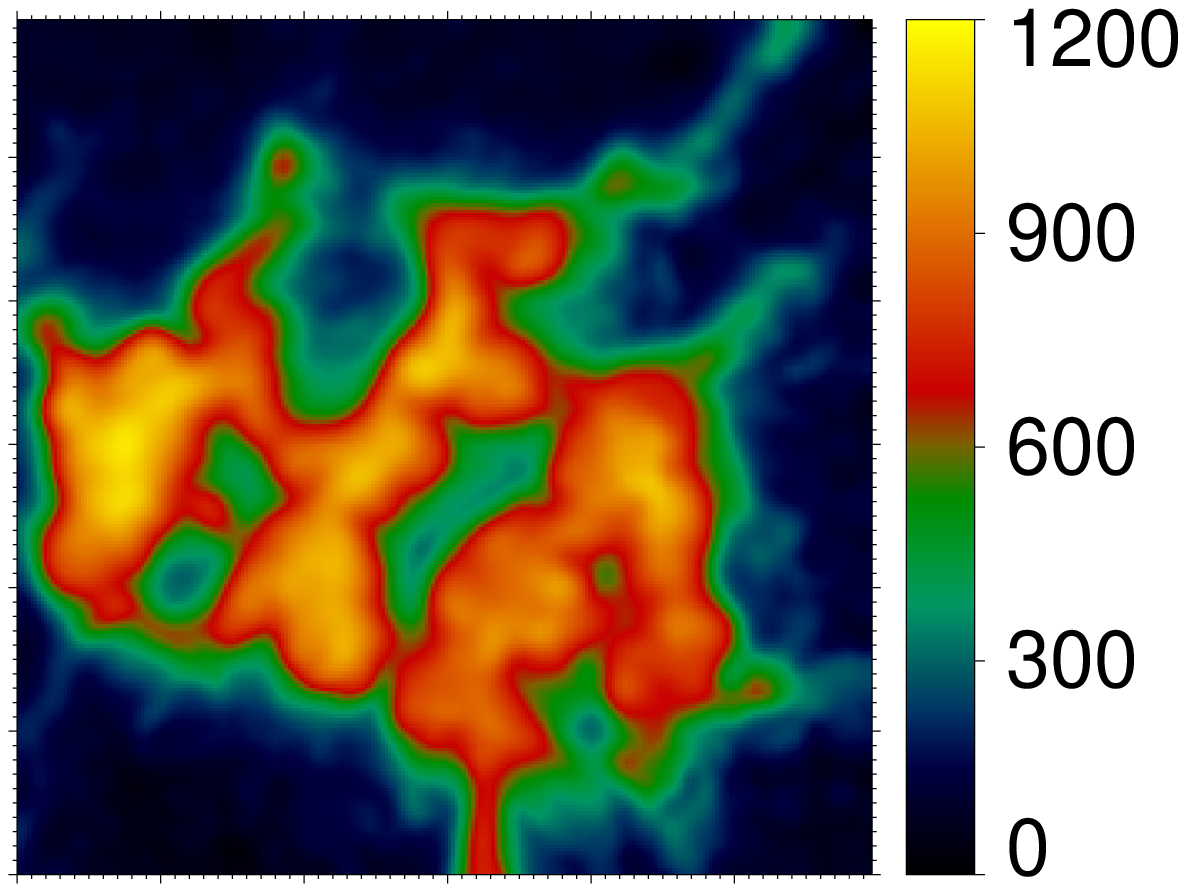} 
\includegraphics[scale=.3, trim=3.6cm 2.2cm  1.6cm  1.2cm,clip=true ]{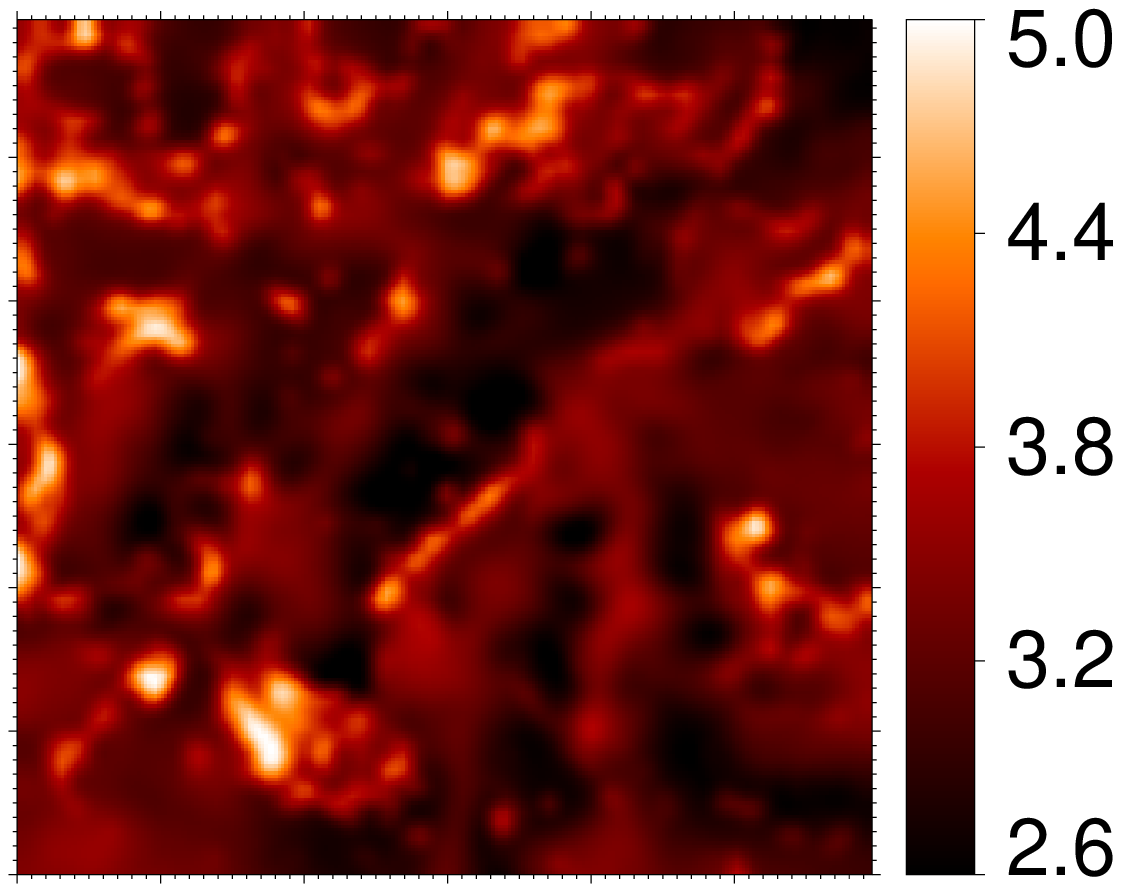} 
\includegraphics[scale=.3, trim=3.6cm 2.2cm  1.6cm   1.2cm,clip=true ]{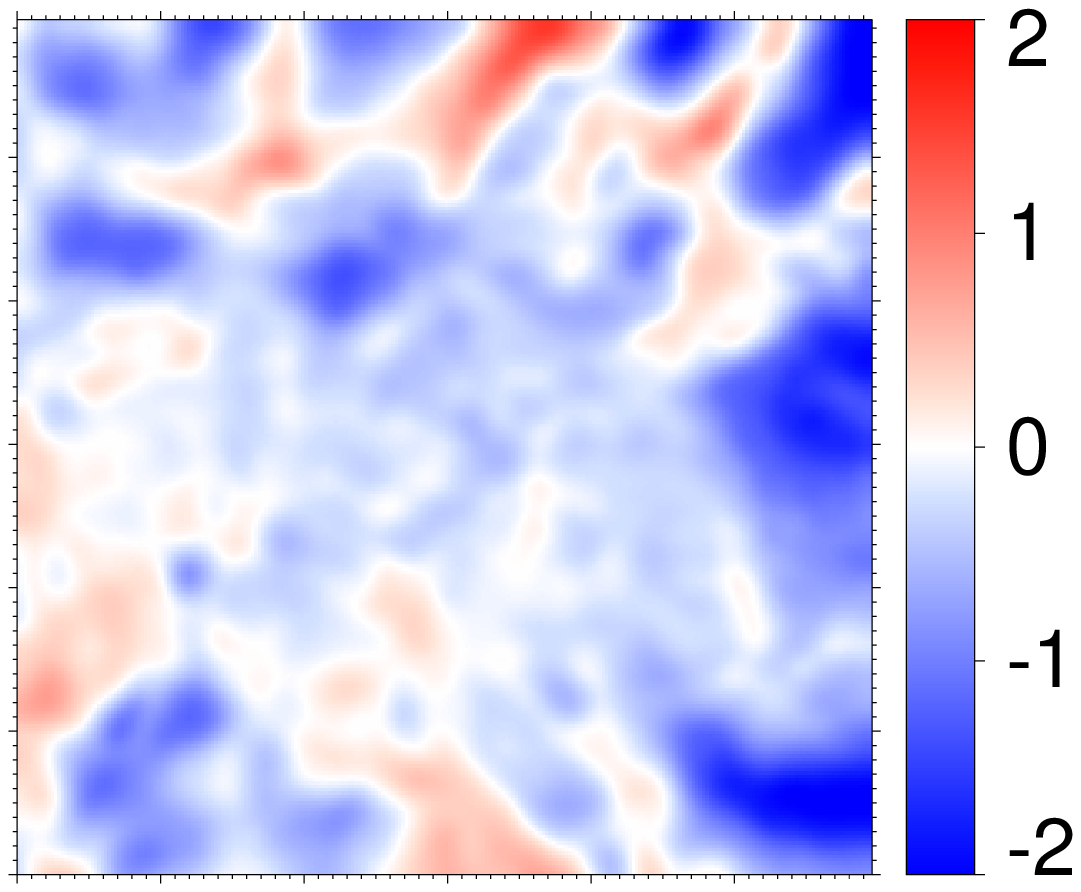}\\
\includegraphics[scale=.3, trim=1.8cm 2.2cm  1.cm  1.2cm,clip=true ]{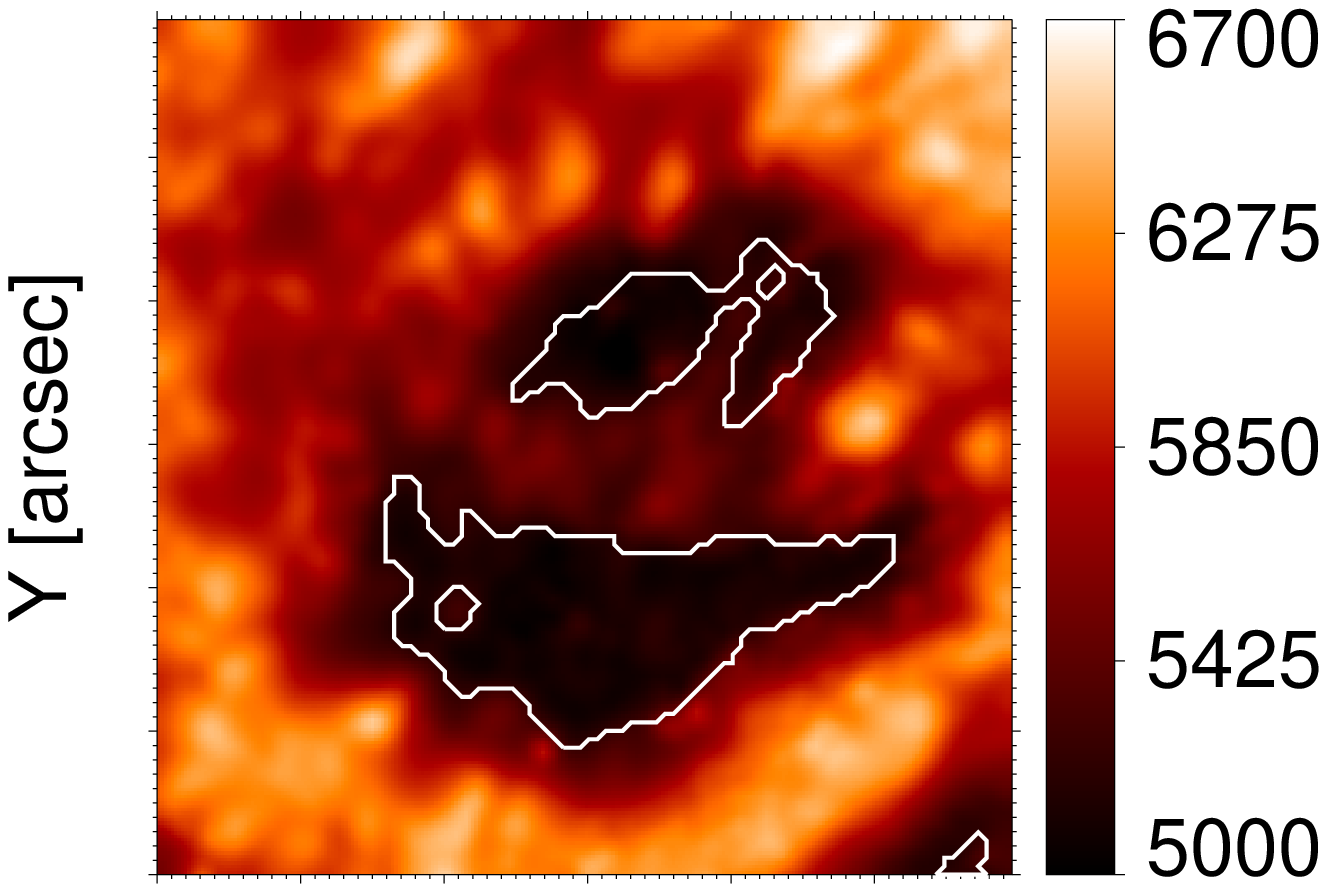} 
\includegraphics[scale=.3, trim=3.6cm 2.2cm  1.6cm  1.2cm,clip=true ]{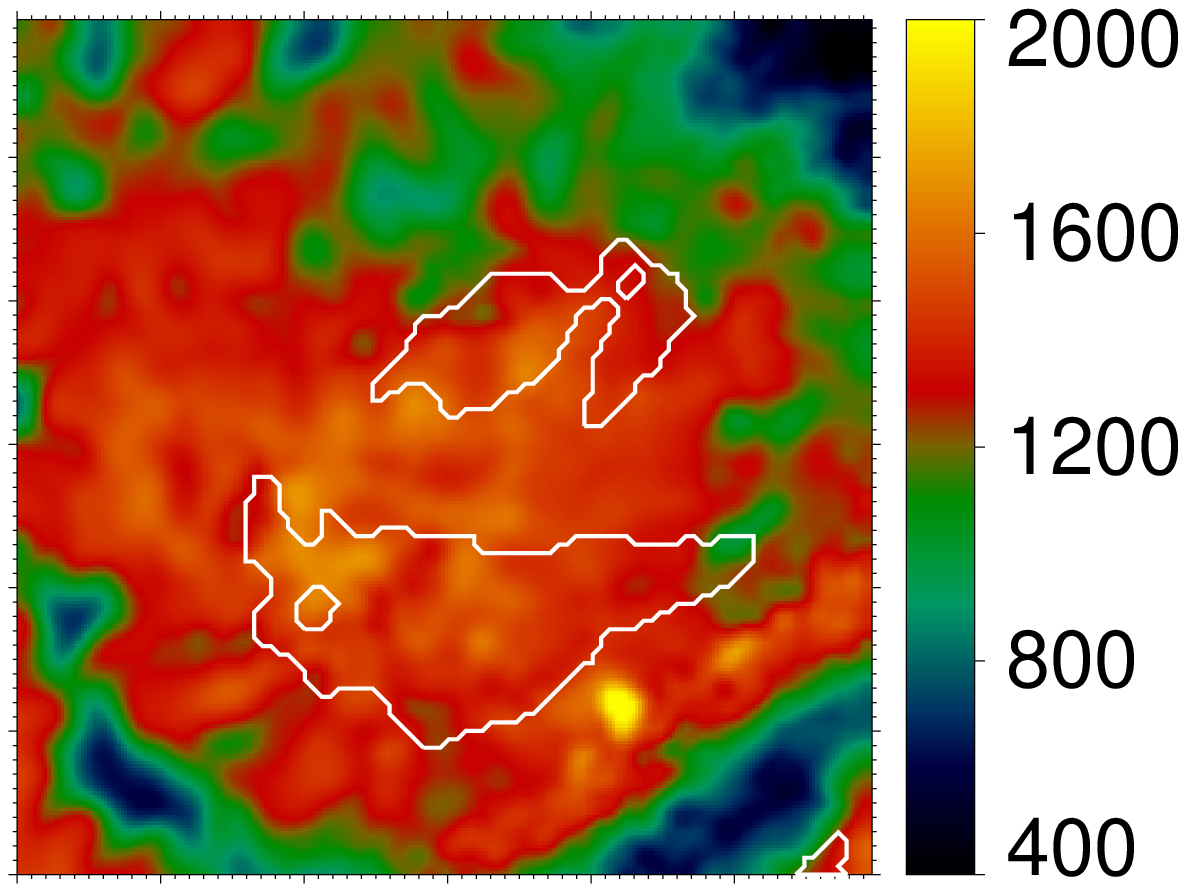} 
\includegraphics[scale=.3, trim=3.6cm 2.2cm  1.6cm  1.2cm,clip=true ]{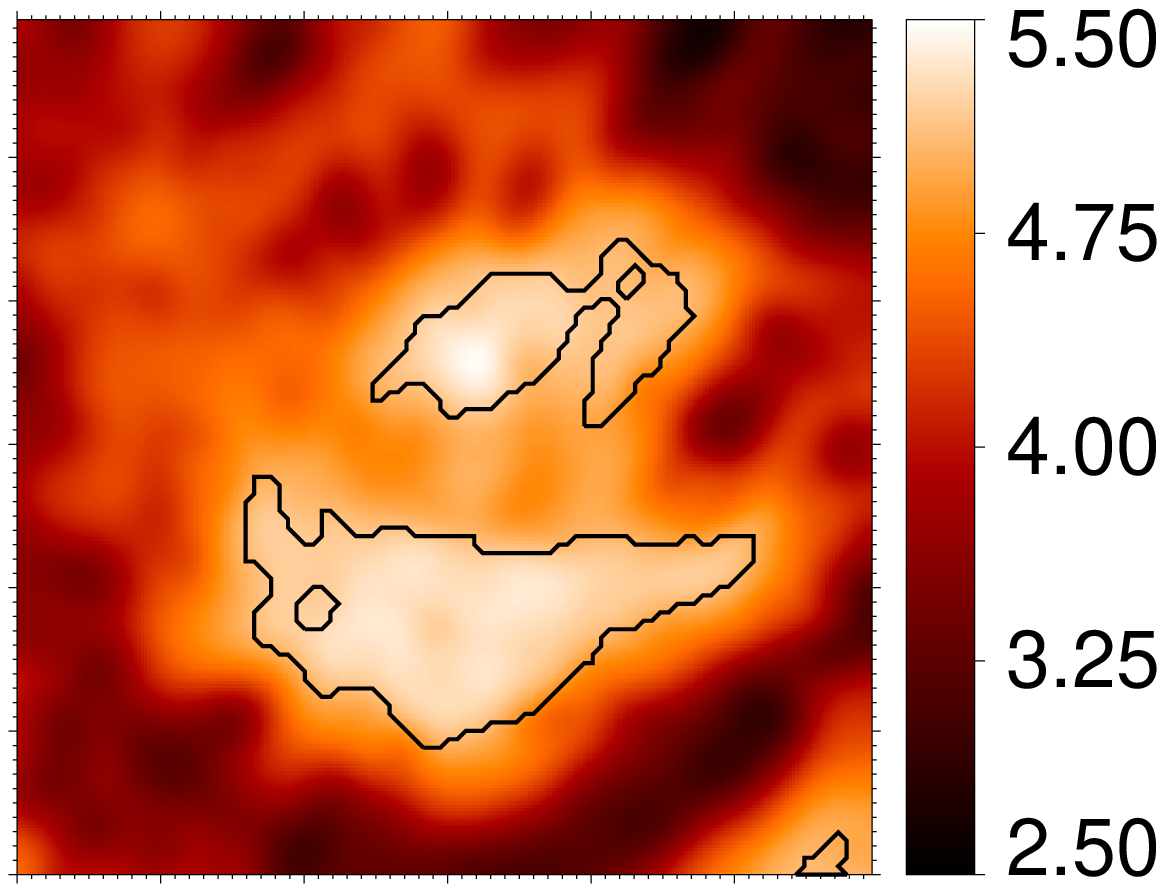}
\includegraphics[scale=.3, trim=3.6cm 2.2cm  1.6cm  1.2cm,clip=true ]{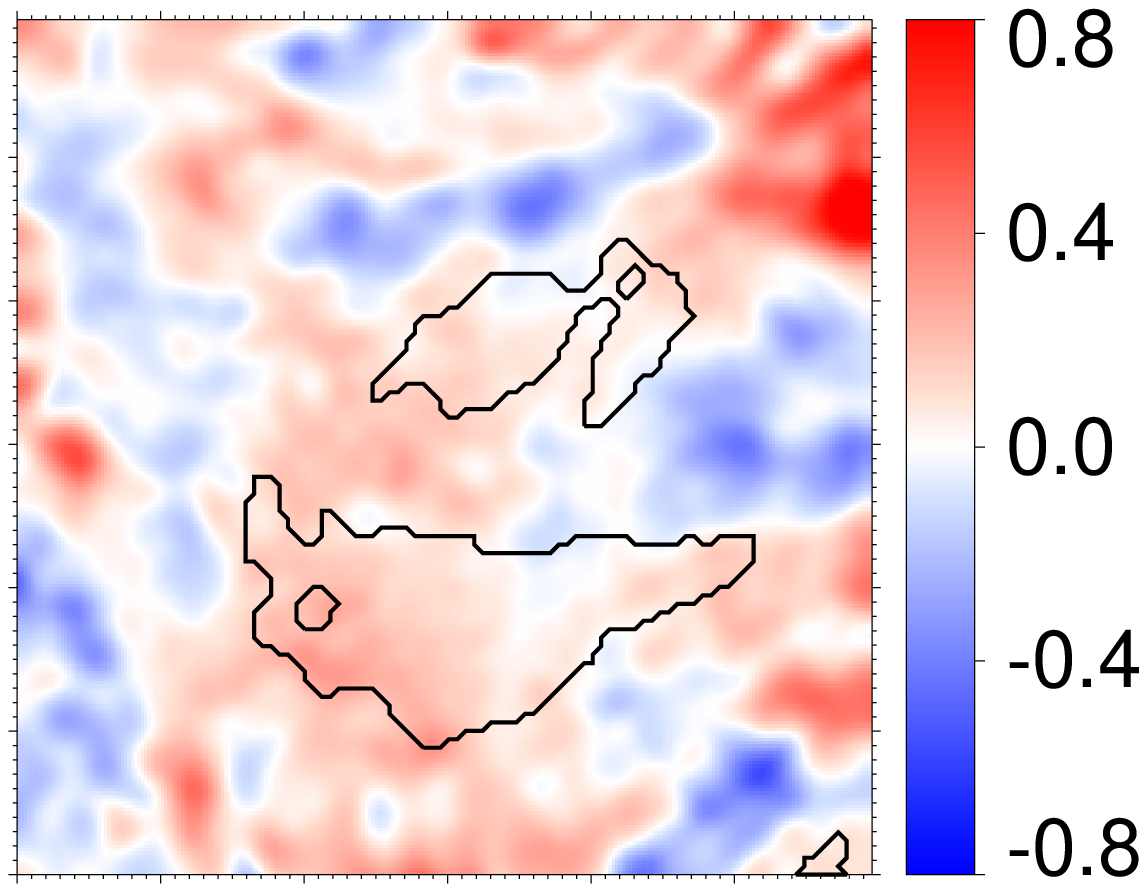}\\		
\includegraphics[scale=.3, trim=1.5cm .3cm  1.cm  1.2cm,clip=true ]{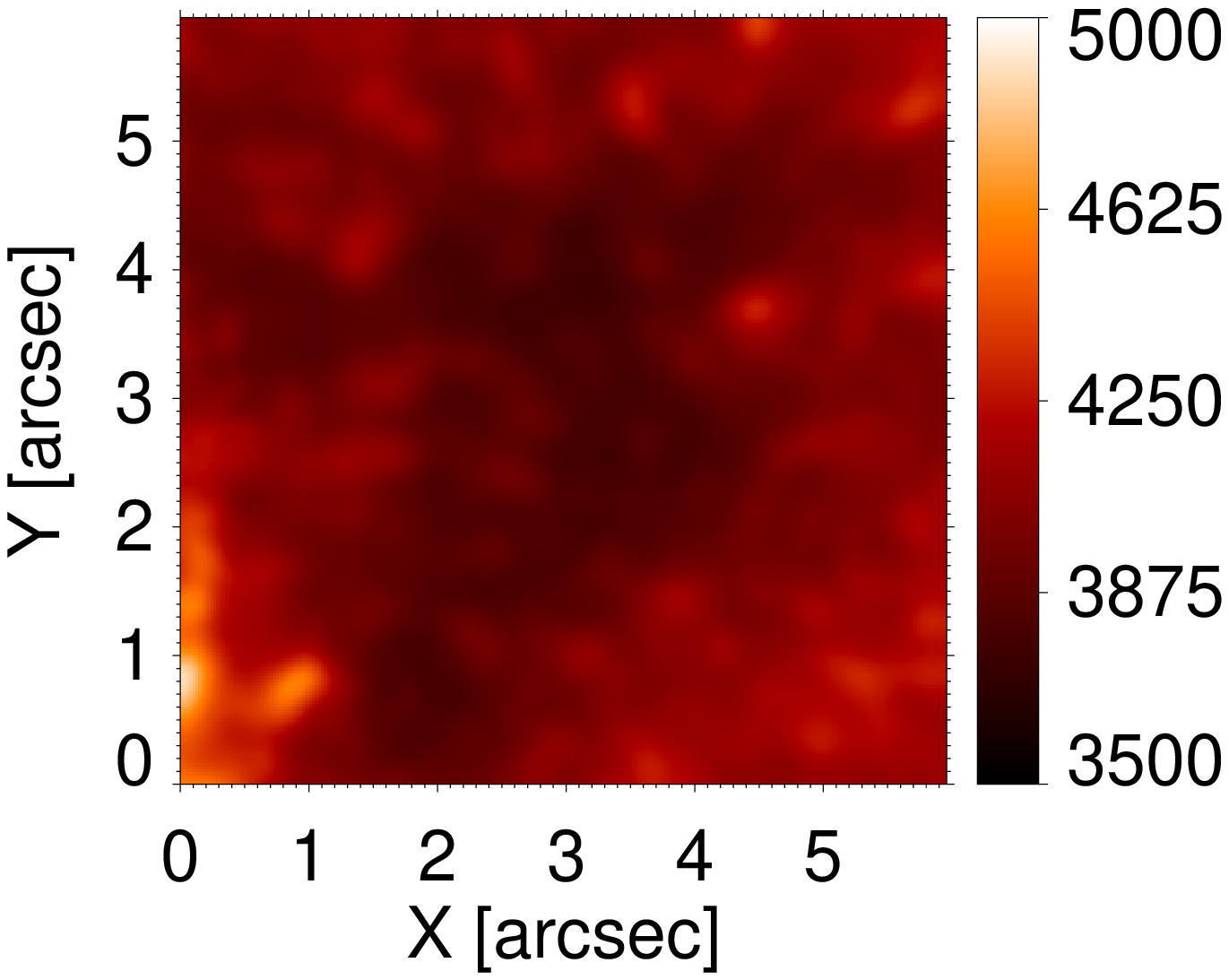} 
\includegraphics[scale=.3, trim=3.4cm .3cm  1.6cm  1.2cm,clip=true ]{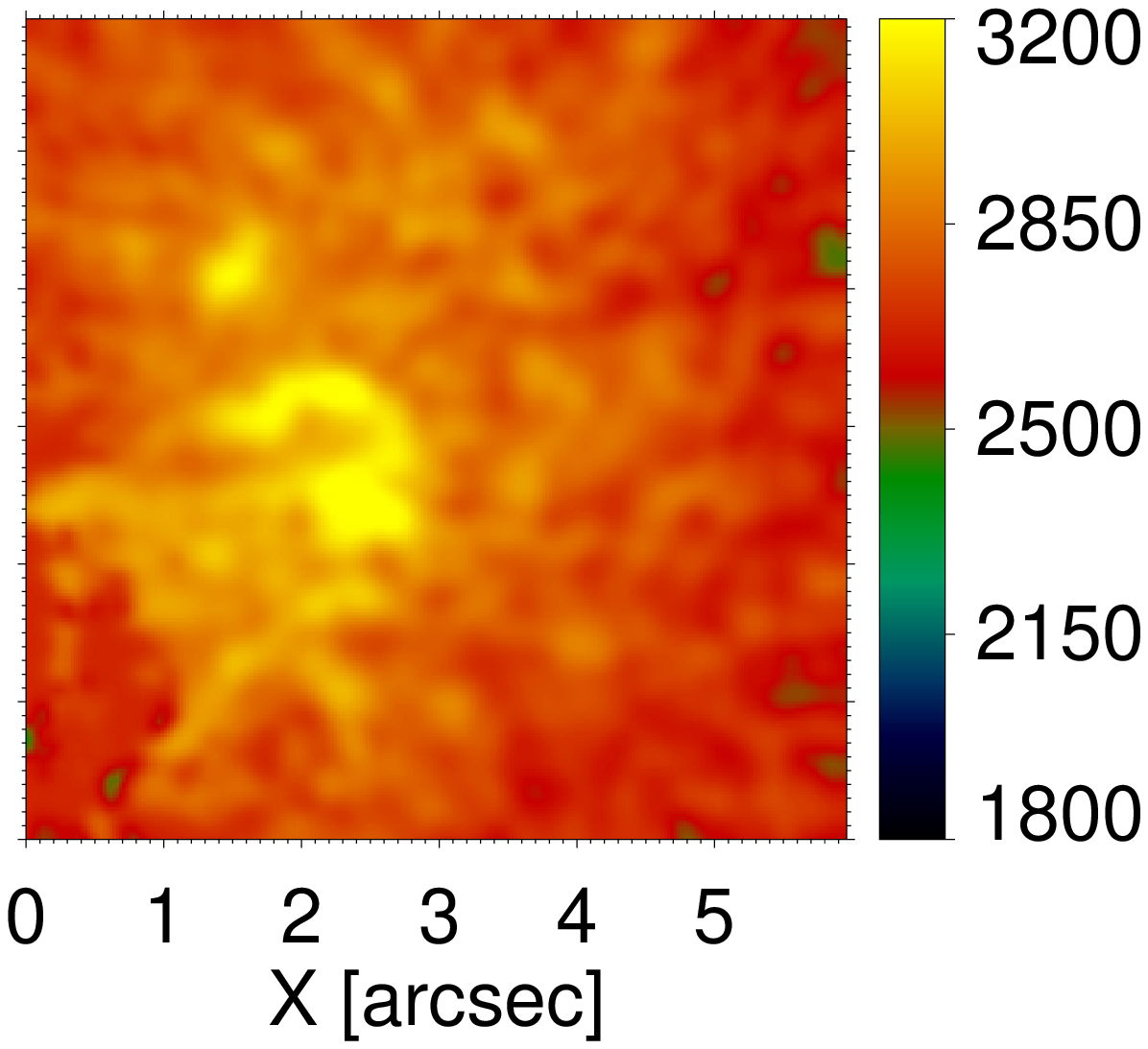} 
\includegraphics[scale=.3, trim=3.5cm .3cm  1.6cm  1.2cm,clip=true ]{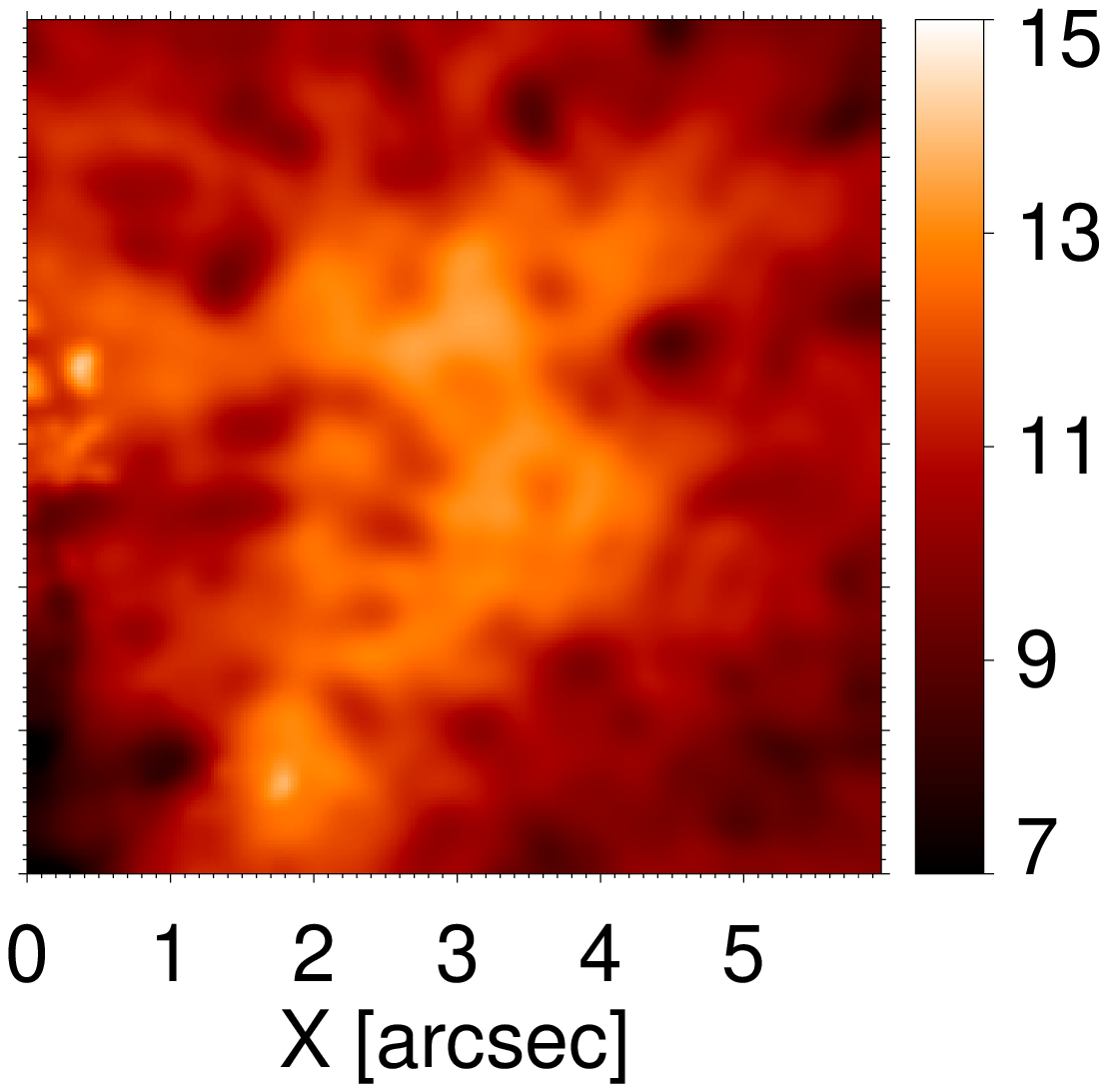}
\includegraphics[scale=.3, trim=3.5cm .3cm  1.6cm  1.2cm,clip=true ]{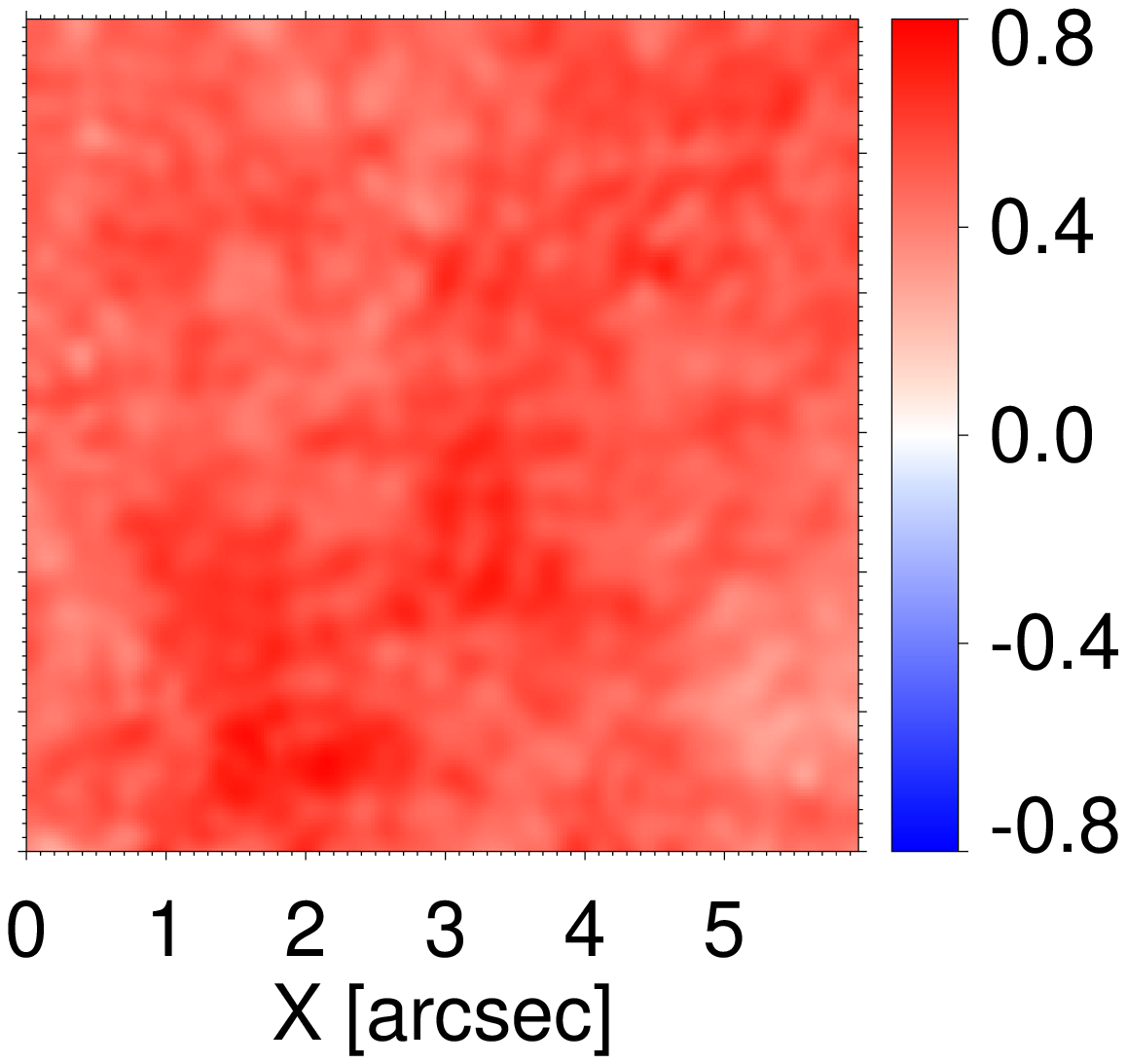}
}
\caption{From left to right, top to bottom: horizontal cuts of the temperature, magnetic field strength, plasma density, and LOS velocity at $log\tau_{500}$=0 on the atmosphere models derived from inversion of the QS, BPs, PL, PO, and UM regions. White and black contours mark the PO region considered to compute the average temperature profiles labeled (a) and (b) in Fig. \ref{fig7bis_c4}.}
\label{fig3_c4}
\end{figure*}

\begin{figure*}
\centering
{
\includegraphics[scale=.3, trim=1.8cm 2.3cm  1.cm  .2cm,clip=true ]{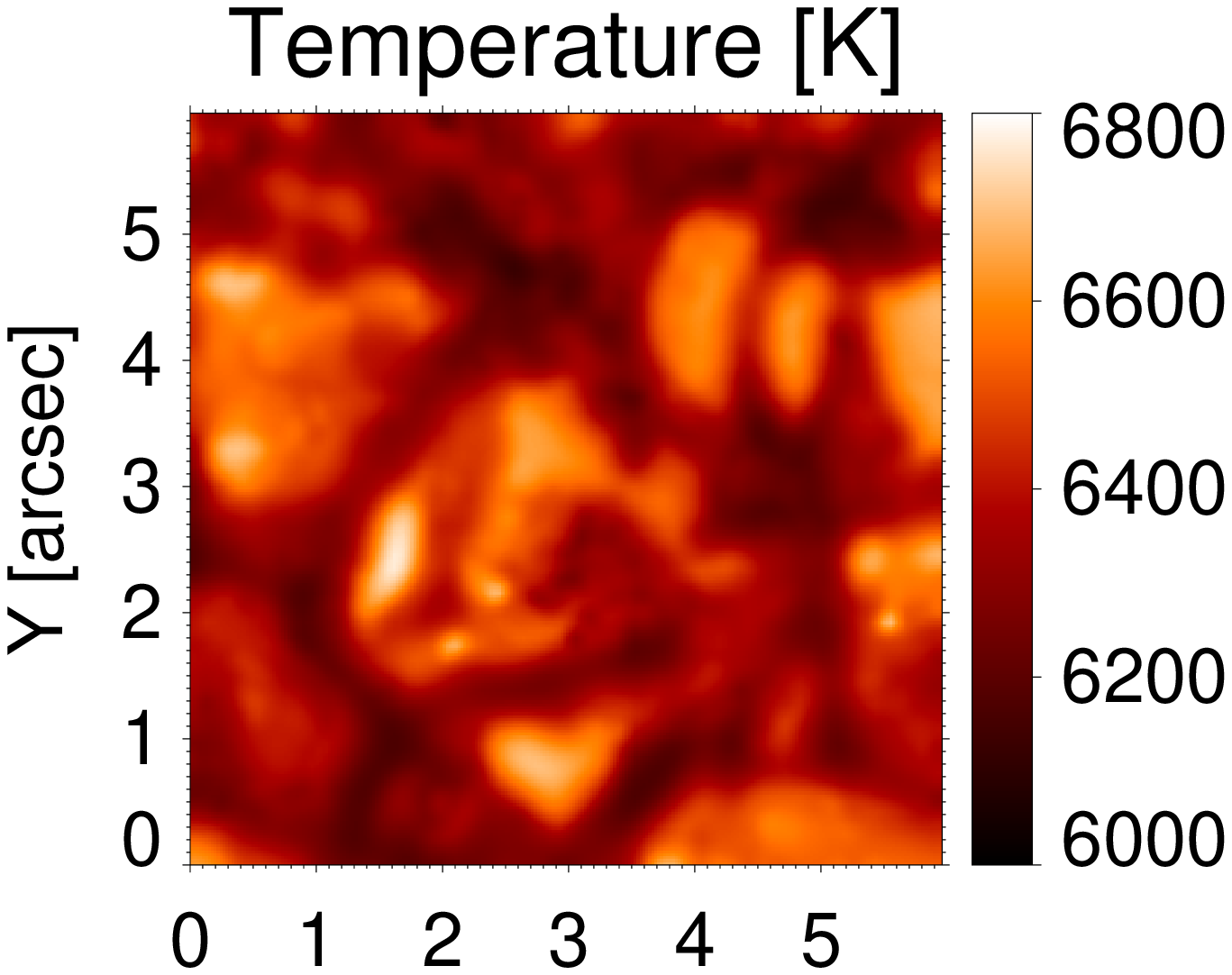}
\includegraphics[scale=.3, trim=4.2cm 2.3cm  1.6cm  .2cm,clip=true ]{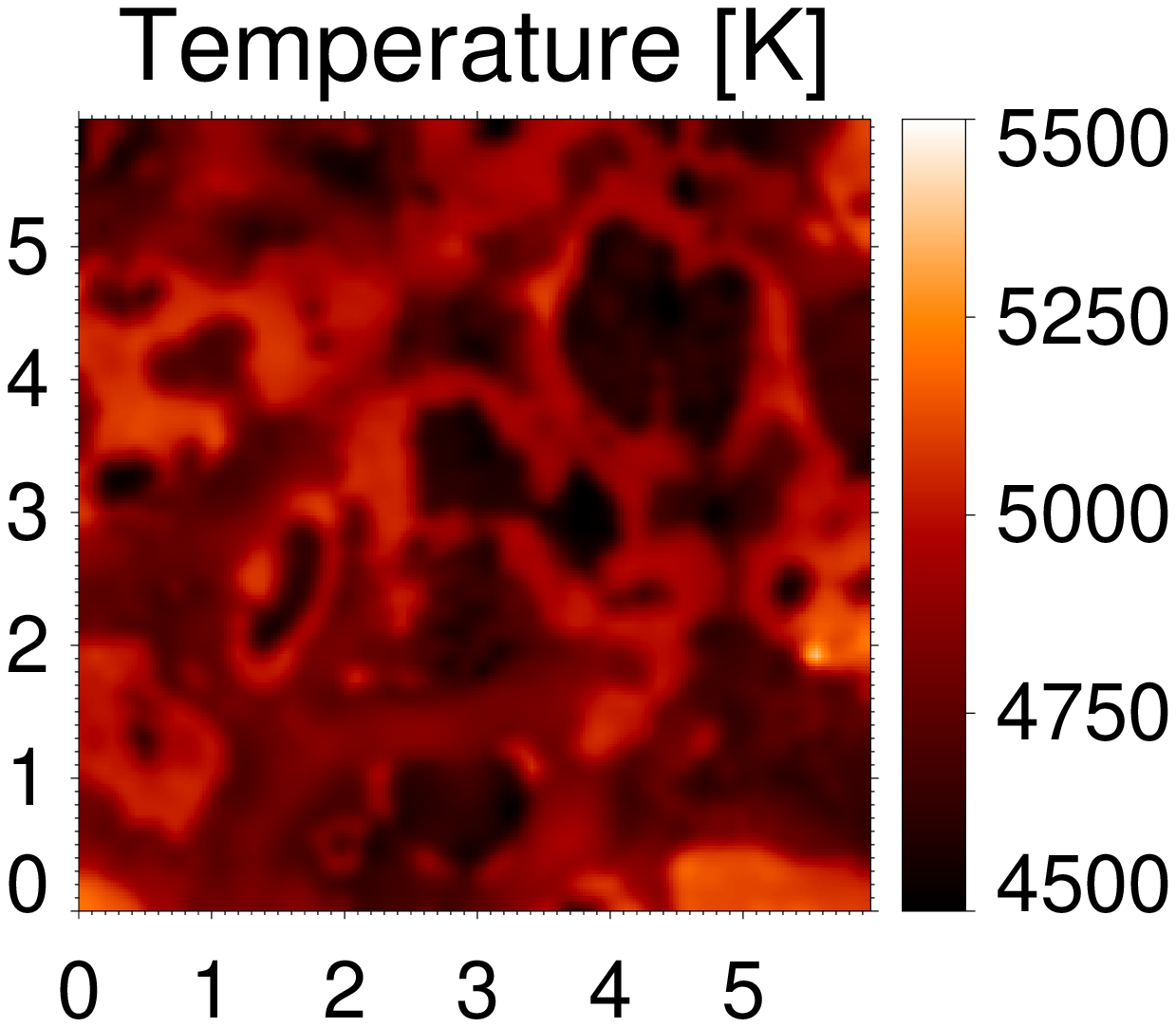}
\includegraphics[scale=.3, trim=4.2cm 2.3cm  1.6cm  .2cm,clip=true]{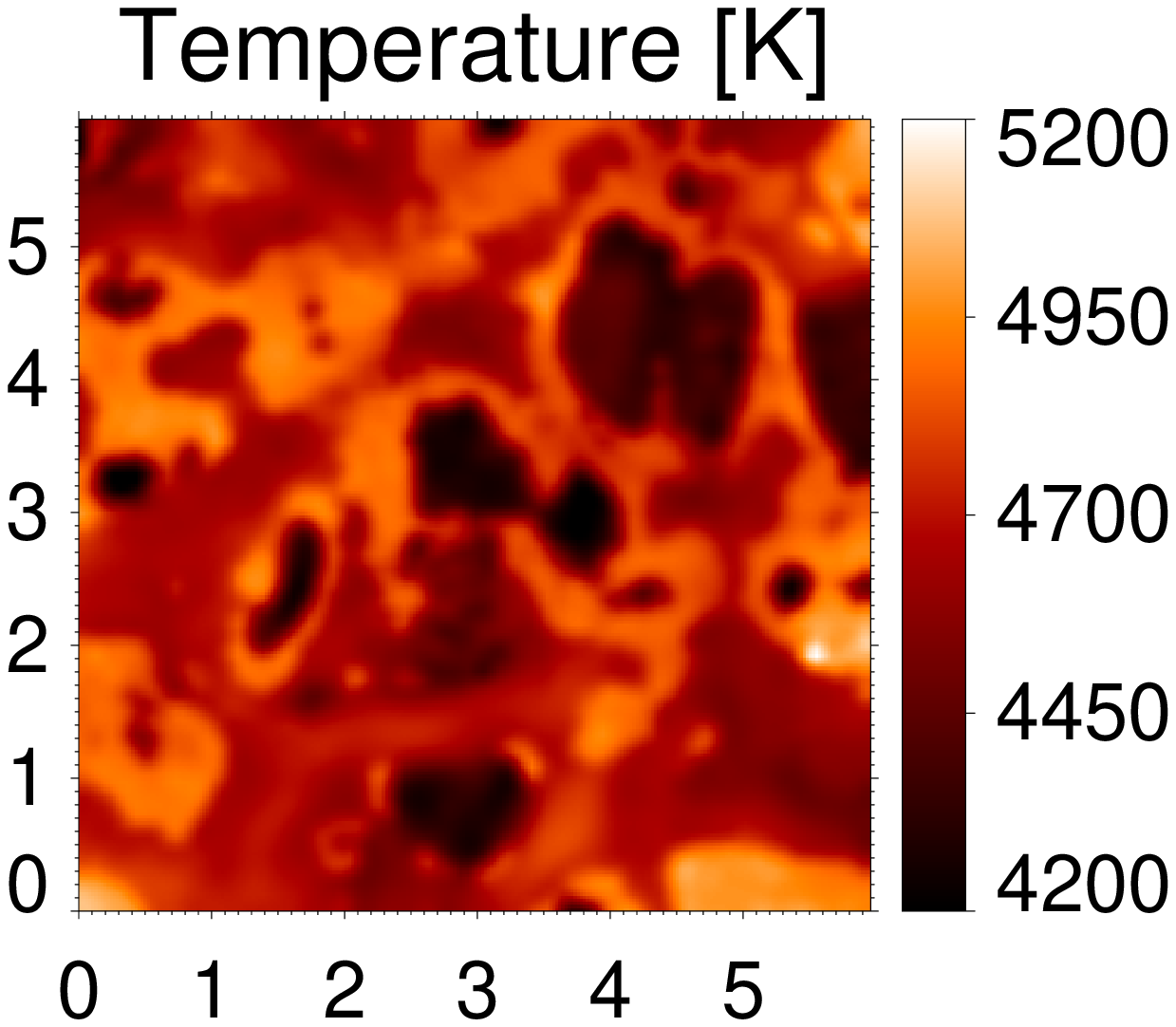}
\includegraphics[scale=.3, trim=4.2cm 2.3cm  1.6cm  .2cm,clip=true ]{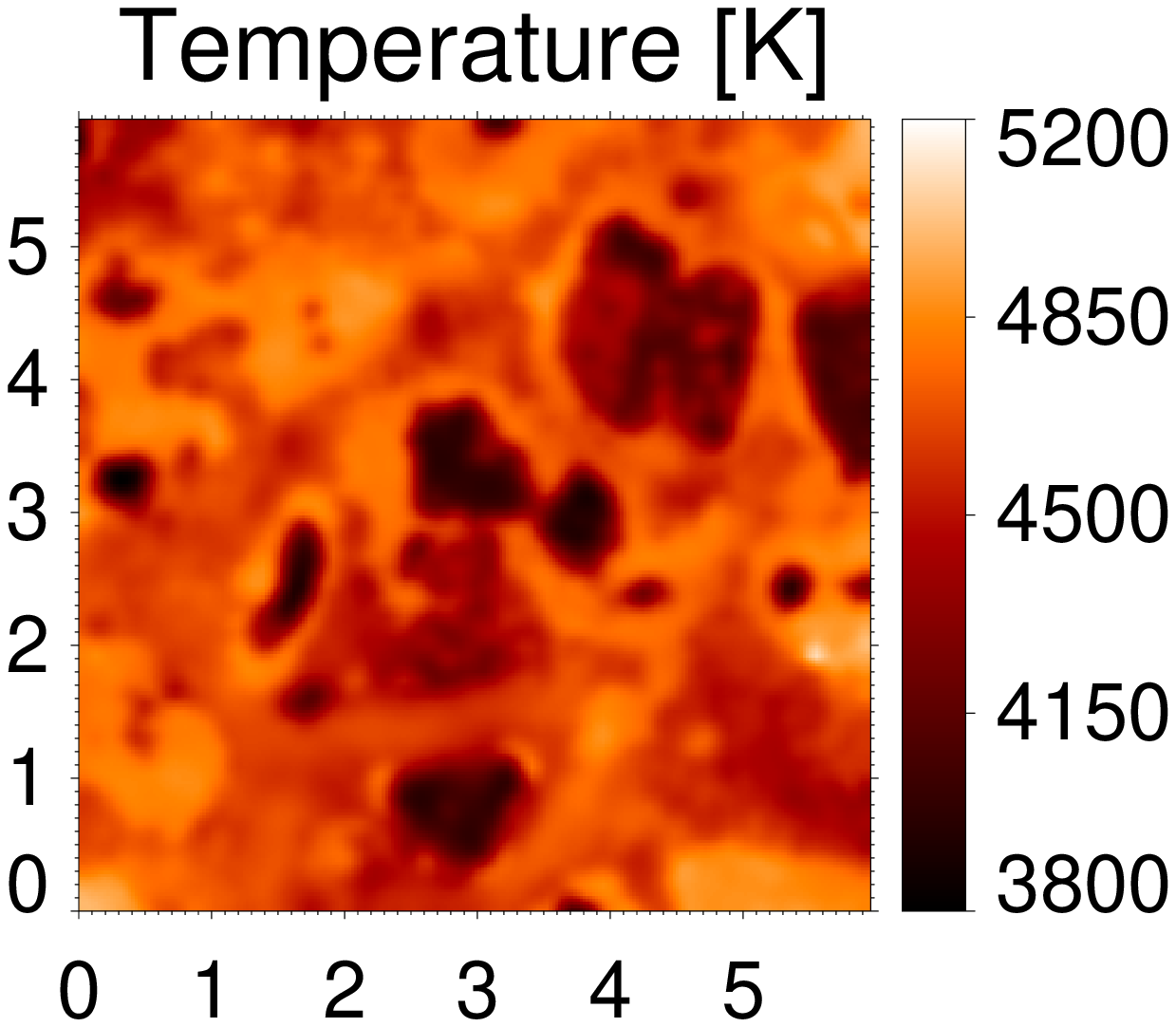} \\
\includegraphics[scale=.3, trim=1.8cm 2.3cm  1.cm  1.3cm,clip=true ]{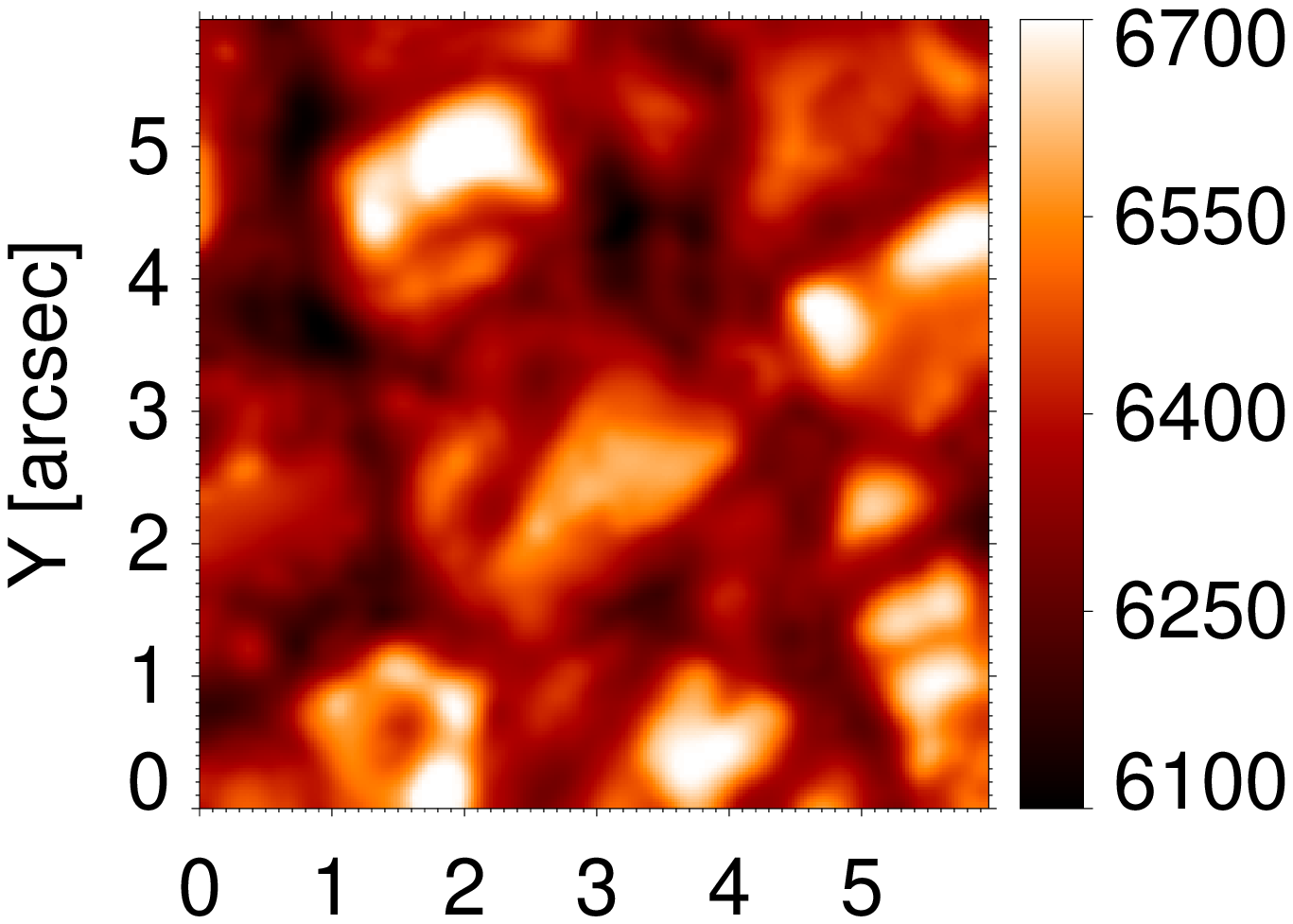}
\includegraphics[scale=.3, trim=4.2cm 2.3cm  1.6cm  1.3cm,clip=true ]{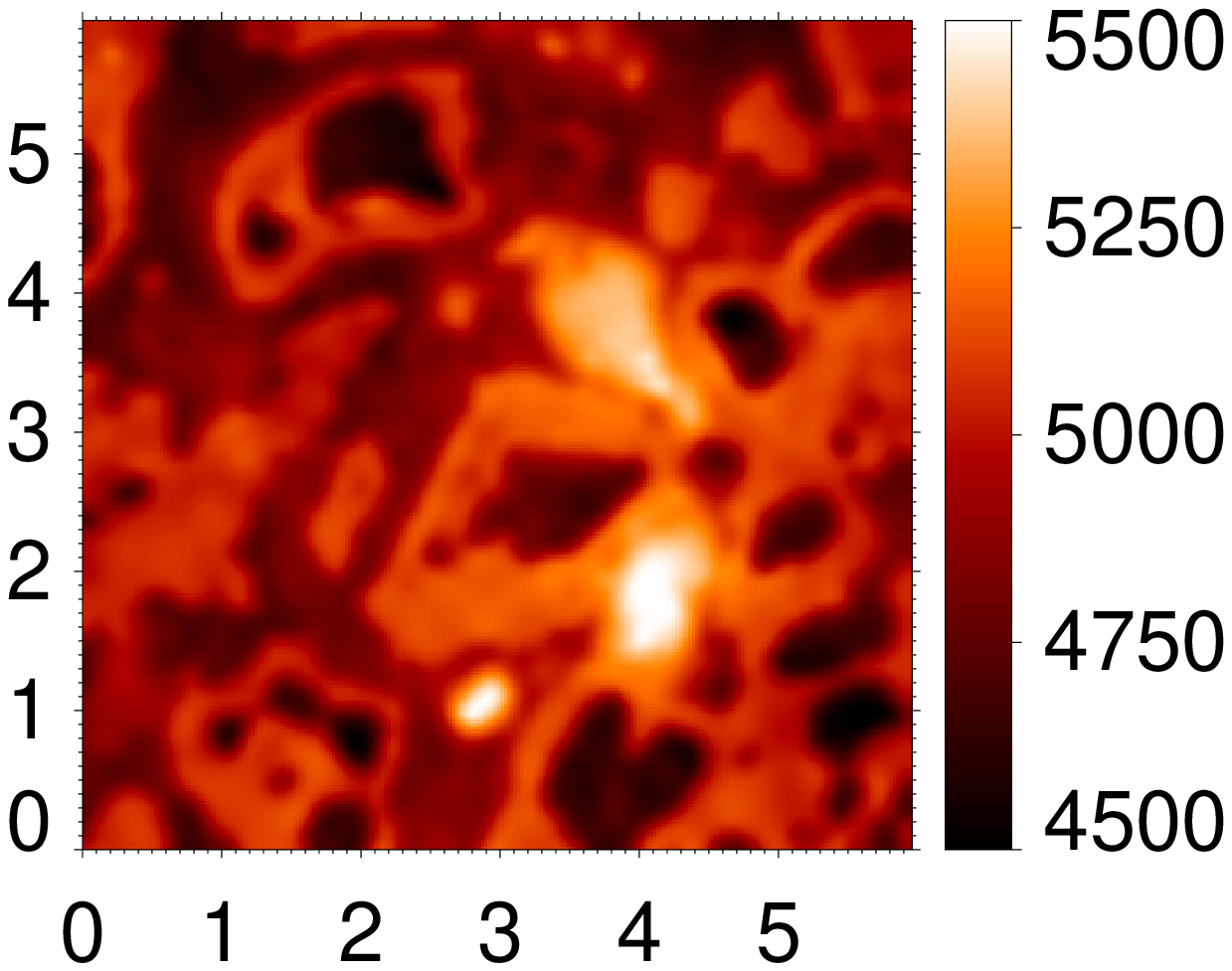}
\includegraphics[scale=.3, trim=4.2cm 2.3cm  1.6cm  1.3cm,clip=true ]{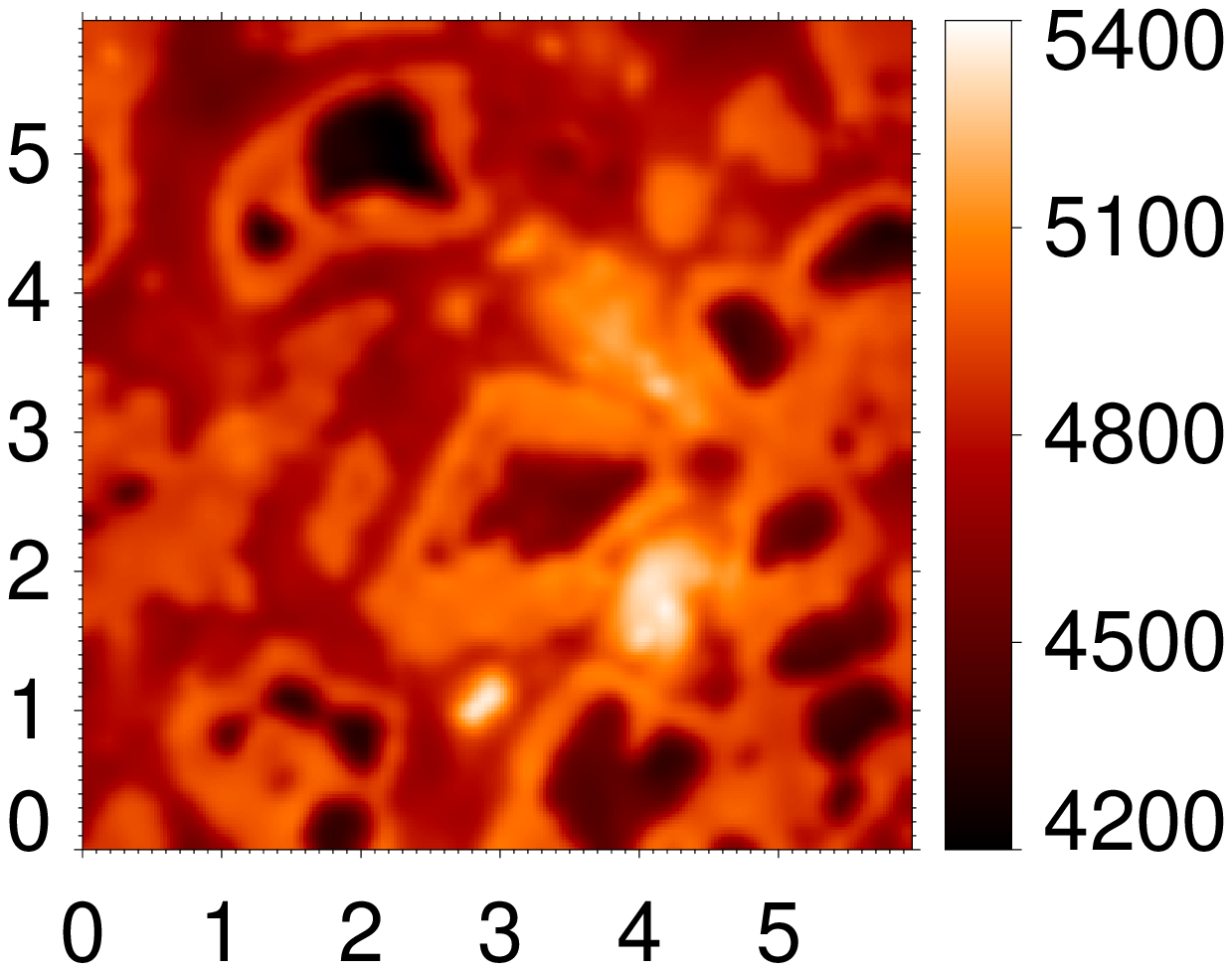}
\includegraphics[scale=.3, trim=4.2cm 2.3cm  1.6cm   1.3cm,clip=true ]{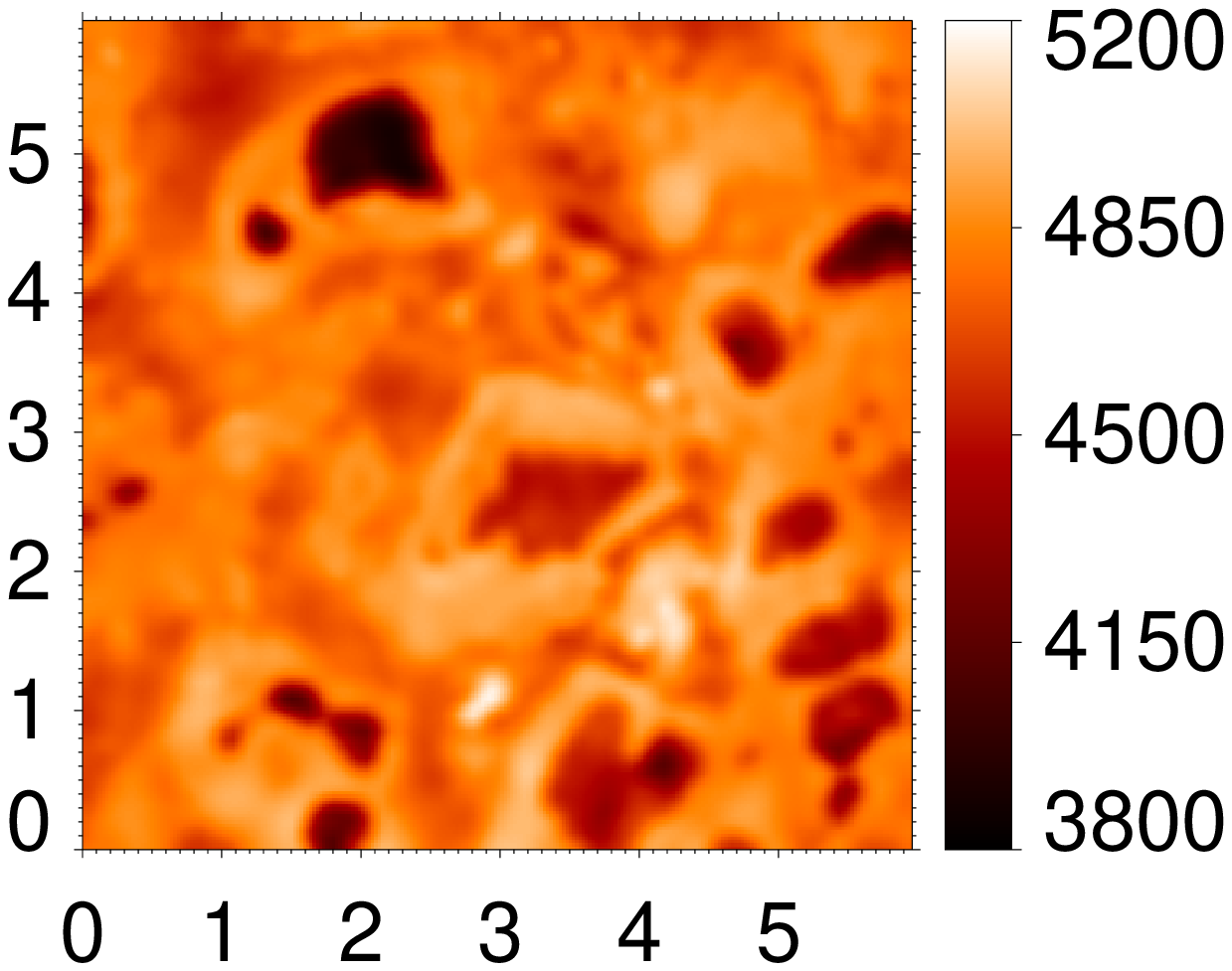}\\
\includegraphics[scale=.3, trim=1.8cm 2.3cm  1.cm  1.3cm,clip=true  ]{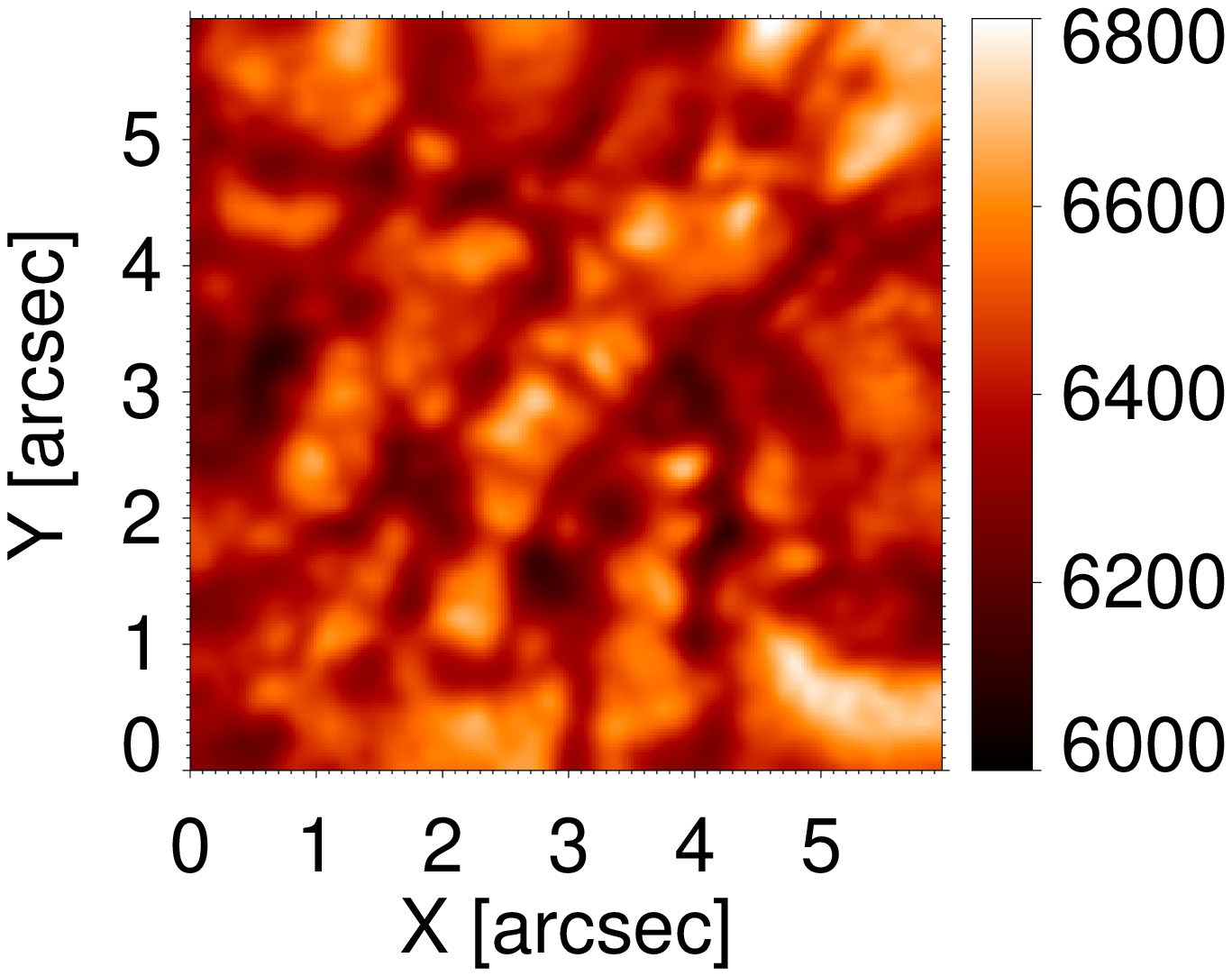}
\includegraphics[scale=.3, trim=4.2cm 2.3cm  1.6cm  1.3cm,clip=true ]{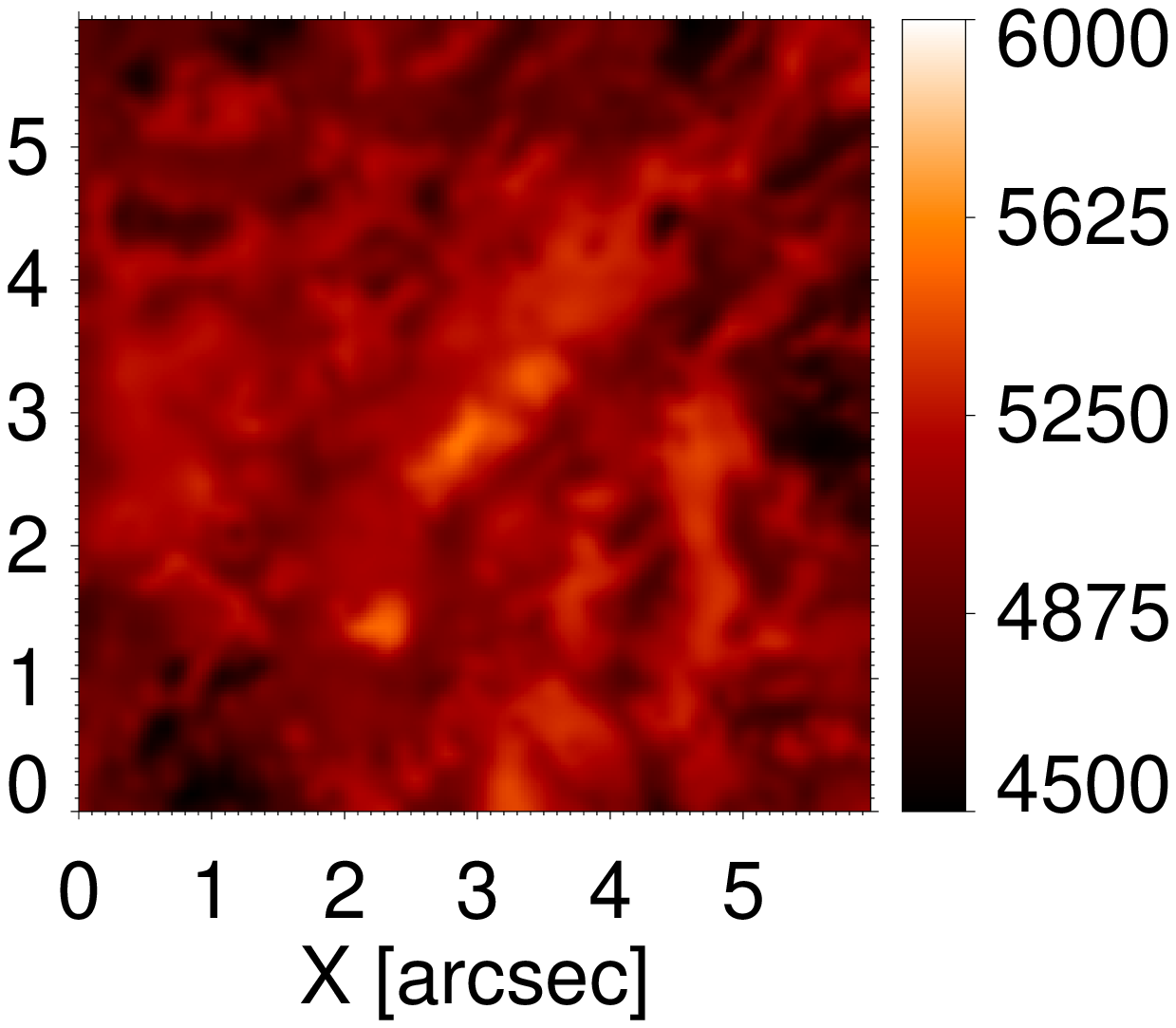}
\includegraphics[scale=.3, trim=4.2cm 2.3cm  1.6cm  1.3cm,clip=true ]{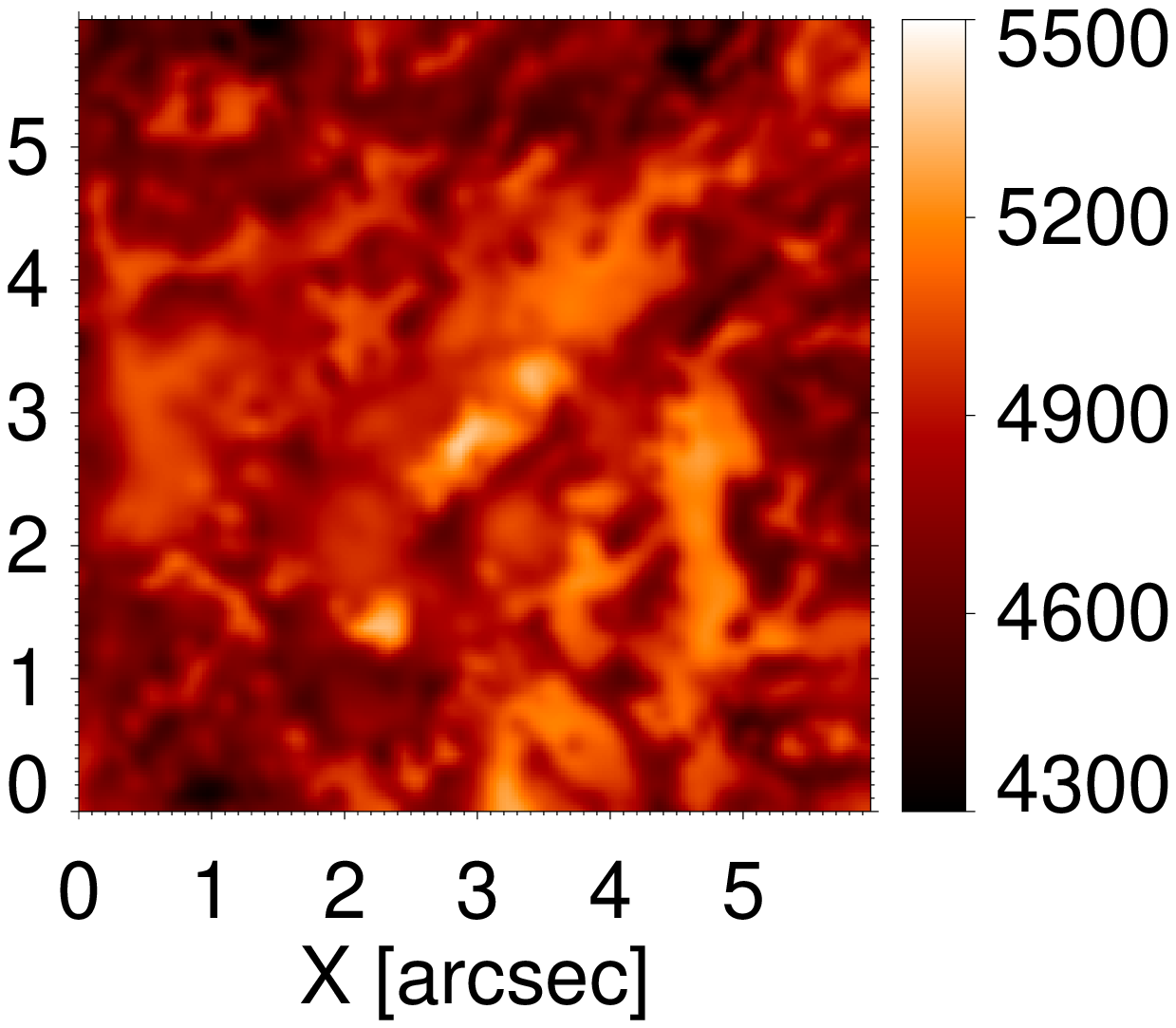}
\includegraphics[scale=.3, trim=4.2cm 2.3cm  1.6cm   1.3cm,clip=true ]{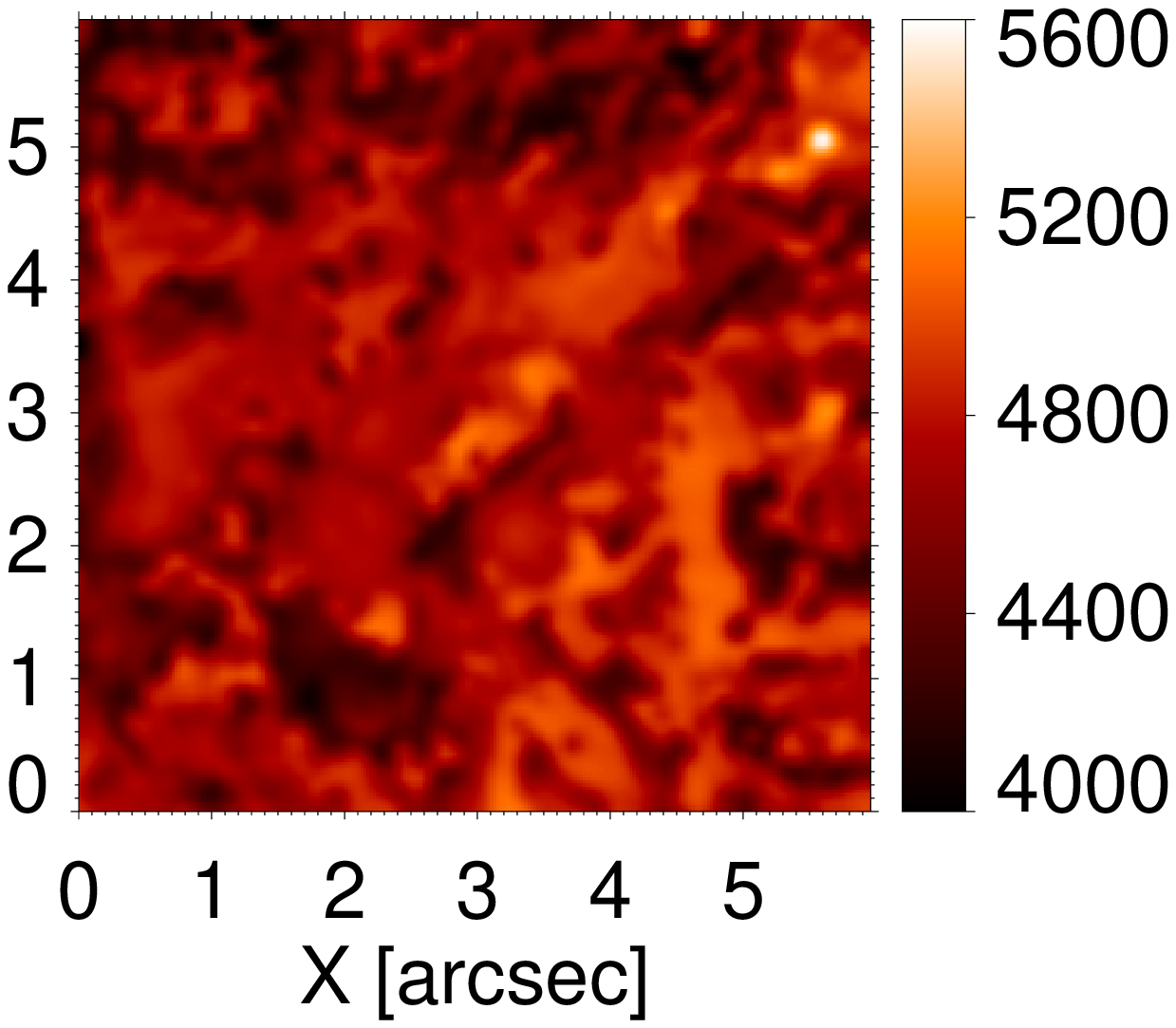}\\
\includegraphics[scale=.3, trim=1.8cm 2.3cm  1.cm  1.3cm,clip=true ]{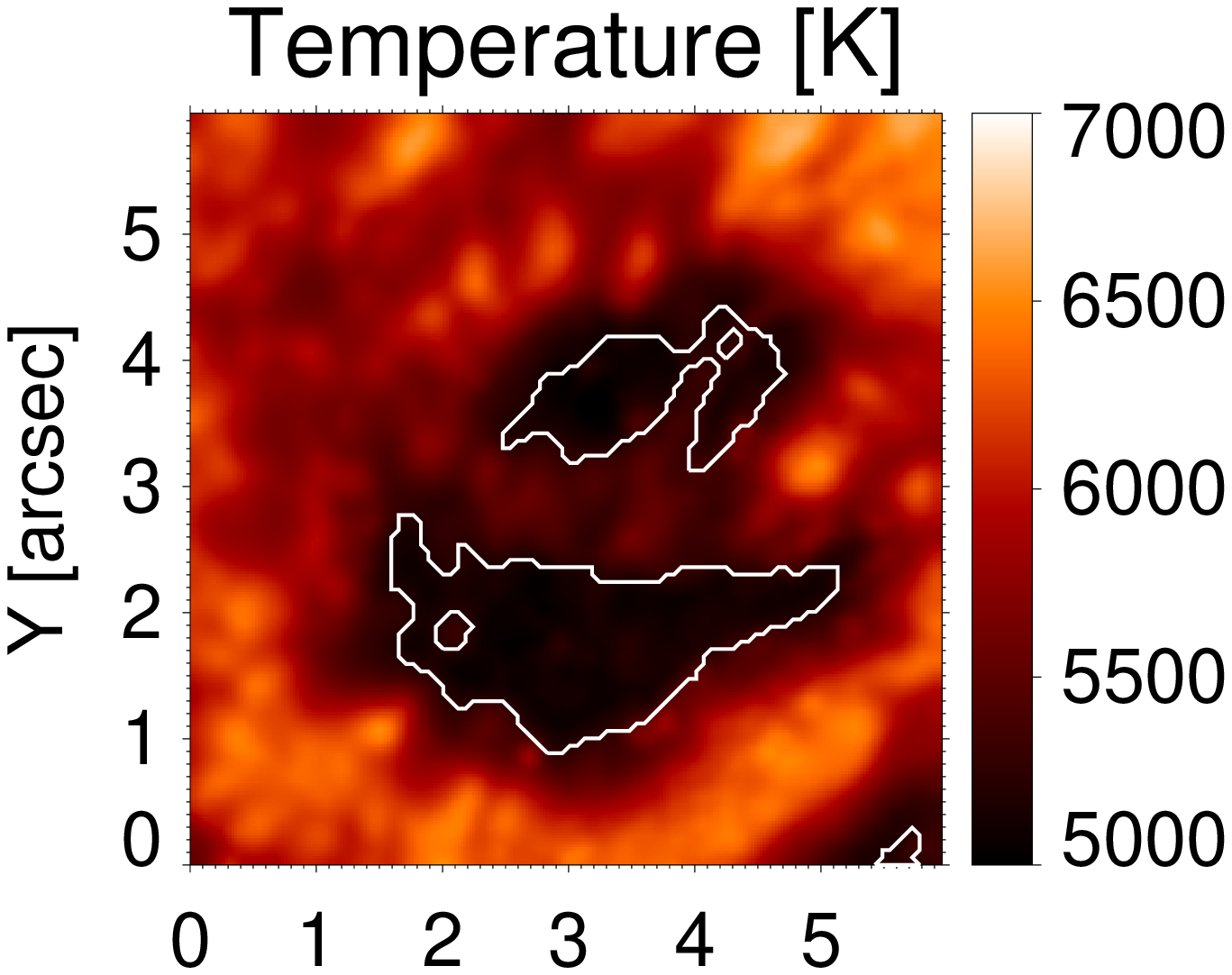}
\includegraphics[scale=.3, trim=4.2cm 2.3cm  1.6cm  1.3cm,clip=true ]{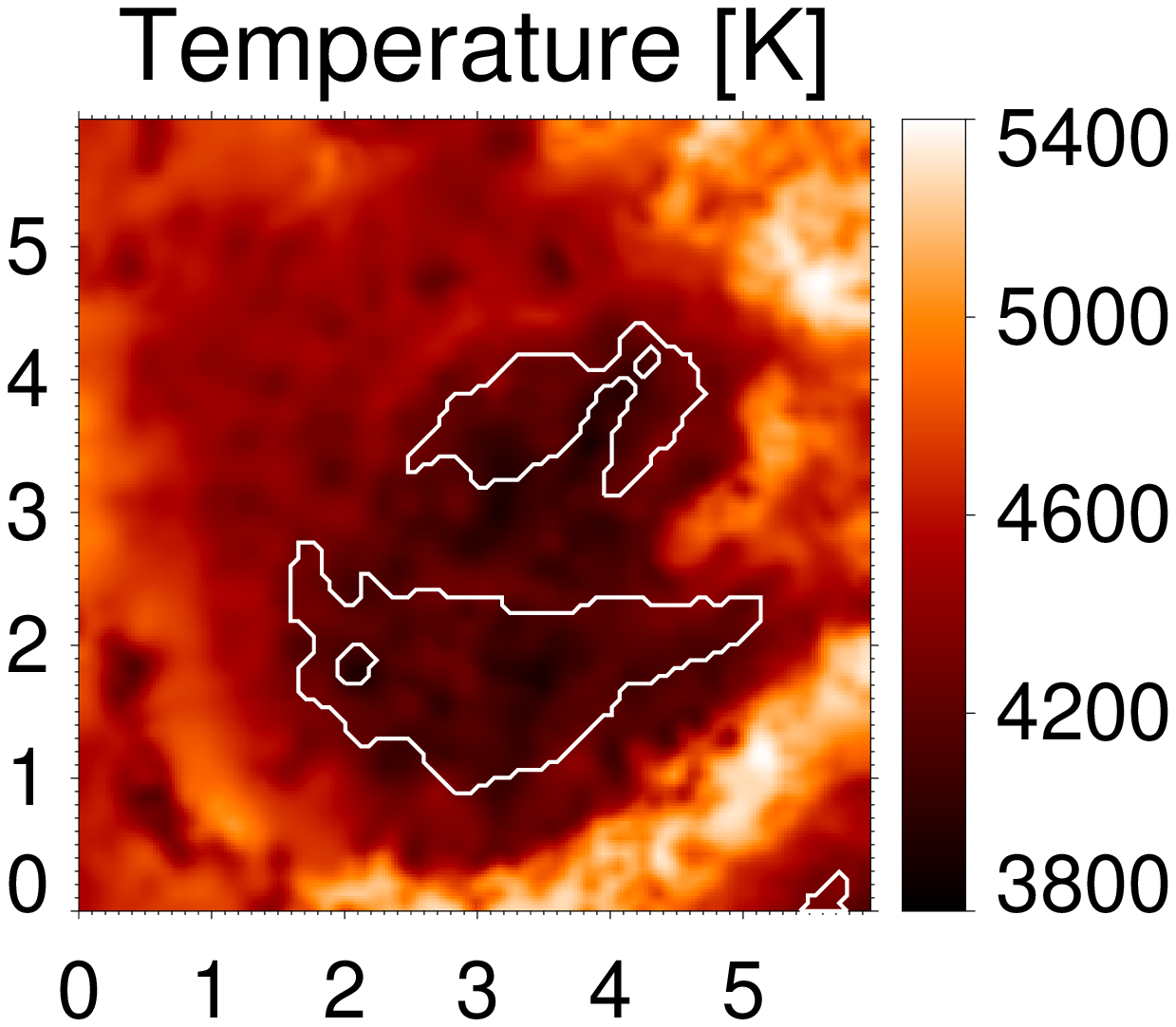}
\includegraphics[scale=.3, trim=4.2cm 2.3cm  1.6cm  1.3cm,clip=true ]{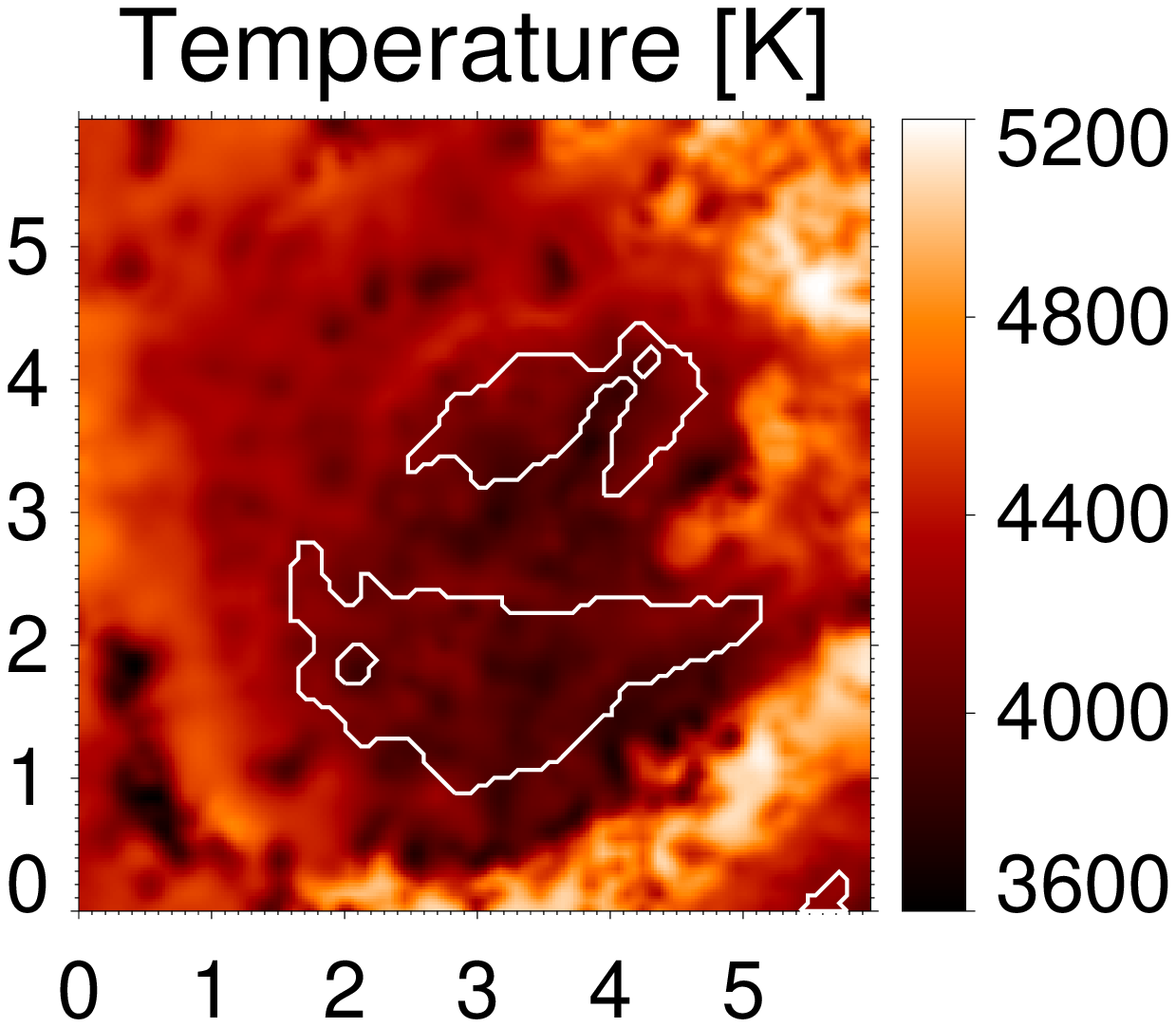}
\includegraphics[scale=.3, trim=4.2cm 2.3cm  1.6cm  1.3cm,clip=true ]{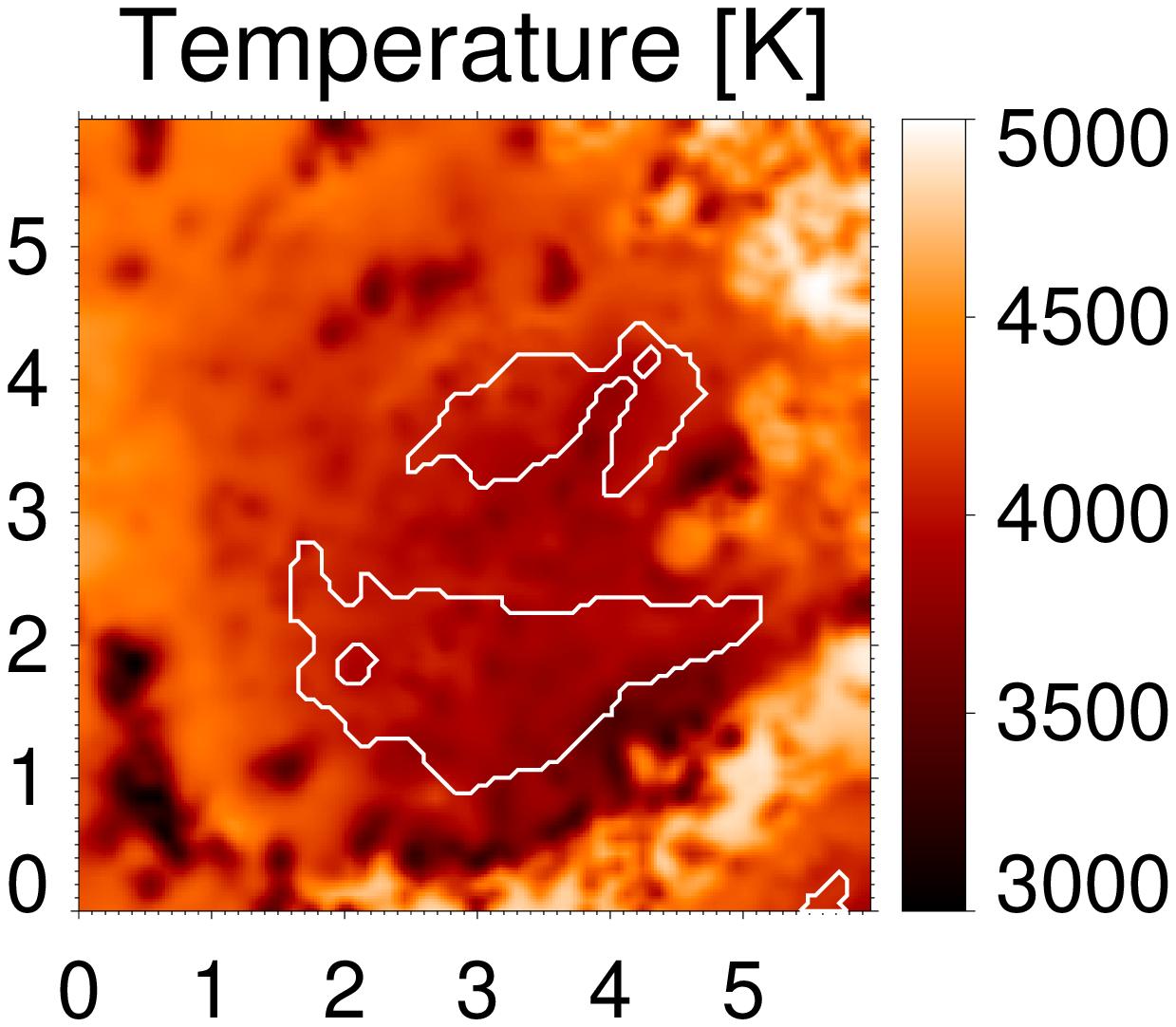}\\		
\includegraphics[scale=.3, trim=1.8cm .3cm  1.cm  1.3cm,clip=true ]{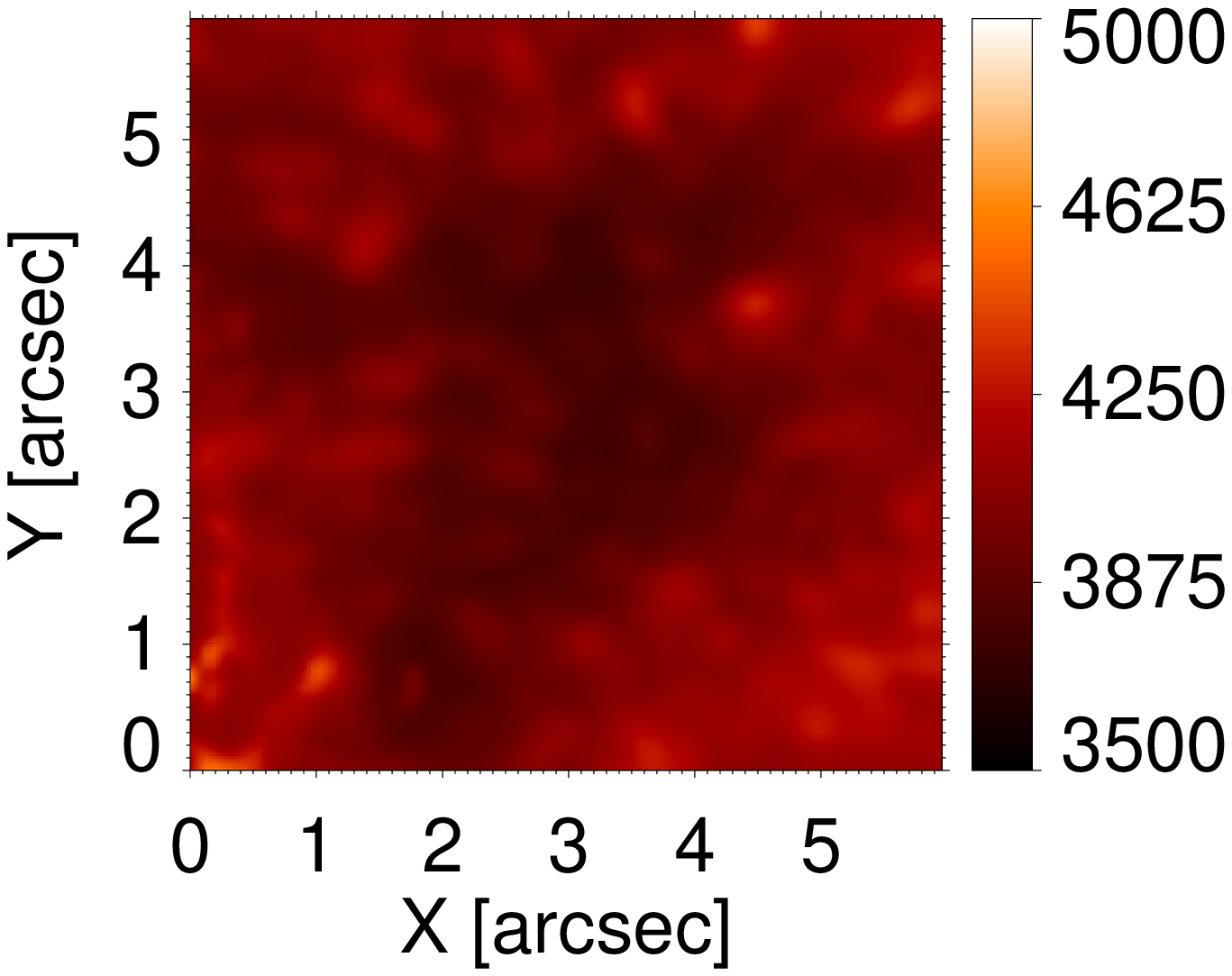}
\includegraphics[scale=.3, trim=4.2cm .3cm  1.6cm  1.3cm,clip=true ]{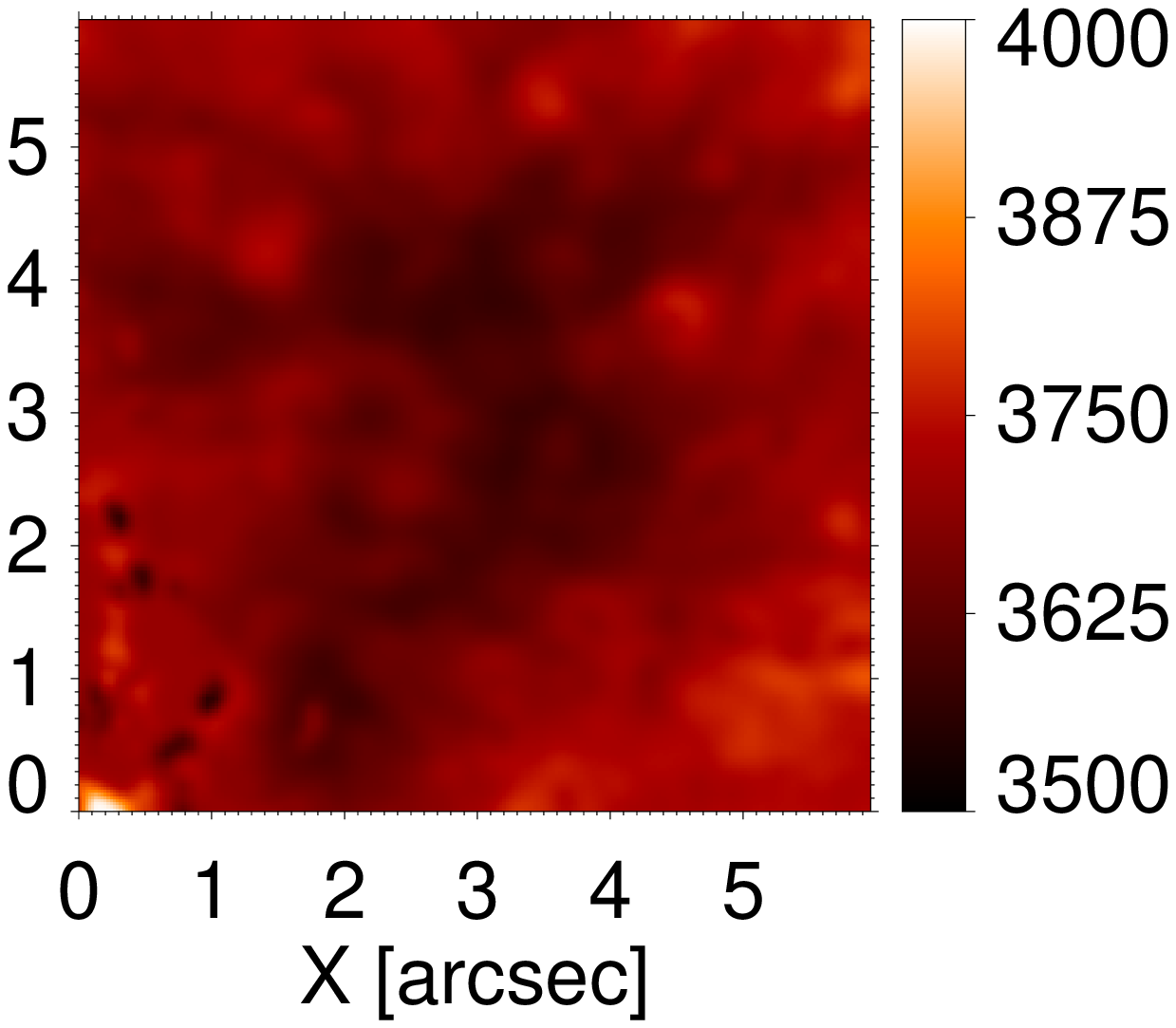}
\includegraphics[scale=.3, trim=4.2cm .3cm  1.6cm  1.3cm,clip=true ]{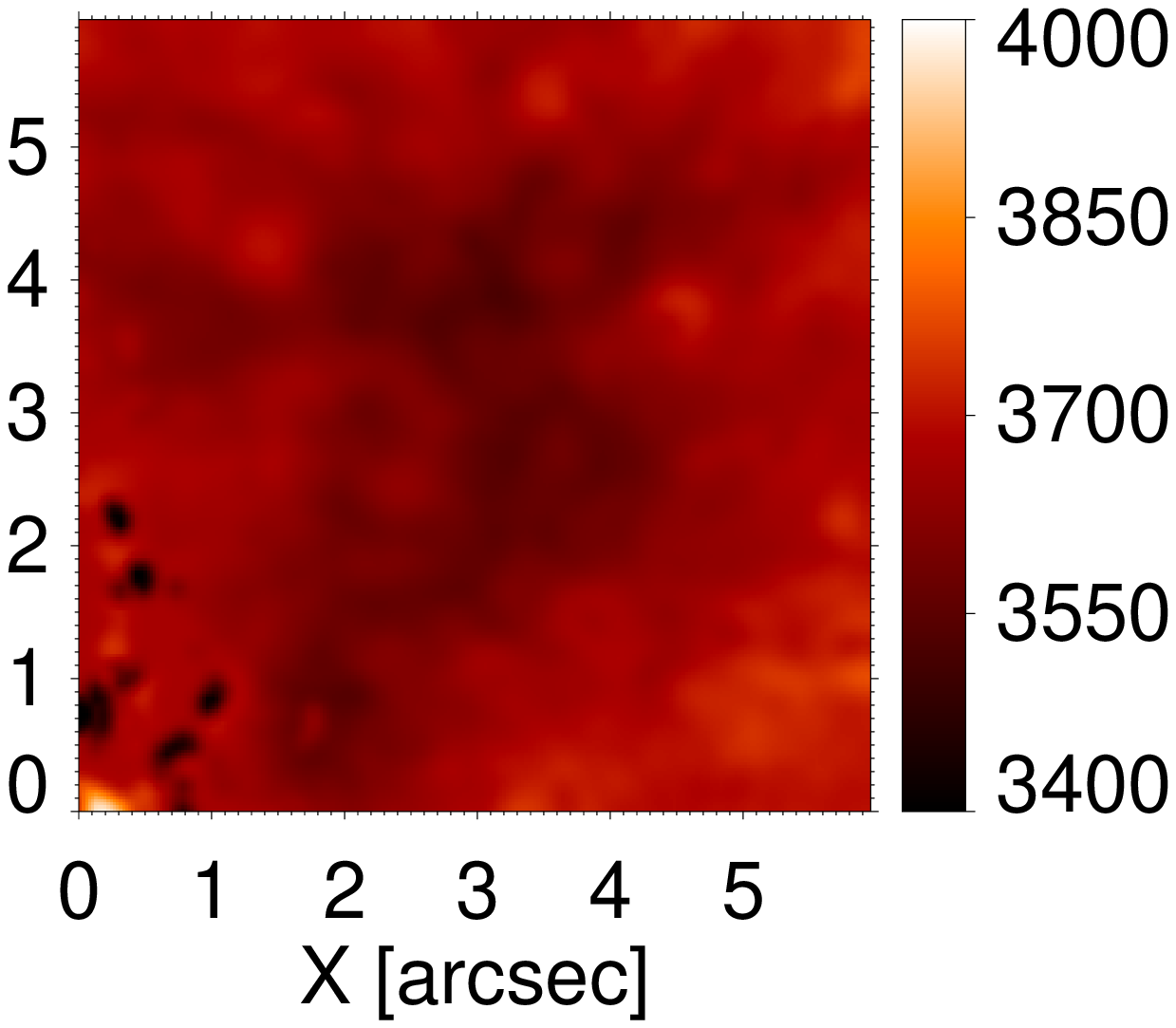}
\includegraphics[scale=.3, trim=4.2cm .3cm  1.6cm  1.3cm,clip=true ]{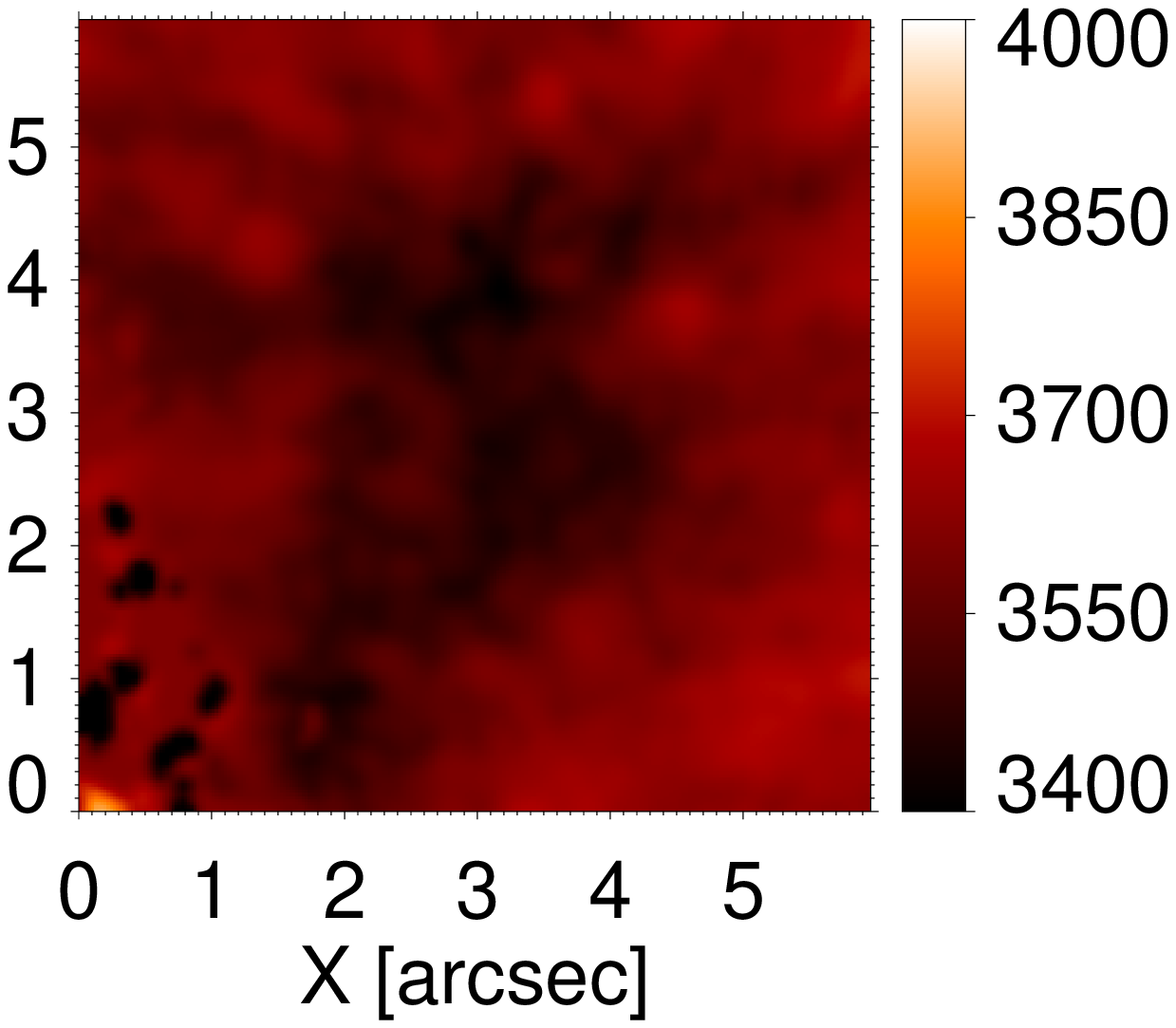}
}
\caption{From left to right, top to bottom: horizontal cuts of the  temperature at four different heights, specifically at $log\tau_{500}$=0, -1.5, -2, -2.5, derived from the inversion of the QS, BPs,  PL, PO and UM 
data. See caption of  Fig. \ref{fig3_c4} for more details.}
\label{fig4_c4}
\end{figure*}

Figure \ref{fig3_c4}  shows examples of the atmospheric models returned by the  data inversion  of the various observed regions.
Each panel displays horizontal cuts at $log\tau_{500}$=0, with $\tau_ {500}$ representing the continuum optical depth at 500 nm. For each region, we show maps of  various physical parameters: temperature, magnetic field strength, gas density, and LOS velocity. 

Temperature values in all maps range from 3500 to 6800 K, the lowest value found in the UM and the highest one in PL and QS regions.
Magnetic field strength reaches 200 G in QS areas, with higher values located within intergranular lanes; in BPs region, it ranges from 0 to 800 G; in  PL and PO 
regions from 0 to 1200 G and from 400 to 2000 G, respectively;  inside the UM region from 1800 to 3200 G.

The LOS velocity in the maps ranges from -2 to 1 km s$^{-1}$  in QS, from -1.5 to 1.5 km s$^{-1}$  in BPs, from -2 to 2  km s$^{-1}$ in PL, and from -0.8 to 0.8 km s$^{-1}$ in the PO and UM regions. For these latter regions, we show the velocity field with respect to the plasma velocity in QS regions. 
 Regions characterized by highest magnetic field strengths, such as central PO and UM regions, and intergranular lanes visible in the QS, display highest density values, as expected when looking at lower geometrical heights due to the higher magnetic field concentration.

Figure \ref{fig4_c4}  shows horizontal cuts of the  plasma temperature in the various observed regions at four different heights, specifically at $log\tau_{500}$=0, -1.5, -2, -2.5. The various panels  display  
plasma temperatures that decrease with atmospheric height for all the analysed regions. 
Top panels show the reversed  granular pattern already  at  atmospheric height $log\tau_{500}$=-1. The same applies to the pattern of BPs, and to less extent also to pattern of PL regions. 

\subsection{Response Functions and uncertainty}
\label{rf}

\begin{figure}
\centering
{
\includegraphics[scale=.45, trim=0.5cm 15.cm 0.2cm 0.cm,clip=true] {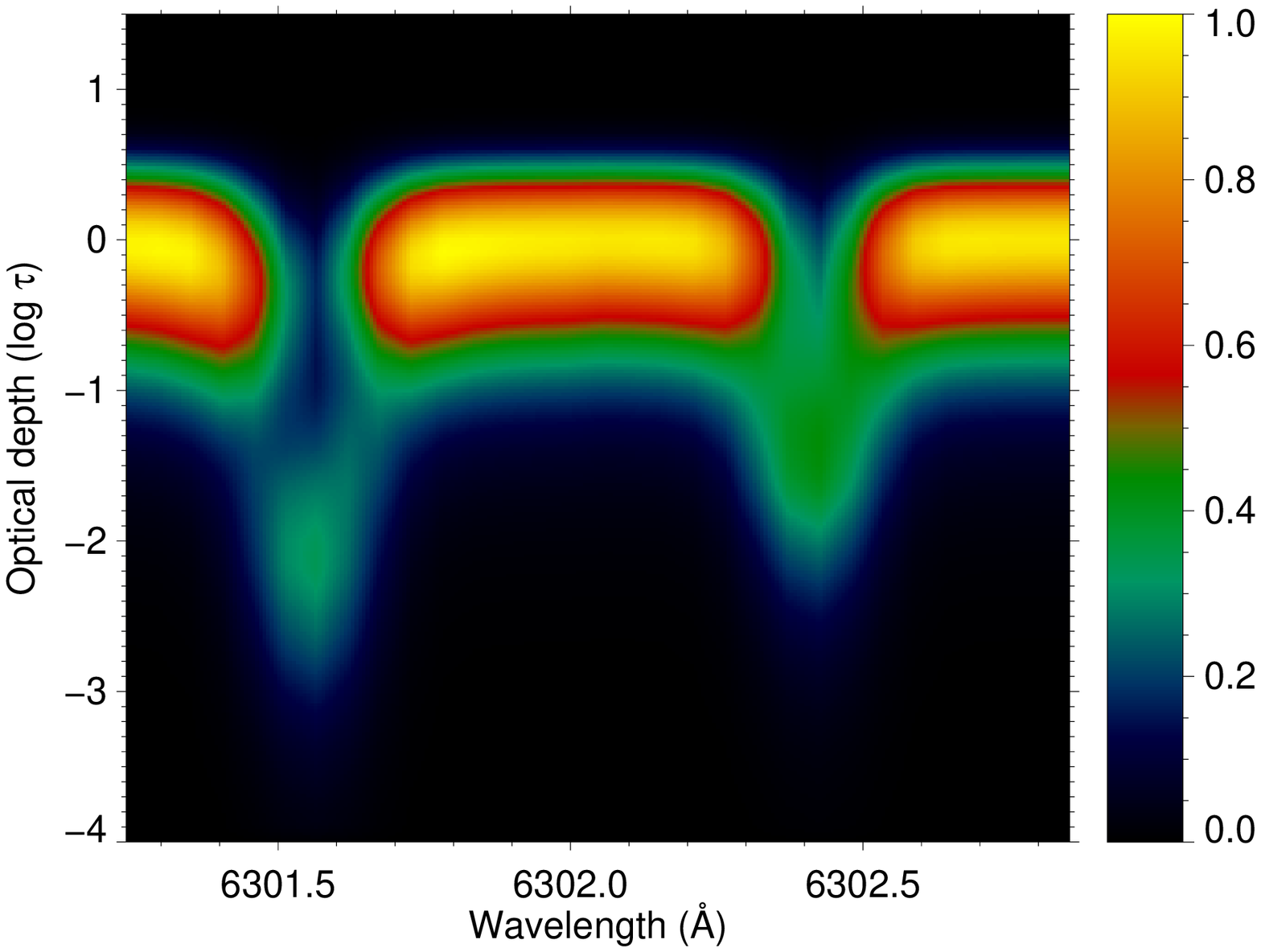}
}
\caption{Normalized RFs of the Stokes-I to temperature perturbation derived from analysis of the QS observations.}
\label{fig5_rf}
 \end{figure}

\begin{figure}
\centering
{
\includegraphics[scale=.45, trim=0.8cm 14.cm 0.cm 1.cm,clip=true] {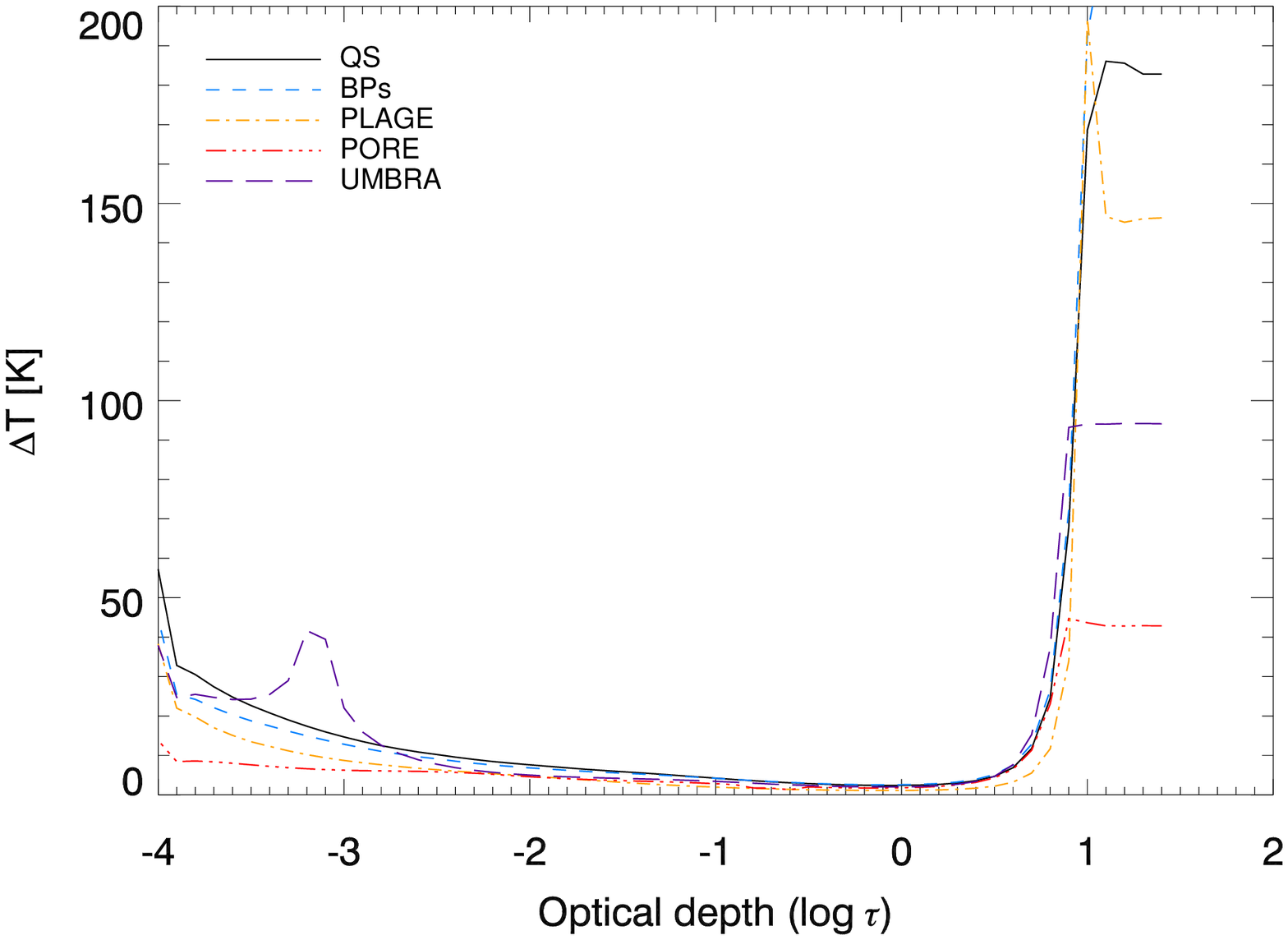}
}
\caption{Temperature uncertainties for each analysed region, as specified in the legend.}
\label{fig5bis_rf}
 \end{figure}

In order to assess the range of atmospheric heights in which the analysed data are sensitive to temperature perturbations, thus to specific properties of the observed atmosphere, we computed the so-called \textit{response functions} \citep[RFs, e.g.][]{Caccin_1977,Landi_1977} 
by applying the mathematical procedure described in \citet{Socasnavarro_2011}. 
Figure \ref{fig5_rf} shows the  RFs  based on the results of the SA inversion of QS data,     
i.e. inversion of data averaged over the studied subFOV, normalized to the maximum value. For a given Stokes parameter, optical depth, and wavelength, RFs values close to unity indicate that the corresponding Stokes-parameter measurements are quite responsive to perturbations of the line-forming atmosphere, while low or null RFs values point out that the Stokes-measurements are unaffected by atmospheric inhomogeneities of temperature and fields. This implies that the data inversion cannot provide reliable information about the physical quantities in the line-forming atmospheric regions  characterized by low RFs values; these regions lie outside the sensitivity range of analysed data. Figure \ref{fig5_rf} shows that, for the observations considered in our study, the sensitivity range  
 spans from $log\tau_{500}$=0  to $log\tau_{500}$=-3.  
The variation of the emergent intensity due to temperature perturbations is always positive: the emergent intensity increases both in the continuum and in the line core, with most of the contribution to the analysed spectra coming from the continuum.


Following  \citet{Socasnavarro_2011}, we also computed the uncertainty in the atmosphere models derived from the data inversion, by weighting the average of the temperature stratification ($T(\tau)$, hereafter) obtained from the inversion at each observed spectral point 
by the above estimated RFs. 
 Figure \ref{fig5bis_rf} shows the  uncertainty estimated for the $T(\tau)$ derived from the five inverted subFOVs. This uncertainty is as low as $\sim$15  K between $log\tau_{500}$=0 and $log\tau_{500}$=-3, i.e. the range of atmospheric heights in which the data inversion returns more reliable results. 
The uncertainty of data inversion results  increases at higher atmospheric heights  and below $log\tau_{500}$=0, up to 50 K and 200 K, respectively, as well as with decreasing the spatial scale of the magnetic feature represented by the inverted subFOV.

\subsection{Temperature stratification}
\label{1Dcomp}

 \begin{figure}
\centering
{
\includegraphics[scale=.4, trim=0.cm 8.cm 0.2cm 0.cm,clip=true] {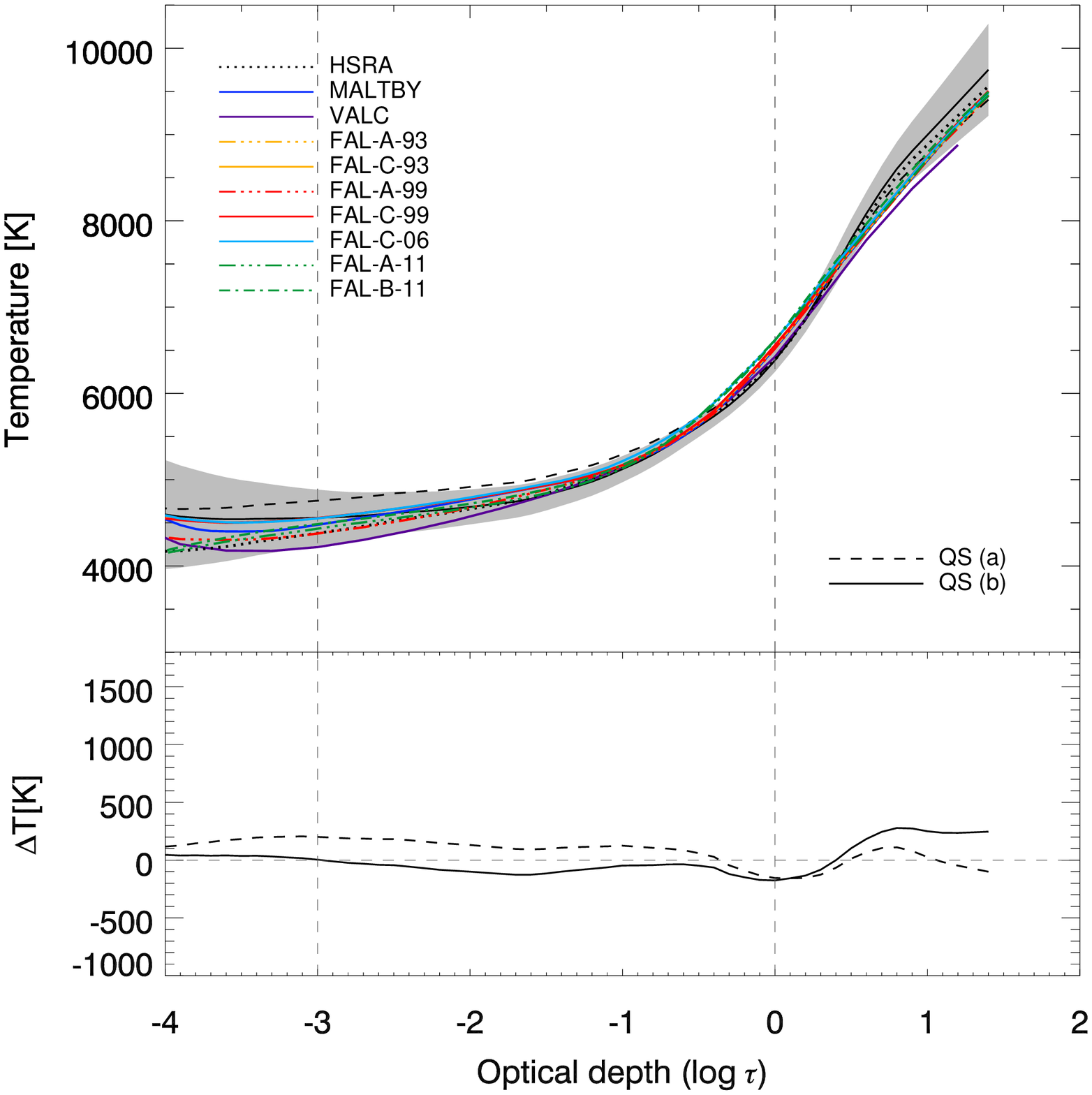}
}
\caption{Top: Comparison among the $T(\tau)$ of several 1D models (co\-lou\-red lines, as specified in the legend) and in the model derived from the inversion of the QS observations. Grey-shaded area represents the 1$\sigma$ confidence interval of data inversion results. Dashed and solid, black lines refer to the $T(\tau)$ retrieved from the SA and FR computations, labelled (a) and (b), respectively. Bottom: Relative difference between the $T(\tau)$ retrieved from the data inversion and that in the FAL-C-99 model. The horizontal, dashed line marks zero values of these differences; vertical, dashed lines in both panels mark the sensitivity range defined by the RFs.}
\label{fig6_c4}
 \end{figure}


\begin{figure}
\centering
{
\includegraphics[scale=.4, trim=0.cm 8.cm 0.2cm 0.cm,clip=true]{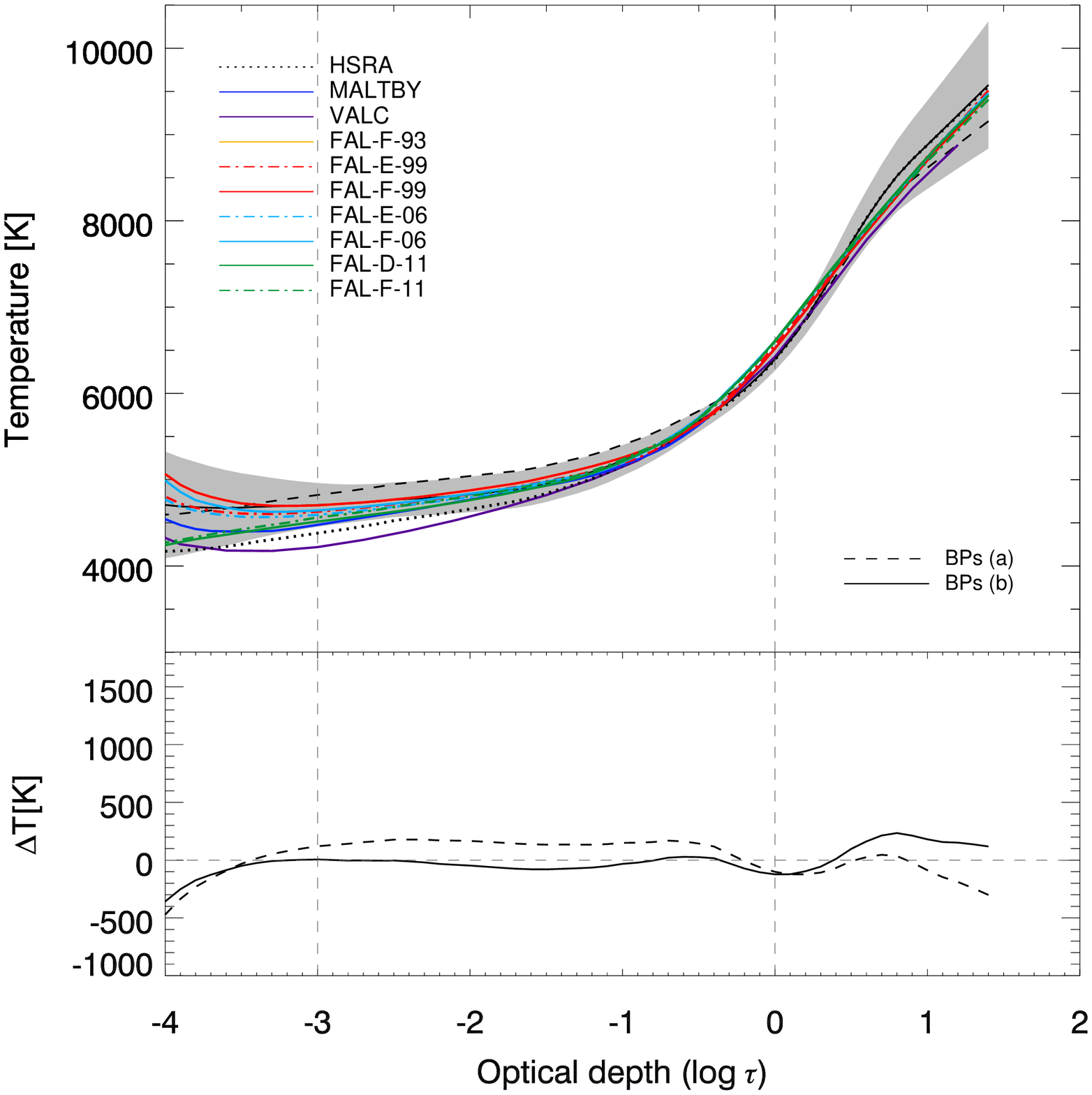}\
\includegraphics[scale=.4, trim=0.cm 8.cm 0.2cm 0.cm,clip=true]{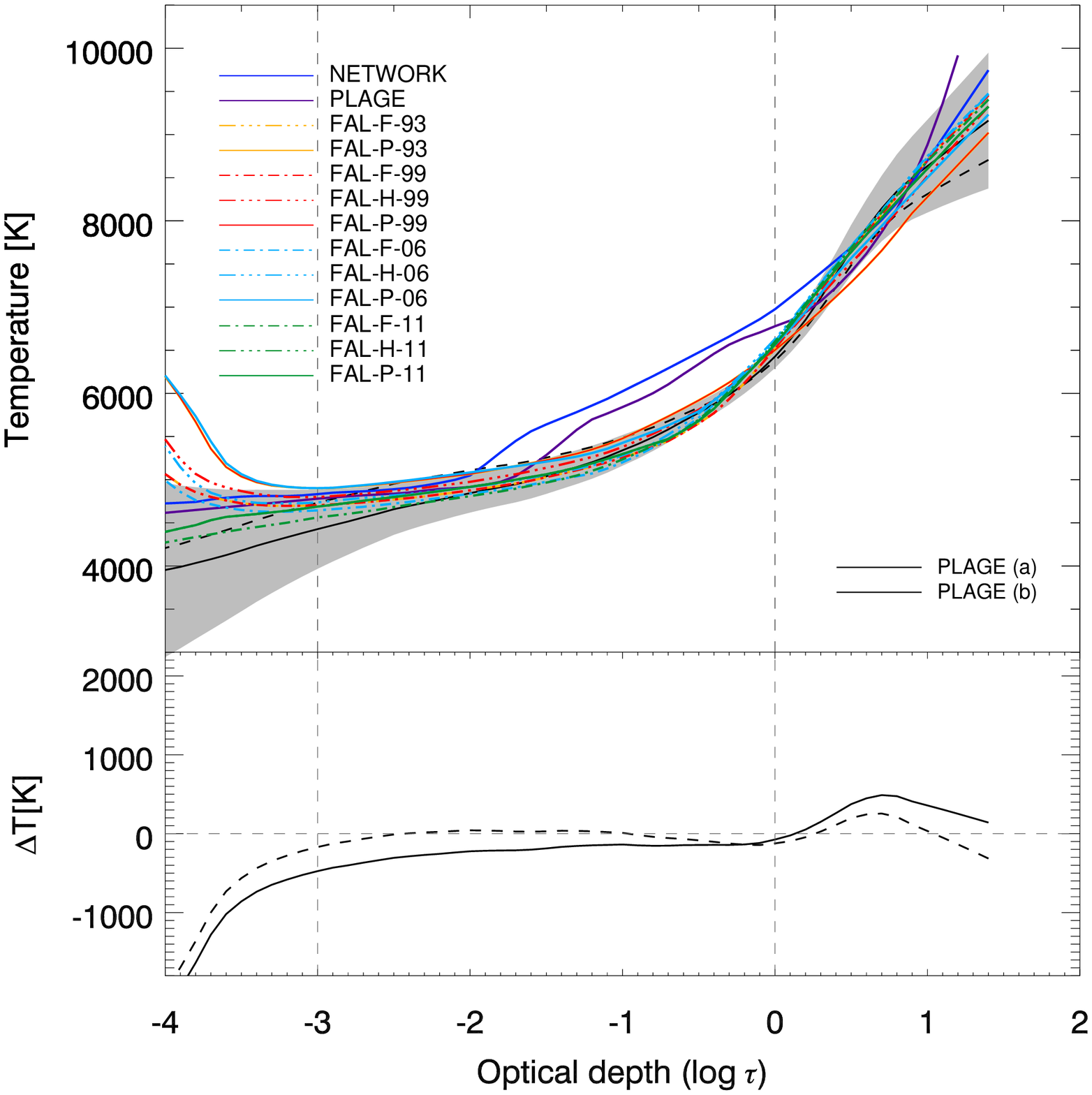}}
\caption{As in Fig. \ref{fig6_c4} but for data representative of small-scale (BPs, top) and large-scale  (PL, bottom) bright magnetic regions. The relative difference is computed with respect to the temperature stratification of the FAL-F-99 (top) and the FAL-P-99 (bottom) models. See caption of Fig. \ref{fig6_c4} for more details.}
\label{fig7_c4}
\end{figure}

\begin{figure}
\centering
{
\includegraphics[scale=.4, trim=0.cm 8.cm 0.2cm 0.cm,clip=true]{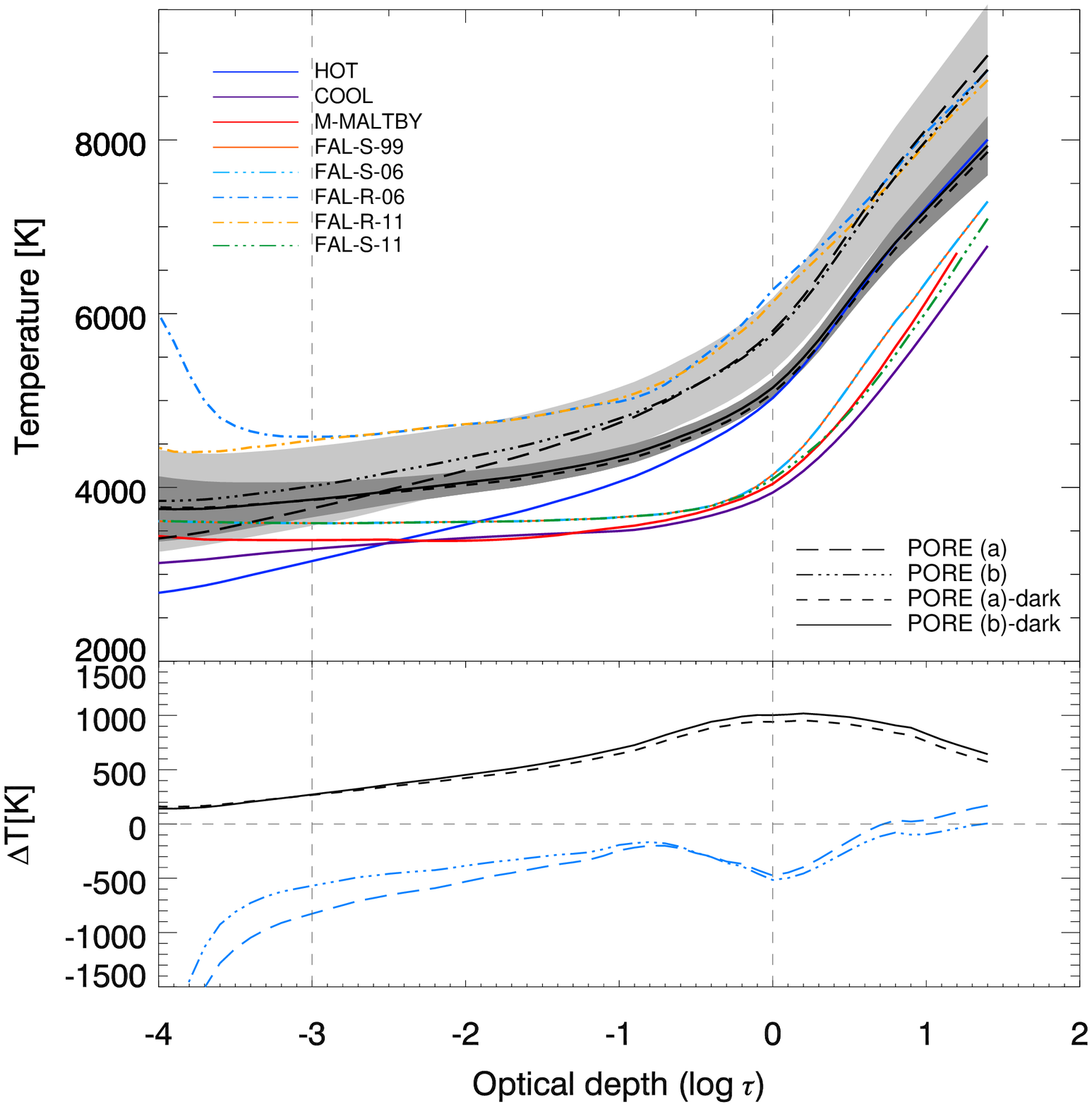}\\
\includegraphics[scale=.4, trim=0.cm 8.cm 0.2cm 0.cm,clip=true]{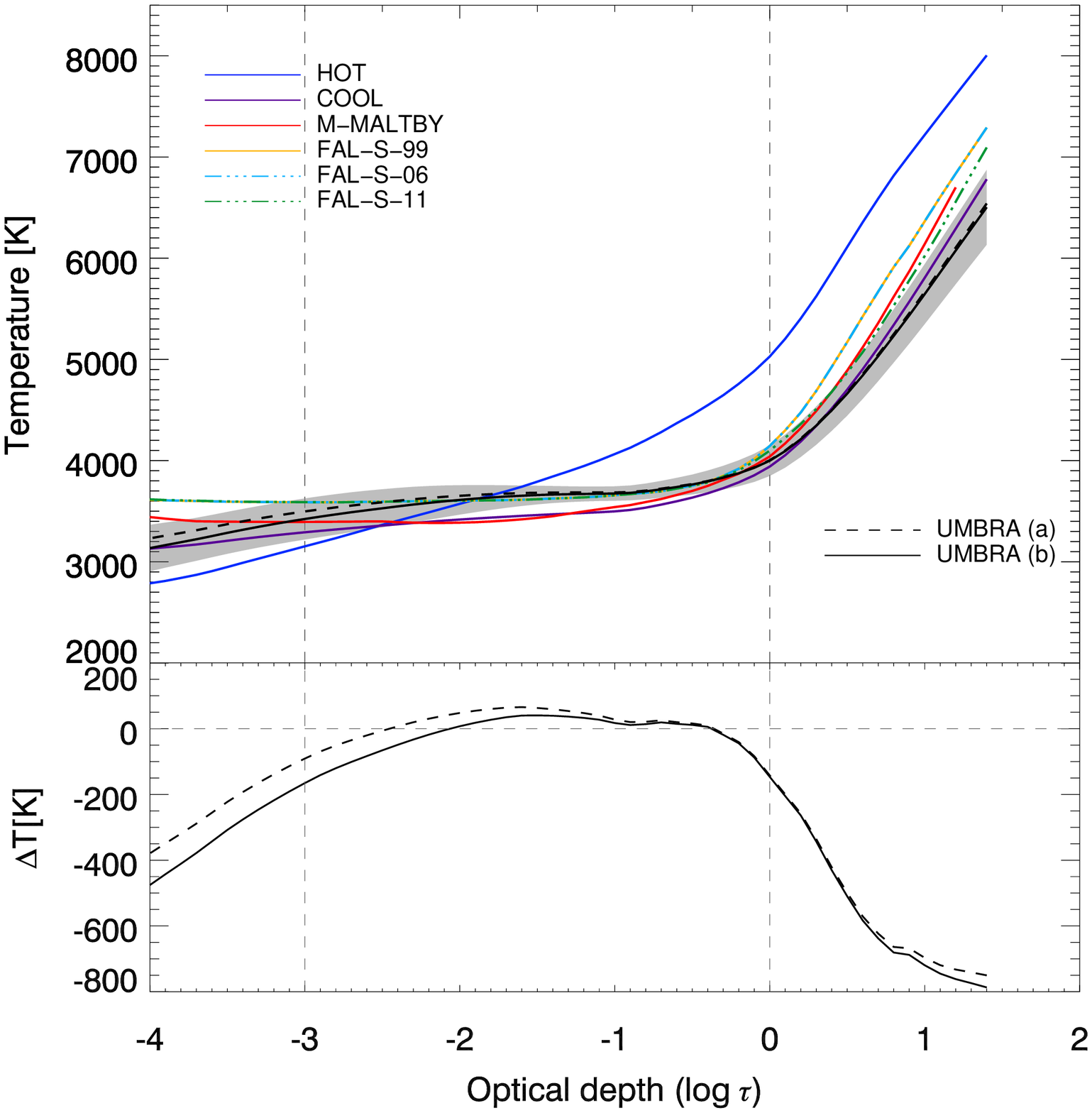}}
\caption{As in Fig. \ref{fig6_c4} but for data representative of dark magnetic regions, PO (top) and UM (bottom). For PO data, labels  (a) and (b) refer to the $T(\tau)$  computed over the whole subFOV, while labels  (a)-dark and (b)-dark refer to the $T(\tau)$  over the region with I$_c<$0.4. The relative difference is computed with respect to the temperature stratification of the FAL-R-06 (dot-dashed and long-dashed lines) and the FAL-S-99 (solid and dashed lines)  models, respectively. See caption of Fig. \ref{fig6_c4} and Sect. 3 for more details.}
\label{fig7bis_c4}
 \end{figure}

We compared the $T(\tau)$  derived from the data inversion of  the various studied regions  to the ones described by  the several 1D models listed in Table \ref{tab:table1}. We here consider  results obtained from inversion of data taken at best seeing conditions for each analysed region and for both the SA and FR computations.

Figure \ref{fig6_c4} (top panel) shows this comparison for QS data.
Dashed- and solid-black lines display the $T(\tau)$ obtained from SA  and FR, labelled (a) and (b), respectively.  Coloured lines correspond to the 1D models employed for comparison as specified in the legend. Grey-shaded area represents the 1$\sigma$ confidence interval of data inversion results.  Figure \ref{fig6_c4} (bottom panel)  shows the relative difference between the  $T(\tau)$ derived from the observations and the one in  a 1D model used as reference, specifically the FAL-C-99 model.  

The panels in Figs. \ref{fig7_c4} and \ref{fig7bis_c4} show the same content  as Fig. \ref{fig6_c4}, but based on the results obtained from the inversion of the BPs,  PL, PO, and UM observations. The bottom part in each panel  shows the relative difference between the $T(\tau)$    derived from the observations and the one in the 1D model used as reference, which is the FAL-(F, P, S)-99 to represent network, plage, and umbral regions, respectively; we also 
analysed the FAL-R-06 model for penumbral regions.

Concerning the PO data, we considered results from SA and FR computations on the image pixels with I$_c<$0.4  (labeled (a)-dark and (b)-dark, respectively), and over the whole subFOV  (labeled (a) and (b), respectively).  
We then estimated the relative difference between the $T(\tau)$ computed over the whole subFOV and the FAL-R-06, as well as between the $T(\tau)$ computed over the image pixels with I$_c<$0.4 and the FAL-S-99.

Figures \ref{fig6_c4}, \ref{fig7_c4}, and  \ref{fig7bis_c4} show that
the various $T(\tau)$  derived from the data inversion agree  quite well with those in most of the compared models, both qualitatively and quantitatively, but for PO observations.  The agreement between compared models decreases outside the sensitivity range defined by the  RFs; we recall, however, that outside the sensitivity range  the physical quantities returned by the data inversion are  uncertain.
 For all studied regions, the $T(\tau)$ obtained from SA and FR computations slightly differ. 
 We discuss this difference in the following and mostly focus here on results from FR computations only. 

At $log\tau_{500}$=0,  the average of the temperature values obtained from the inversion of the QS, BPs, PL, UM data  agree with those in the FAL-(C, F, P, S)-99 models within the deviation of results on  the analysed  subFOV, being the average and standard deviation of values  
6383$\pm$132 K, 6397$\pm$132 K,  6427$\pm$132 K,  and 3998$\pm$150 K with respect to the values 6520 K, 6520 K, 6502 K, 4170 K in the FAL-(C, F, P, S)-99 models, respectively. At same atmospheric height,  the value of the plasma temperature estimated  by the  inversion of PO data is  5147$\pm$109 K, $\sim$1000 K higher and $\sim$1100 K lower than the values envisaged in the FAL-S-99 and FAL-R-06 models, respectively. The relative difference between our $T(\tau)$ derived from analysis of the whole subFOV and the FAL-R-06 model is less pronounced;  the $T(\tau)$  of the FAL-R-06 model lies within the deviation of values derived from our analysis, in the atmospheric range between $log\tau_{500}$=-0.5 and $log\tau_{500}$=-2.

 Within $log\tau_{500}$=-1 and $log\tau_{500}$=-3, i.e. from the middle to the high photosphere, the  $T(\tau)$ returned from the observations of QS, BPs, PL, and UM regions agree within  $\sim$10\% with all the $T(\tau)$ in the models by \citet[][]{Fontenla_etal1999} employed for comparison, but with slightly different results for the various compared sets;  for QS, BPs,  and UM regions, the agreement is within 5\%, while for the PL regions within 10\%.  
Overall, most of the $T(\tau)$ derived from the data inversion  are slightly lower than the ones in  the compared 1D models in the middle photosphere (about 100 K at $log\tau_{500}$=-1) and slightly higher in upper layers (about 150 K at $log\tau_{500}$=-3) and below $log\tau_{500}$=0 (about 200-400 K), but for the UM data, which show lower plasma temperatures  (down to $\simeq$700 K) below $log\tau_{500}$=0 than those displayed by all compared models, and the PL data which exhibit lower values (about 50$-$100 K) above  $log\tau_{500}$=-2.5, than all other models. In particular, 
the $T(\tau)$ obtained from QS and BPs  data are, on average,  up to $\sim$100 K lower than in the corresponding FAL-C-99 and FAL-F-99  models, respectively.  
In the range between $log\tau_{500}$=0 and $log\tau_{500}$=-2, i.e. in the lower  and middle  photosphere, the $T(\tau)$  obtained from PL data is, on average, up to $\sim$400 K lower than represented in the corresponding FAL-P-99 model. 
On the other hand,  at these atmospheric heights, the $T(\tau)$  from PO observations  is up to $\sim$1000 K higher than reported by the FAL-S-99 model; for the  UM data,  it  is close (within $\sim$50-100 K) to that described in  the FAL-S-99 model, but it is 
$\sim$150-200 K lower at $log\tau_{500}$=0 and $log\tau_{500}$=-3.
 
All the atmosphere models derived from the observations exhibit higher plasma  temperatures at higher atmospheric heights than those represented by the earlier HSRA and VAL-C models, 
 except for the model derived by the PL data.  Besides, the $T(\tau)$ obtained from  BPs and PL data also do not reproduce the  temperature enhancement represented in the SOLANNT and SOLANPL models, neither considering SA, nor FR results.  The $T(\tau)$  from SA analysis of PL data most closely follows the FAL-P-93 and FAL-H-99 models.
 

Atmosphere models derived  from observations seem to reproduce better former models by \citet[][]{Fontenla_etal1993,Fontenla_etal1999} than more recent sets by  \citet[][]{Fontenla_etal2006,Fontenla_etal2011}, at least in the lower photosphere up to $log\tau_{500}$=-2, and mainly for QS and BPs regions. The opposite seems to occur in  upper atmospheric layers, where however the level of confidence of our data inversion results is lower.


\subsection{Effects of spatial averaging and temporal evolution} 
\label{averaging}

 \begin{figure}
\centering
{
\includegraphics[scale=.65, trim=1.5cm 1.5cm 6cm 1.cm,clip=true]{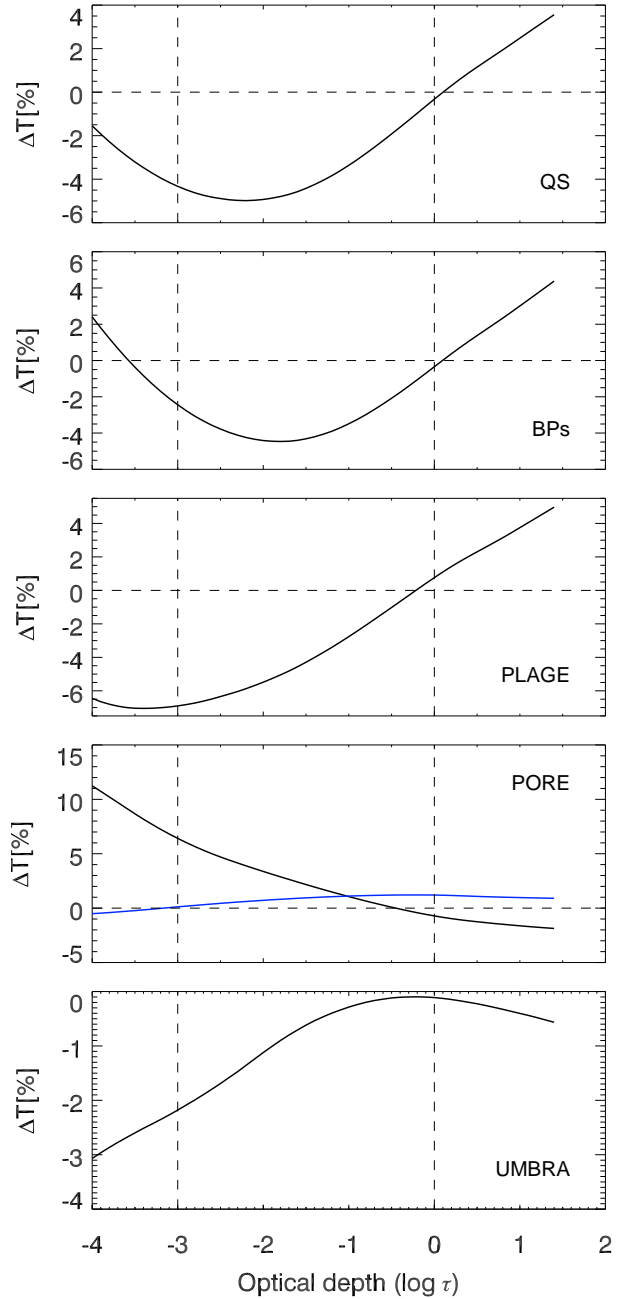}
}
\caption{From top to bottom: Relative difference between the average $T(\tau)$  obtained from analysis of fully-resolved (FR) and spatially-averaged (SA) results from inversion of the QS, BPs, PL, PO, and UM data. The blue line in the panel of PO data shows results obtained by considering only image pixels with I$_c<$0.4.}
\label{fig10_c4}
 \end{figure}

 \begin{figure}
\centering
{
\includegraphics[scale=.65, trim=0.5cm 1.5cm 6cm 1.cm,clip=true]{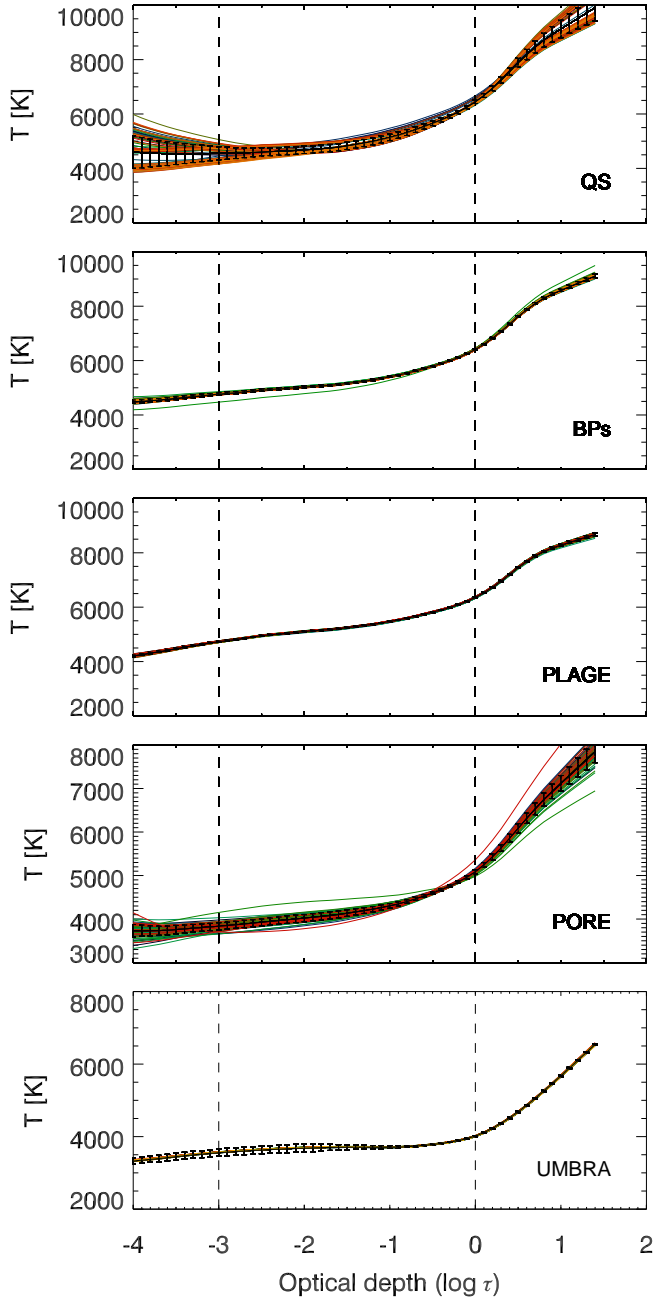}
}
\caption{From top to bottom: $T(\tau)$ obtained from the analysis of the whole series of available QS, BPs, PL, PO, and UM observations. Each temporal step in the observational series is shown with different colour. Error bars indicate the standard deviation of values with respect to the  $T(\tau)$ averaged over the whole time series, shown with black, solid line. }
\label{fig10bis_c4}
 \end{figure}

Figures \ref{fig6_c4}, \ref{fig7_c4} and \ref{fig7bis_c4} show slight differences between the $T(\tau)$  obtained from  SA and FR computations, i.e. the spatially-averaged and fully-resolved  observations.
In Fig. \ref{fig10_c4} we quantify this difference, by showing relative percentage values between the $T(\tau)$ obtained under the two computations applied; for PO regions we also show results from analysis of image pixels with I$_c<$0.4. At $log\tau_{500}$=0,  the difference between $T(\tau)$  obtained from SA and FR lies within 1\% for all the analysed regions. For PO and UM data, the difference is  within  2\% at all the investigated atmospheric heights, if we restrict our analysis to image pixels with I$_c<$0.4.

Within $log\tau_{500}$=0 and $log\tau_{500}$=-3, i.e. from the lower  to the higher photosphere, for QS, BPs, and PL regions, the $T(\tau)$  computed from SA on the whole subFoV has up to 6\% higher values than obtained from FR, while for the PO  the $T(\tau)$ has up to $\simeq$6\% lower values; for UM, the $T(\tau)$  computed from SA on the whole subFOV has only up to 2\% higher values than obtained from FR.  Therefore, the results obtained from bright and dark magnetic regions are  affected by the method applied  similarly, but  the sign for PO observations; this holds  if the analysed data are characterized by a spatial resolution of $\approx$5-6 arcsec as considered in our study.
 
The above results indicate that the method applied affects less the modeled atmosphere in homogeneous magnetic regions. 
This is in  agreement with  \citet{Uitenbroek_Criscuoli_2011}, who showed that spatially-averaging the properties of an inhomogeneous atmosphere returned from MHD simulations and evaluating physical quantities after the averaging operation, does not give the same result as estimating the physical quantities in the inhomogeneous atmosphere and then averaging it. 

We also investigated the possible effects  due to the temporal evolution of the observed features on the obtained results, and other possible processes occurring on the analysed regions (waves, seeing, etc.). To this purpose, we analysed  inversion results derived from SA computations on the whole series of data available for each observed region.
In Fig. \ref{fig10bis_c4} we show the $T(\tau)$  derived from the inversion of all averaged Stokes spectra for each observed region. 
The $T(\tau)$ retrieved from analysis of each observation available for the studied region is displayed with different colours.
 The $T(\tau)$  averaged over the whole time series is displayed as a solid, black line with error bars representing the 1$\sigma$ confidence interval; this interval is larger for QS and PO regions. Figure \ref{fig10bis_c4} shows that the dispersion of results due to the effects of the temporal evolution of the studied region and other possible processes lies within the confidence interval of results estimated for all the analysed regions. This finding proves that the results presented in Sects. 3.1-3.3 can be assumed to be quite representative of the studied regions, at least for the dataset considered in our study.

\section{Comparison with results in  the literature} 
\label{liter}

The literature presents a number of atmosphere  models derived from inversion  of spectro-polarimetric data acquired with both ground-based and space-borne instruments. 
We now discuss the $T(\tau)$  derived from our analysis with respect to those reported by former studies of QS, PL, and UM regions. We focus on the models presented since year 2000, and derived from analysis of data taken with similar characteristics, in terms of spatial and spectral resolution, than the ones considered in our study. 

\citet{Borrero_2002} presented a two-components model of the quiet solar photosphere, representative of typical granular and intergranular regions, derived  from inversion performed with the SIR code on the intensity of 22 {Fe\,{\footnotesize I} lines, 
 observed at the Fourier Transform Spectrometer (FTS) installed at the McMath telescope of the Kitt Peak Observatory. 
 The data consist of 1579 spectral points that sample the 22 selected {Fe\,{\footnotesize I} lines at intervals of 6 m{\AA}.
At $log\tau_{500}$=0, the plasma temperature derived from our analysis of the QS observations is comparatively close to the values in both their models, with  relative differences of $\sim$100 K. Given that the granular component is statistically predominant in QS regions, we expect that our QS model, which was obtained without distinguishing between the two components considered by \citet{Borrero_2002},  is closer to their model for the granular region than the one for intergranular areas. Indeed, our $T(\tau)$  from FR computations lies close  to that in their granular model at all the optical depths; values are within the deviation of results on our analysed subFOV.

\citet{Socasnavarro_2011} inverted full-Stokes spectro-polarimetric observations of a QS region observed at the {Fe\,{\footnotesize I} line pair at 630 nm with the slit spectro-polarimeter onboard the Hinode satellite, by using the 
 NICOLE code \citep{Socasnavarro_etal2015} that allows  inversions under non-LTE (NLTE) conditions. 
 The observed FoV was close to disc center. The data consist of three consecutive scans of 315 slit positions on the FoV, each position with 112 wavelength samples of the {Fe\,{\footnotesize I} line pair taken with $\simeq$21 m{\AA}  sampling.
 The inversion was performed on a sub-array of about 30$\times$30 arcsec$^2$, 200$\times$200 pixels wide. The retrieved average temperature stratification was compared to the HSRA and the model by \citet[][]{Asplund_2004}, which resulted to be warmer in the middle layers compared to the model presented in \citet{Socasnavarro_2011} and cooler upwards; it is very close to that obtained from our analysis of FR QS data, but for the  elbow at  $log\tau_{500}$=-1 that is not found neither in our results, nor in  the HSRA model. 
 The uncertainties derived from  RFs by  \citet{Socasnavarro_2011}  exhibit a similar trend than  the ones derived from our study, at least for image pixels with continuum brightness close to the average of the whole observed region. For those pixels, the uncertainties estimated by \citet{Socasnavarro_2011} are lower than 50 K in the middle atmospheric layers, up to  $log\tau_{500}$=-3.4, and reach values up to more than 500 K in the upper layers. These values are sensitively higher than those derived from our study.

\citet{BellotRubio_2000} analysed averaged Stokes-I and Stokes-V spectra of the {Fe\,{\footnotesize I} line pair at 630 nm emerging from a facular region observed at $\mu$=0.96 with the slit Advanced Stokes Polarimeter (ASP) at the Sacramento Peak Observatory.
The observations, which covered a FoV of about 110$\times$90 arcsec$^2$, were taken in almost 20 minutes. The spatial resolution of the acquired data was $\simeq$1-3 arcsec and the spectral sampling was $\simeq$13 m{\AA}. The analyzed data consist of averaged Stokes profiles constructed by accounting for the contribution of all pixels within facular regions whose degree of polarization was lower than 0.4\%. 
The model they presented for the central part of the studied region  is hotter than the model we obtained from the PL data, $\sim$500 K hotter at $log\tau_{500}$=0. Our $T(\tau)$   better agrees with the one reported by \citet{BellotRubio_2000}  for  plage outlying regions. However, the $T(\tau)$  obtained from our study  agrees with their results within the standard deviation of values  in our studied  subFOV.

\citet{Buehler_2015}  analysed the full-Stokes spectra of a plage region observed at the {Fe\,{\footnotesize I} line pair at 630 nm with  the slit spectro-polarimeter aboard the Hinode mission, by using the revised version of the SPINOR inversion code \citep{Frutiger_2000} that allows to account for  the instrumental point-spread-functions \citep[][]{Vannoort_2012}. 
The observations covered a FoV of about 50$\times$150 arcesec$^2$, acquired with a spatial resolution of $\simeq$0.32 arcsec and spectral sampling of $\simeq$21 m{\AA} over about 5 minutes. The $T(\tau)$  they reported for core regions, defined as the image pixels  with magnetic field strength decreasing with height and absolute value $>$1000 G, shows comparatively higher values (up to $\sim$600 K) than obtained from our study; their values are closer to the empirical plage flux-tube model derived by \citet[][]{Solanki_etal1992}, at least up to $log\tau_{500}$=-1, compared to ours. Nevertheless, the results they obtained  for the average temperature of pixels representative of QS and magnetic field concentrations at $log\tau_{500}$=0, -0.9, -2.3 are in good agreement 
with the ones we obtained from the inversion of QS and PL data, within the deviation of results in the analysed subFOV.

\citet{Westendorp_2001} studied full-Stokes spectra taken at the  {Fe\,{\footnotesize I} line pair at 630 nm on  a sunspot region  with the slit ASP, by using the SIR code. 
They computed the RFs of Stokes-V to the perturbation of the magnetic field strength, and deduced a sensitivity range spanning between $log\tau_{500}$=0 and $log\tau_{500}$=-2.8, in good agreement with our estimation of the sensitivity range of the studied data. They also derived the confidence limits for the retrieved stratification of physical parameters from the computation of RFs, and verified that the errors obtained were in good agreement with Monte Carlo simulations. Their estimated formal errors are comparable to, but sligthly larger than our computed uncertainties for results from the PO and UM data.

\textbf{Analysis of key aspects of the studies described above 
shows that our work benefits from a higher spectral and spatial resolution of the analysed observations, and a wider dataset taken under excellent seeing conditions than considered in all previous studies. We also discuss our results with respect to a significantly larger set  of 1D models than   earlier presented in the literature.
Besides, within the computational uncertainty of results, all findings derived from our study are consistent with results of the above earlier works. Thus,  the results  derived from our study  can reasonably be assumed to  represent   the analysed regions.}

\section{Application to SI studies}
   
  \begin{figure}
\centering
{
\includegraphics[scale=.45, trim=.3cm 0.cm 0.cm 0.5cm,clip=true]{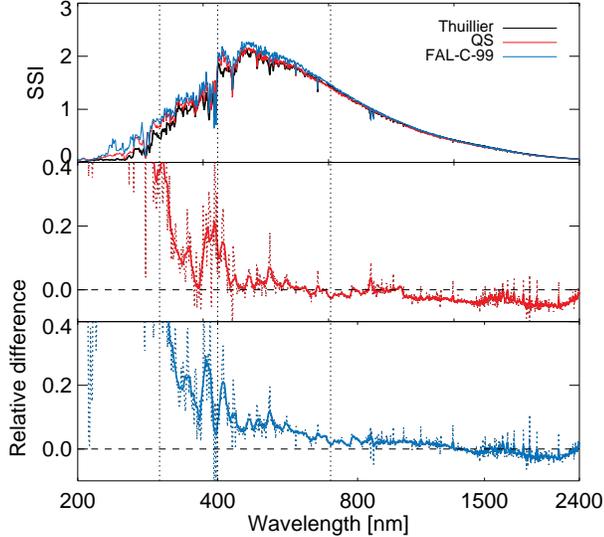}
}
\caption{Top panel: SI from 200 to 2400 nm calculated with our spectral synthesis performed on the QS and FAL-C-99 atmospheres representative of quiet Sun regions and measured reference data by \citet[][]{Thuillier_2004}. Middle and bottom panels: relative difference between the SI derived from the synthesis on the QS (middle) and FAL-C-99 (bottom) atmospheres with respect to the  reference data. SI data are given in [W m$^{-2}$ nm$^{-1}$] units. 
 All spectra were convolved with a 1 nm Gaussian kernel to account for the spectral resolution of available measurements in the visible range. The solid lines in middle and bottom panels show relative differences between the data convolved with a 10  nm Gaussian Kernel. Vertical lines mark the UV, NUV, Vis, and NIR bands.}
\label{figN}
 \end{figure}

  \begin{figure}
\centering
{
\includegraphics[scale=.45, trim=.3cm 0.cm 0.cm 0.5cm,clip=true]{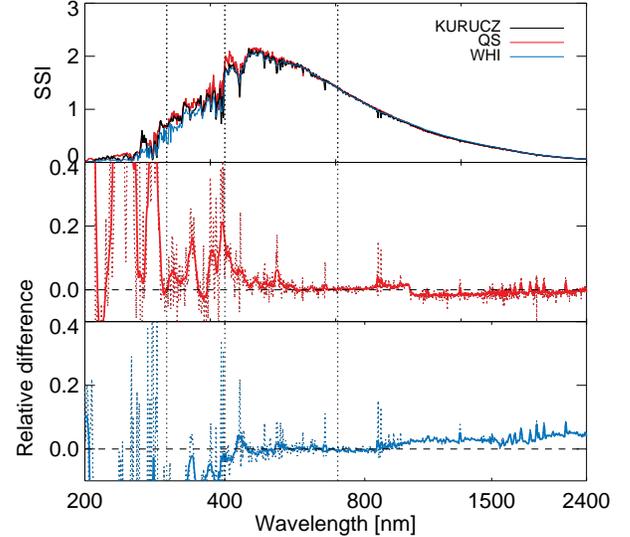}
}
\caption{Top panel: SI from 200 to  2400 nm derived from our spectral synthesis performed on the QS atmosphere,   calculated with the  Kurucz quiet Sun model employed in the SATIRE-S  SI model, and given by the  WHI reference data considered in the NLRSSI SI model. Middle and bottom panels: relative difference between the SI derived from the synthesis on the QS  atmosphere (middle) and considered in the NRLSSI model (bottom)  with respect to the one calculated with the Kurucz quiet Sun model employed in the SATIRE-S.  
See caption of Fig. \ref{figN}  for more details.}
\label{figN3}
 \end{figure}

  \begin{figure}
\centering
{
\includegraphics[scale=.5, trim=.3cm 2.8cm 0.cm 0.5cm,clip=true]{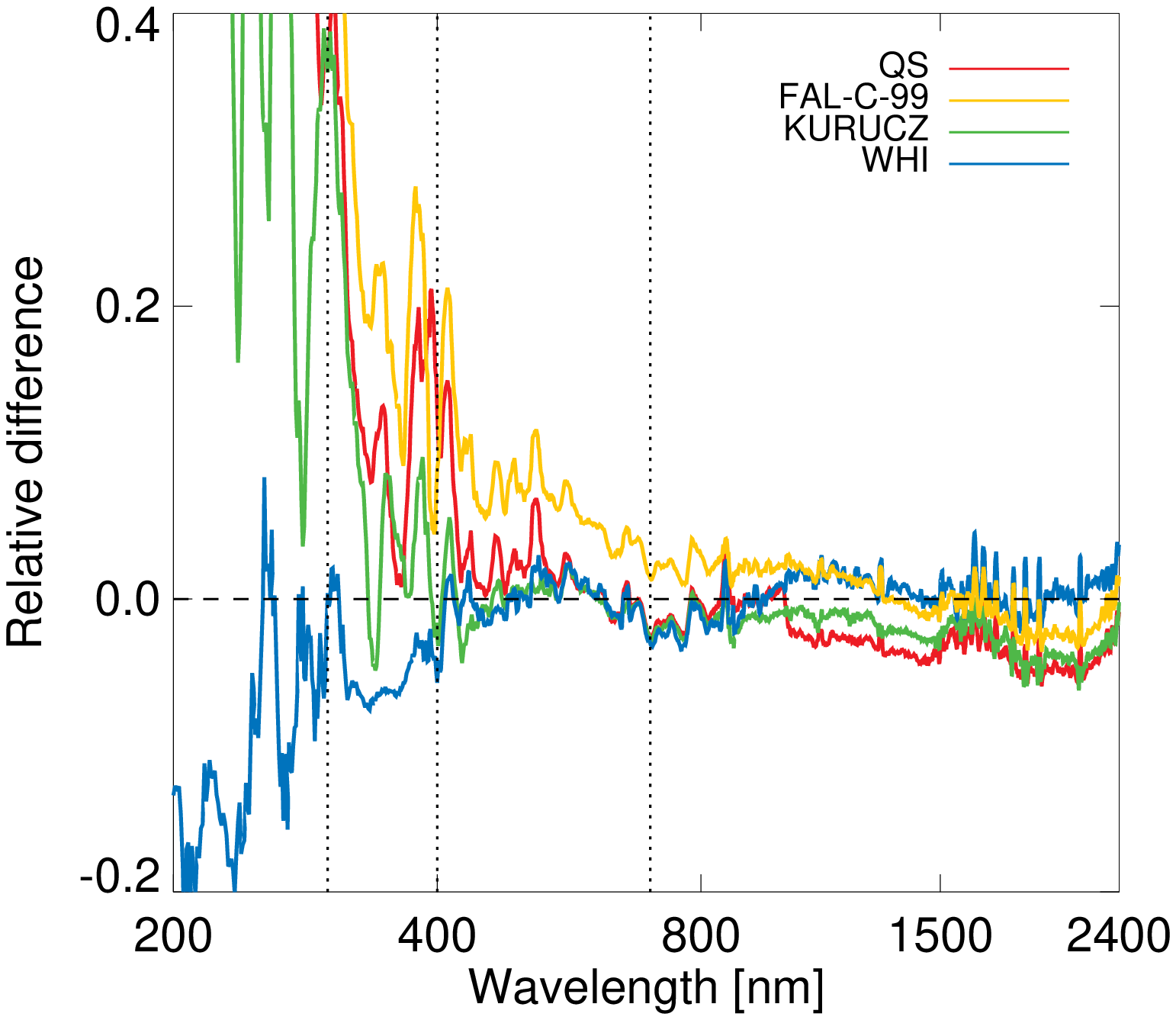}
}
\caption{Relative difference between the SI spectra derived from our synthesis on the  QS and FAL-C-99 atmospheres, and computed with  the Kurucz quiet Sun model employed in the SATIRE-S,   with respect to the reference data from  \citet{Thuillier_2004} and WHI by \citet{Woods_etal2009}, the latter considered in the NRLSSI model. See caption of Fig. \ref{figN} and Sect. 5 for more details. }
\label{figN4}
 \end{figure}
The set of 1D  models employed in semi-empirical SI reconstructions is of pivotal importance to  reproduce measured SI variations accurately \citep[see e.g.][]{Ermolli_etal2013}. 
It is thus interesting to test the accuracy  of the  observational-based atmosphere models derived from our study for possible application in SI models. 

In order to provide a preliminary  assessment of such accuracy, we computed the radiative flux emerging from our observational-based atmospheres and compared it with the one resulting from other 1D models employed in semi-empirical SI models, and with other available data that are described below. We performed the spectral synthesis for the wavelength range from 200 to 2400 nm  on the various analysed atmospheres  with the one-dimensional version of the RH code 
\citep{Uitenbroek_2001}, which solves the RT and statistical equilibrium equations under general NLTE conditions. We computed the emergent spectrum  with a 0.01 nm spectral resolution at nine lines of sight, spaced according to the zeroes of the Gauss-Legendre polynomials as a function of $\mu$. We then   convolved the spectra derived from the synthesis with a Gaussian kernel 1 nm wide, to roughly account for the spectral resolution of the SI data considered for comparison.

We applied standard NLTE  RH computations,  which consider contributions of Thomson scattering by free electrons, Rayleigh scattering by neutral hydrogen and helium atoms, and H$_2$ molecules. 
Other background opacity sources included were bound-free and free-free transitions of H$^-$ and neutral hydrogen, free-free transitions of H$_2$ and bound-free transitions of different metals. 
  The synthesis  was performed by computing populations of several atoms\footnote{Including H, C, O, Si, Al, Ca, Fe, He, Mg, Na, N,  S.} and of more than 10  molecules\footnote{Including H$_2$, H$^+_2$, C$_2$, N$_2$, O$_2$, CH, CO, CN, NH, NO, OH, SiO,  LiH, MgH.}.  We assumed the atomic line data  from Kurucz\footnote{kurucz.harvard.edu}.


Figure \ref{figN}  (top panel) shows the SI spectra derived from the synthesis performed on the QS and FAL-C-99 models representative of quiet Sun regions, compared to the ATLAS-3 reference spectrum by \citet{Thuillier_2004}; Figure \ref{figN}  (middle and bottom panels) display  the relative difference of the synthesized spectra with respect to the above reference data. The ATLAS-3 is a composite solar spectrum derived from analysis of various available measurements; it  is considered a standard reference for SSI covering the  UV (200 to 300 nm), near UV (NUV, 300 to 400 nm), visible (Vis, 400 to 700 nm) and near IR (NIR, 700 to 2400 nm) spectral regions. 
In the Vis and NIR bands, the median (standard deviation) relative difference between the SI spectrum derived from our QS atmosphere and reference data is $\simeq$0.8\% (4\%)  and -3\% (1.9\%), respectively; the median (standard deviation) relative difference between the compared spectra, however,  increases up to 13\% (14\%) and 85\% ($>$100\%) in the NUV and  UV regions, respectively. In the Vis and NIR bands, the median (standard deviation) relative difference between the SI spectrum derived from our synthesis on the FAL-C-99 model and reference data is $\simeq$6\% (4\%)  and -0.3\% (2\%), respectively; the median (standard deviation) relative difference  between these compared spectra increases up to 22\% (17\%) and 140\% ($>$100\%) in the NUV and  UV regions, respectively. 

Other available spectra from SI models show relative difference with respect to the ATLAS-3 data such as those reported above.   Among the various models developed to reproduce the SI variability, we considered the ones most representative of the two classes introduced in Sect. 1, the  proxy NRLSSI model \citep[][]{Lean_etal2000,Coddington_etal2016}  and semi-empirical SATIRE-S model \citep[e.g.][ and references therein]{Yeo_etal2014,Yeo_etal2014b}. 

  Figure  \ref{figN3}  (top panel) shows the SI spectrum derived from our synthesis performed on the QS atmosphere, compared to the SI spectrum computed with the  Kurucz quiet Sun model and to the WHI reference data by  \citet[][]{Woods_etal2009}, which give the quiet Sun spectrum in the SATIRE-S and NRLSSI SI models, respectively. 
  Figure \ref{figN3}  (middle and bottom panels) display  the relative difference between the SI spectrum from the QS atmosphere and the WHI data,  with respect to the SI spectrum calculated with the Kurucz quiet Sun model,  which  is employed} in the SATIRE-S,  the present-day, most-advanced semi-empirical SI model.  The SI spectrum derived from the QS atmosphere (described by WHI) shows median  relative difference to the spectrum  computed on the Kurucz atmosphere model of $\simeq$34\%, 6\% 0.7\%, -0.6\% (-32\%, -11 \%, -0.2\%, 3\%) in the  UV, NUV, Vis, NIR bands, respectively. 
 
 In order to highlight the main features of  the SI data compared above, we also show in Fig.  \ref{figN4} the relative difference between the various available spectra with respect to the ATLAS-3, after spectral convolution of the data with a  Gaussian kernel 10 nm wide, in order to display average trends over the various spectral regions. 
 In the Vis, the agreement among the compared spectra is very good, ranging from  -0.14\% (Kurucz model in SATIRE-S) to 6\% (FAL-C-99) for all the data analysed; median relative difference is 0.8\%, 6\%, -0.1\% -0.3\% for QS, FAL-C-99, Kurucz, and WHI, respectively.  In the NUV range, such  agreement decreases to $\simeq$13\% and 22\% for our QS and FAL-C-99 computations, while it decreases only to $\approx$-6\% and  6\% for the data considered in the NRLSSI and SATIRE-S models.
It is worth nothing that these latter  models  estimate the time- and wavelength-dependent contribution to SI from bright and dark magnetic features in quite different ways, but both  
 apply intensity offsets and some scaling in order that the reconstructed SI spectra match the absolute levels of some observed reference spectra. Besides, the spectral synthesis performed in  the SATIRE-S assumes LTE that fails to reproduce the SI below $\simeq$300 nm accurately. Outcomes from the SATIRE-S synthesis are  rescaled to  reference data  by \citet[][]{Woods_etal2009}.
 In contrast, the results of the spectral synthesis performed on our observational-based QS atmosphere  and the FAL-C-99 model  shown in Fig.  \ref{figN4} are taken as they are from our RH calculations, without applying any scaling to improve the match to  the reference data by  \citet{Thuillier_2004}.

 We also analysed the spectra derived from the synthesis on the observational-based BPs, PL, PO, and UM models,  with respect to the ones obtained from the synthesis on the FAL-(E, F, H, S)-99 and FAL-R-06 models. This comparison  shows higher relative differences in the UV than in the other spectral bands. Specifically, in the UV and NUV, the various spectra differ, on average, $\simeq$15\% and 80\% for models of bright and dark features, respectively; in the Vis (NIR)  they differ, on average, from 0.3 to 2\% (0.04 to 1.6\%) for models of bright features, and from 11 to $>$250\% (7 to 60\%)  for models of dark features. These differences affect calculations of the SSI based on the various compared models as summarised  in Table \ref{tab:tabssi}. We report in  Table \ref{tab:tabssi}
 the SSI  computed by integrating the various synthesized spectra from 200 to 2400 nm; we also show the relative difference between SI computed for models corresponding to same surface feature. The relative difference between the computed SSI ranges from -0.5\% (PL with respect to FAL-F-99 computations) to 93\% (P0 with respect to FAL-R-06 computations). Apart from these extremes,  
the best (worse) agreement between the compared quantities is  found for the SSI computed on the BPs and the FAL-E-99 atmospheres (PO and FAL-S-99 atmospheres).

\begin{deluxetable}{ccccc}
\tablecaption{Spectrally integrated flux from  200 to 2400 nm  computed from the observational-based  atmospheres and the FAL-(C, E, F, H, S)-99 and FAL-R-06 models.\label{tab:tabssi}}
\tablecolumns{5}
\tablenum{2}
\tablewidth{0pt}
\tablehead{
\colhead{Obs } & \colhead{SSI obs}  & \colhead{Model} & \colhead{SSI FAL}  & \colhead{Rel. diff.} \\
\colhead{region} &  \colhead{[$W/m^2$]}  & \colhead{label} &  \colhead{ [$W/m^2$]} & \colhead{[\%]}}
   \startdata 
QS & 1354.60 & C & 1416.85    &   4.6 \\
BP & 1404.10 & E & 1426.06    &  1.6 \\
PL & 1453.55 & F & 1446.77    &   -0.5 \\
PL & 1453.55 & H & 1510.46    &   3.9 \\
PO & 624.24 & R & 1205.53    &  93 \\
PO & 624.24 & S & 303.36    &  -51 \\
UM & 273.57 & S & 303.36    &  11 \\  
\enddata
\end{deluxetable}

It is worth nothing that the values in Table   \ref{tab:tabssi} result from the synthesis performed on the atmosphere models presented in Sect. 3. In our study, we assumed that these  data represent atmosphere regions employed in SI reconstructions satisfactorily.   However, the subFOVs analysed in our observations represent only a minute fraction of the solar disc at a given time, while the solar regions the data  are assumed to represent can cover significant fractions of the solar disc and show different brightness in time.
For some surface features,  the results derived from our study  may not reflect the properties of the modeled atmosphere accurately. This is  especially the case when rather inhomogenous regions are considered. For example, we show in Fig. \ref{figN2} results from the synthesis performed on the three PL regions marked with red and blue boxes in Fig. \ref{fig1_fov}. The $T(\tau)$  derived from the blue marked  regions show  slightly different average values than obtained from analysis of the red marked region. In particular, the $T(\tau)$  of the PL region discussed in Sect. 3  lies between the ones obtained from the other two analysed PL areas. 
The SI spectra derived from the synthesis on  the three PL atmospheres differ on average  $\simeq$10\%, 5\%, and $<$5\%  in the UV, NUV, and Vis ranges, respectively. When entered in SI models, these differences translate in   SSI estimated values that differ from 1.1\% to 2.6\%.

    \begin{figure}
\centering
{
\includegraphics[scale=.58, trim=2.7cm 14.7cm 0.5cm 3.cm,clip=true]{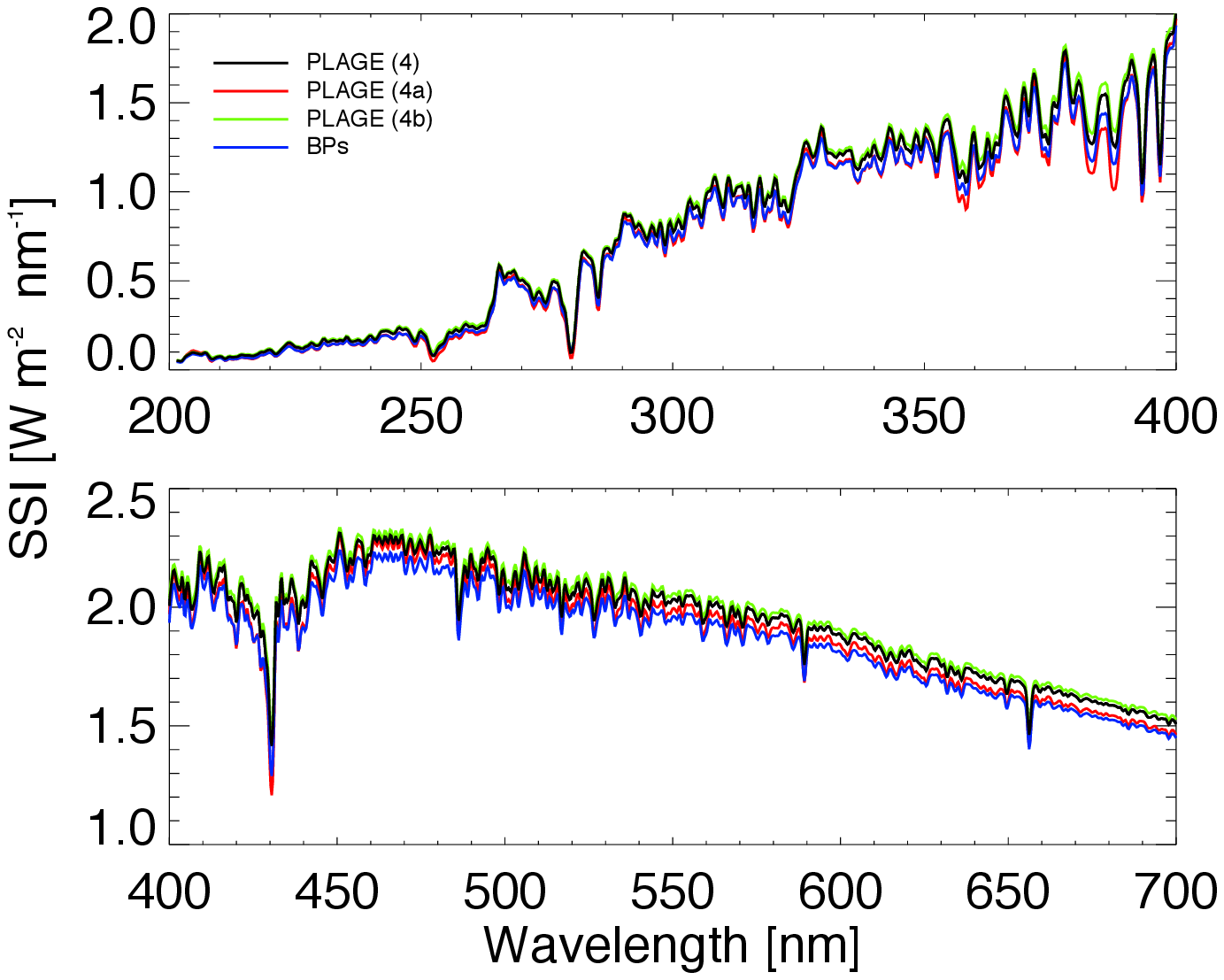}
}
\caption{SI  from  200 to 400 nm (top), and from 400 to 700 nm (bottom), derived from the synthesis on the atmosphere models of three PL regions and BPs area  shown in Fig. \ref{fig1_fov}. 
}
\label{figN2}
 \end{figure}

\textbf{The above results encourage us to further 
investigate the accuracy of entering   atmosphere models derived from spectro-polarimetric observations  in SI estimates. Indeed, 
the Vis and NIR SSI  
synthesized on the atmosphere model derived from our QS observations differs on average from the ATLAS-3 and WHI reference data less than 2.5\%, and -0.14\% from the spectrum  computed on the  Kurucz quiet Sun atmosphere employed in the SATIRE-S model. 
 Besides, the lower  agreement  reported above for synthesis results of the NUV and UV bands, below 400 nm,  is fully consistent with  
the limited range of atmospheric heights sampled by the data analysed in the present study, which  spans from the low to the high photosphere only, thus limiting  the reliability of our spectral synthesis results for the SI originating from higher atmospheric  heights.
On the other hand, 
in the 1000 to 2400 nm spectral region, the SI derived from our synthesis on the QS atmosphere underestimates (-3.5\%$\pm$1.5\%) the reference data, as a consequence of the clear SSI drop seen at about 1000 nm. This drop  challenges our synthesis calculations of the H$^-$ opacity for the NIR range. Indeed, in the same spectral region, results derived from our synthesis on the FAL-C-99 atmosphere by assuming the H population data available for that  model differ, on average, only  -0.7\%   with respect to the reference. 
However, it is also worth noting that several recent NIR SI measurements  show a systematic negative difference (of about 8\%) with respect to the ATLAS-3 reference composite, see e.g. \citet[][]{Weber_2015}, in agreement  with our findings.}

\section{Discussion and Conclusion}
\label{discus_c4}

We found that the average temperature stratification derived from the data inversion of the various analysed regions agrees well with that represented by the corresponding 1D model, both qualitatively and quantitatively,  but for pore data, which  exhibits a different trend at all atmospheric heights compared to the 1D models representative of umbral regions. 
%
This result is not surprising, since previous studies already strengthened the linear dependence of umbral core brightness on their size \citep[e.g.,][]{Collados_etal1994, Mathew_2007}. These latter results however suggest that the 1D models employed in current SI reconstructions, 
may inaccurately represent the temperature stratification of dark, magnetic regions which are neither spatially extended nor characterized by strong magnetic fields as typical umbral regions. Moreover, such features are not accounted for in the   atmosphere models of dark structures employed in SI reconstructions.  
 Our results  also suggest that pixel-by-pixel inversion of high-resolution observations allows to retrieve atmosphere models that possibly better account for the contribution of the smaller-scale features in the studied  FoV, than obtained from analysis of less resolved observations. This is particularly interesting, since SI cyclic variations are closely linked to the evolution of small-scale, strong-field magnetic features. 
 


Our preliminary investigation of the accuracy of potentially entering  the various atmosphere models derived from our study in SI estimates gave encouraging results. Indeed, 
the SI spectrum from 400 to 2400 nm synthesized on the atmosphere model derived from our QS observations differs on average $\simeq$2.2\%  from the ATLAS-3 reference data by \citet[][]{Thuillier_2004}, and  $\simeq$-0.14\% from the spectrum  computed on the  Kurucz quiet Sun atmosphere employed in  the  SATIRE-S SI model.  In the same spectral range,  the median difference between the quiet Sun spectra  considered in   the SATIRE-S and  NRLSSI SI models is   2.7\%. 
It is worth recalling that the NRLSSI is a regression-proxy model, while the SATIRE-S is a more physics-based model that  includes spectral synthesis computations. 
At all wavelengths  analysed in our study, the spectrum derived from the Kurucz atmosphere employed  in the SATIRE-S is closer to the one derived from our QS observations than the spectrum considered in the NRLSSI SI model.

 The  results presented above  encourage us to refine our RH calculations on  observational-based atmospheres for  potential use in  SI models. In particular, the significantly lower agreement we found between our synthesis results and reference data in the  NUV and UV bands, than in the Vis and NIR, shows that further work is needed to improve e.g. some atomic data employed in our calculations. 
Besides,  to properly enter synthesis results of  observational-based models in SI reconstructions, a more detailed study is also required to account for  the center-to-limb dependence of the  intensity emerging from features observed at different positions on the solar disc, and for  
the different brightness of each magnetic feature depending on the magnetic filling factor.  
However, 
preliminary  tests of the accuracy of the outcomes derived from the present study, by using data representative of other solar  regions that also cover wider ranges of atmospheric heights than discussed above, have given promising  results that motivate us to
further work for the exploitation of atmosphere models derived from inversion of spectro-polarimetric observations in SI reconstructions.

\acknowledgments

The authors  wish to thank  the referee, Kok Leng Yeo, for fruitful comments that helped them improve the manuscript.
They are also thankful to
Gianna Cauzzi, Serena Criscuoli, Valentina Penza, Hector Socas Navarro, Sami Solanki, and Han Uitenbroek  for useful discussions and advice for the data inversion and synthesis computations, and  to the International Space Science Institute, Bern, for the opportunity to discuss this work with  the team "Towards a Unified Solar Forcing Input to Climate Studies" lead by Natasha Krivova.  
This study
received funding from the European Unions Seventh Programme for Research, Technological
Development and Demonstration, under the Grant Agreements of the SOLARNET (n 312495, www.solarnet-east.eu) and SOLID
(n 313188, projects.pmodwrc.ch/solid/) projects.  This work was also supported by the Istituto Nazionale di Astrofisica (PRIN-INAF-2014) and Italian MIUR (PRIN-2012). 
\bibliography{alice_bib.tex}
\bibliographystyle{aasjournal}

\end{document}